\documentclass[11pt]{article}
\usepackage[dvips]{epsfig}
\usepackage{cite}
\usepackage{wasysym} 
\newcommand{\lsim}{\stackrel{<}{_\sim}}

\newcommand{\ee}{$e^+e^-\: $}

\usepackage{cite}

\begin{document}

\pagestyle{empty}

\vspace*{2.5cm}                


\vspace*{1.5cm}



\begin{center}
\LARGE{\bf\boldmath ILC Beam Energy Measurement by means of 
                    Laser Compton Backscattering \unboldmath}
\end{center}

\vspace{1.0cm}
\large
\begin{center}
N. Muchnoi$^1$,
H.J. Schreiber$^2$ and M. Viti$^2$
\end{center}

\vspace{0.3cm}
\bigskip \bigskip  
\begin{center}
\small
$^1$ Budker Institute for Nuclear Physics, Novosibirsk, Russia \\ [2mm]

$^2$ Deutsches Elektronen-Synchrotron DESY, D-15738 Zeuthen, Germany \\ [2mm]

\normalsize
\end{center}

\vspace{2.0cm}
\pagestyle{plain}

\pagenumbering{arabic}

%
\begin{center}
\section*{Abstract}
\end{center}

\noindent 
A novel, non-invasive method of measuring the beam energy at the
International Linear Collider is proposed. Laser light collides head-on
with beam particles and either the energy of the Compton scattered electrons
near the kinematic endpoint is measured or the positions
of the Compton backscattered $\gamma$-rays, the edge electrons
and the unscattered beam particles are recorded.
A compact layout for the Compton spectrometer is suggested. 
It consists of a bending magnet
and position sensitive detectors operating in a large
radiation environment. Several options for high spatial resolution detectors
are discussed. Simulation studies support the use of an infrared or green laser
and quartz fiber detectors to monitor 
the backscattered photons and edge electrons. Employing a cavity monitor,
the beam particle position downstream of the magnet can be recorded
with submicrometer precision. Such a scheme provides
a feasible and promising method
to access the incident beam energy with precisions 
of $10^{-4}$ or better on a bunch-to-bunch basis
while the electron and positron beams are in collision.

\newpage

\section {\boldmath Introduction \unboldmath}

A full exploitation of the physics potential 
of the International \ee Linear Collider (ILC)
must aim to control the absolute incoming beam energy, $E_b$,
to an accuracy of $10^{-4}$ or better. 
Precise measurements of $E_b$ is a critical component
to measuring the center-of-mass energy, $\sqrt{s}$, as it sets 
the overall energy scale of the collision process. 
Good knowledge of $E_b$, respectively, $\sqrt{s}$ had always been
a tremendous advantage for performing precise measurements of particle masses
and the differential dependence of the luminosity, $d{\cal{L}}/d\sqrt{s}$.
At circular machines, for example at the Large Electron Positron Collider  (LEP),
beam energy determination using resonant depolarization allowed
an exquisite measurement of the $Z$ boson mass, $M_Z$, to an uncertainty
of 2 MeV or 23 parts per million (ppm). At the ILC, however, the resonance
depolarization technique cannot be applied  and different methods 
have to be employed.

\vspace{3mm}
\noindent
A beam position monitor-based magnetic spectrometer is considered
to be a well established and promising device to achieve this goal \cite{ILC_spec}.
By means of this method, the energy is determined by measuring the deflection angle
of the particle bunches utilizing beam position monitors (BPMs)
and the field integral, $B \equiv \int B d\ell$, mapped to high resolution.
The performance of such a spectrometer has been demonstrated at LEP at CERN, where
an in-line spectrometer with button monitors was successfully operated
to cross-check the energy scale for W mass measurements \cite{LEP2}.
A relative error on $\sqrt{s}$ of 120-200 ppm has been achieved, thanks to
careful cross-calibrations using resonant depolarization.
While the primary beam energy determination was based on the NMR magnetic model,
its validity was, after corrections
for different sources of systematic errors,
verified by three other methods: the flux-loop, which is sensitive
to the bending fields of all dipole magnets of LEP, a BPM-based spectrometer
and an analysis of the variation of the synchrotron tune with the total RF voltage.
At SLAC, a synchrotron radiation-stripe (WISRD) based bend angle
measurement in the extraction line 
of the \ee interaction point (IP) was performed to access $\sqrt{s}$ \cite{SLAC264}.
The results obtained were, however, subject to corrections by 46$\pm$25 MeV,
i.e. by 500 ppm, utilizing the precise value of $M_Z$ from LEP.
All these trials to measure the energy evidently emphasize
the following lesson:
more than one technique should be applied for precise $\sqrt{s}$ determinations
 and cross-calibration of the absolute energy scale is mandatory.
In the past, novel suggestions, see e.g. \cite{suggestions},
were proposed for the ILC and some of them were evaluated in detail. 
Within the next years
some consensus should, however, be arrived at as to which methods are most promising
of being complementary to the canonical BPM-based spectrometer technique.

\vspace{3mm}
\noindent
In this note we propose a new non-destructive approach to perform beam energy measurements
using Compton backscattering of laser light by beam particles.
The energy at the kinematic endpoint (edge) of the Compton electrons depends
on $E_b$, and its direct measurement provides the beam energy.
Alternatively, recording the positions
of the Compton backscattered photons and the edge electrons together with 
the position of the unscattered beam particles allows to infer
the primary beam energy with high precision.

\vspace{3mm}
\noindent
Compton backscattering experiments have been performed with great success
at circular low-energy accelerators. At the Taiwan Light Source \cite{Taiwan},
the beam energy of 1.3 GeV was determined with an uncertainty
of 0.13\%. At BESSY I and II \cite{BESSY} with 800 MeV, respectively,
900 or 1700 MeV electron energy, $E_b$ was found to be in very good
agreement with the resonant depolarization values, and
at Novosibirsk \cite{Novosibirsk} an accuracy of 60 keV was obtained
for beam energies between 1.7 and 1.9 GeV. In all these experiments,
beam particles were collided head-on with photons from a $CO_2$ laser.
The maximum energy of the forward going Compton $\gamma$-rays was measured
with high-purity germanium detectors and converted
into the central primary beam energy.

\vspace{3mm}
\noindent
This method, however, is not practicable at the ILC since 
precise $E_b$ measurements require collective
and accurate information on Compton backscattered particles using
large event rates per bunch crossing.
The selection of the photon with highest energy
and its precise measurement out of a large number
of $\gamma$-rays cannot be performed.
In particular, within bunch crossings of picosecond duration
a calorimetric approach (with demanding calibration performance)
to access the maximum $\gamma$-ray energy is unable to
resolve the individual backscattered photons.
Therefore, the method proposed for the linear collider is different 
and can be summarized as follows:
after crossing of laser light with beam electrons,
a bending magnet separates the forward collimated Compton
photons and electrons as well as the non-interacting beam particles 
such that downstream of the dipole high spatial resolution detectors
measure the positions of the backscattered photons
and the edge electrons, i.e. of electrons
with smallest energy or largest deflection. If these measurements
are either combined with the magnetic field integral or
with the position of the unscattered beam particles, the beam energy
can be inferred.

\vspace{3mm}
\noindent                             
At the ILC, laser Compton backscattering off beam particles
is also suggested to probe other properties of the beam,
such as the transverse profile \cite{Blair} or the degree of polarization  \cite{Schueler}.

\vspace{3mm}
\noindent
In the past, laser backscattered $\gamma$-rays off
relativistic electrons were employed as a highly promising alternative
of producing intense and directional quasi-monochromatic 
(polarized) photon beams to investigate
photonuclear reactions \cite{beams}, to calibrate detectors
or to record medical images. 

\vspace{3mm}
\noindent
The paper is organized as follows. 
Sect.2 describes the basic properties of the Compton
scattering process, emphasizing features which are relevant for $E_b$ determinations.
In Sect.3 an overview of the proposed method is presented.
Two schemes to perform beam energy measurements are suggested and
precisions achievable are discussed. This will be followed by a setup proposal,
a layout of the vacuum chamber,
a suitable dipole suggestion, a possible laser system
and detector options to measure
the photon and edge electron positions
as well as that of the unscattered beam. Simulation studies 
support the feasibility and reliability of the concepts proposed.
Processes beyond the Born approximation in the laser crossing region 
such as nonlinear effects, multiple scattering, higher order QED contributions
and pair production background are also discussed.
This is followed by a discussion of potential sources of errors
affecting the measurement of $E_b$.
Possible locations of a Compton energy spectrometer within the
ILC beam delivery system \cite{BDS_present} are summarized at the end of Sect.3.
Sect.4 contains the summary and conclusions.

\section {\boldmath The Compton Scattering Process \unboldmath}

\begin{figure}[ht]
\center
\includegraphics[height=8cm,width=12cm]{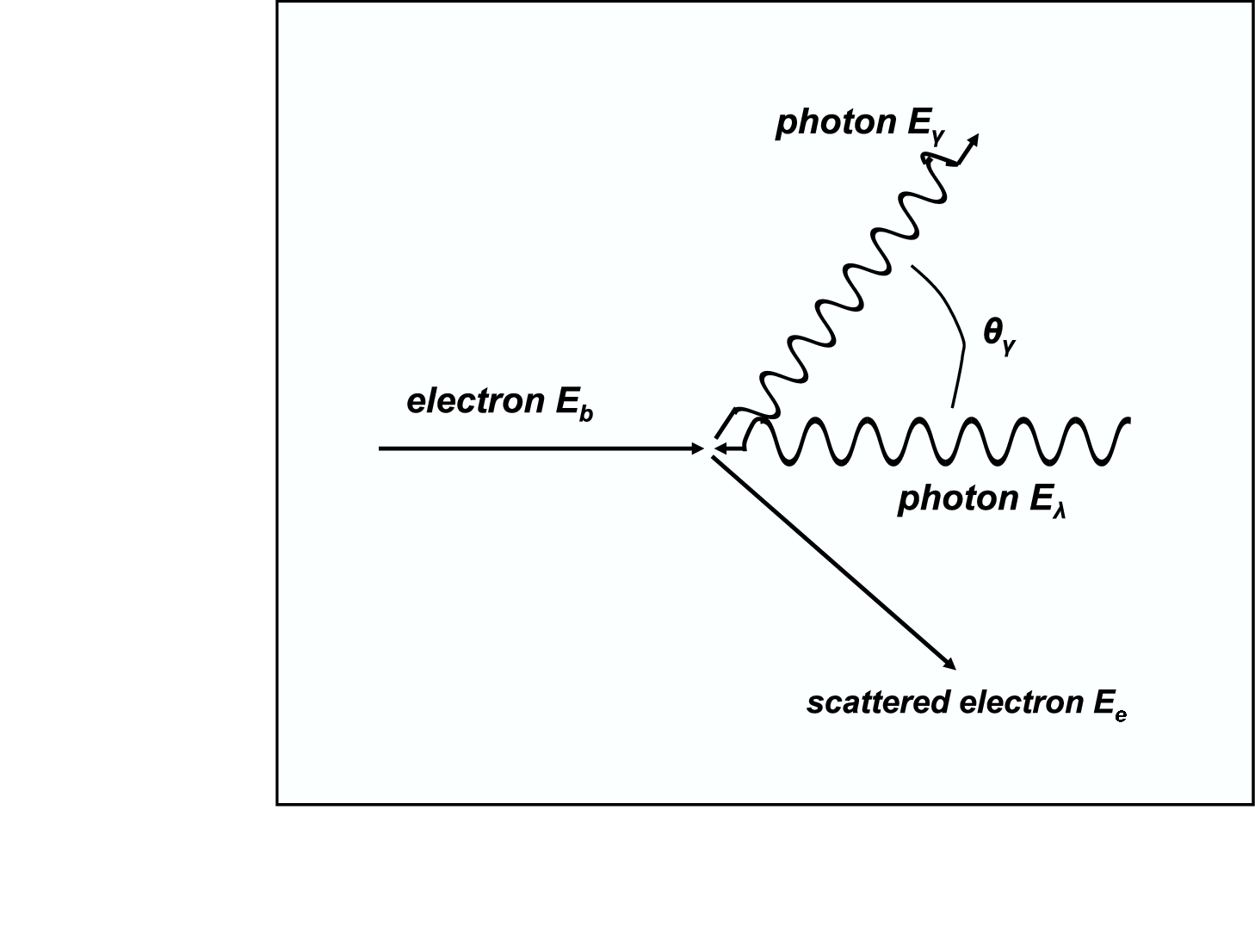}
\vspace{-10mm}
\caption{The kinematics of Compton scattering in the lab frame.
  The energies of the colliding electron
  and laser photon are denoted as $E_b$ and $E_{\lambda}$, respectively.
  $\theta_{\gamma}$ is the scattering angle between the initial electron
  and final state photon. The angle $\alpha$ is not shown.}
\label{fig:kin_principle}
\end{figure}

Feenberg and Primakoff \cite{Feenberg} proposed in 1948 the kinematics formula for
the two-to-two Compton scattering process
\begin{equation}
 e + \gamma \rightarrow e^{\prime} + \gamma^{\prime}~,
\end{equation}
which is shown in Fig.~\ref{fig:kin_principle} in the lab frame.
The initial photon and electron energies are expressed as $E_{\lambda}$
and $E_b$, respectively, while the energy of the backscattered photon is expressed
as $E_{\gamma}$ and that of the electron as $E_e$.
$\theta_{\gamma}$ is the scattering angle between the initial electron
and the laser direction.
The angle $\alpha$, not shown in Fig.~\ref{fig:kin_principle},
is defined between the incident electron\footnote{Throughout the paper,
the incident beam particle denoted so far as electron means either electron
or positron.} and the laser direction.

\vspace{3mm}
\noindent
Throughout this study, the convention is used where
the positive z-axis is defined to be the direction of the incident beam,
the x-axis lies in the horizontal or bending plane and the y-axis points 
to the vertical direction such that a right-handed coordinate system is obtained.

\subsection {\boldmath Compton Scattering Cross Section \unboldmath}

\begin{figure}[ht]
\center
\includegraphics[width=8.5cm]{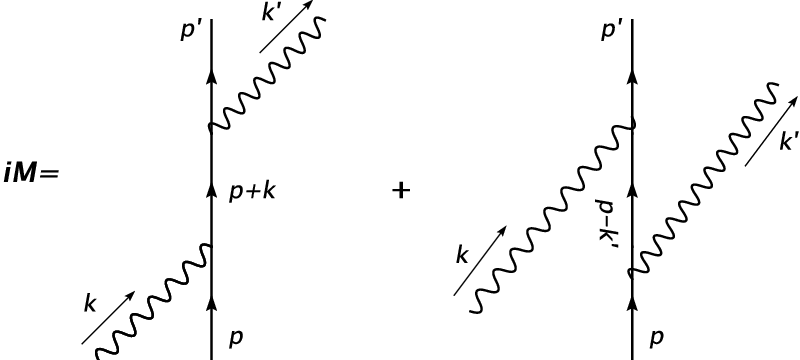}
\caption{Leading order Feynman diagrams contributing 
  to the Compton cross section.}
\label{fig:feyngraphs}
\end{figure}

In order to calculate the cross section for Compton scattering 
(in Born approximation) we start from the matrix element which
involves two Feynman diagrams as shown in Fig.~\ref{fig:feyngraphs}.
Since the ILC is also planned to operate
with polarized electrons/positrons, it is advantageous to consider
the most general case by including possible spin-states of the incident particles.

\vspace{3mm}
\noindent
In the lab frame, the Compton kinematics are characterized by the dimensionless variable
\begin{eqnarray}
 x = \frac{4 E_b E_{\lambda}}{m^2} \cdot \cos^2(\alpha/2) \sim
    \frac{4 E_b E_{\lambda}}{m^2}             \label{equ:x_variabel}
\end{eqnarray}
and the normalized energy variable
\begin{eqnarray}
 y = 1 - \frac{E_e}{E_b} = \frac{E_{\gamma}}{E_b}~.
\end{eqnarray}
Applying QED Feynman rules,
the spin-dependent differential cross section is after summing over the
non-interesting spin and polarization states of the final state particles
\begin{eqnarray}
 \frac{d\sigma}{dy} = \frac{2\sigma_0}{x} [\frac{1}{1-y} + 1 - y -4r(1-r)
                    + P_e \lambda r x (1-2r)(2-y)]~,
\end{eqnarray}
where $P_e$ is the initial electron helicity (-1 $\leq P_e \leq$ +1), $\lambda$ the
initial laser helicity (-1 $\leq \lambda \leq$ +1),
$ r = \frac{y}{x(1-y)}$ and $\sigma_0 = \pi r^2_0$ = 0.2495~barn,
with $r_0$ the classical electron radius.

\begin{figure}[ht]
\center
\includegraphics[height=8cm,width=12cm]{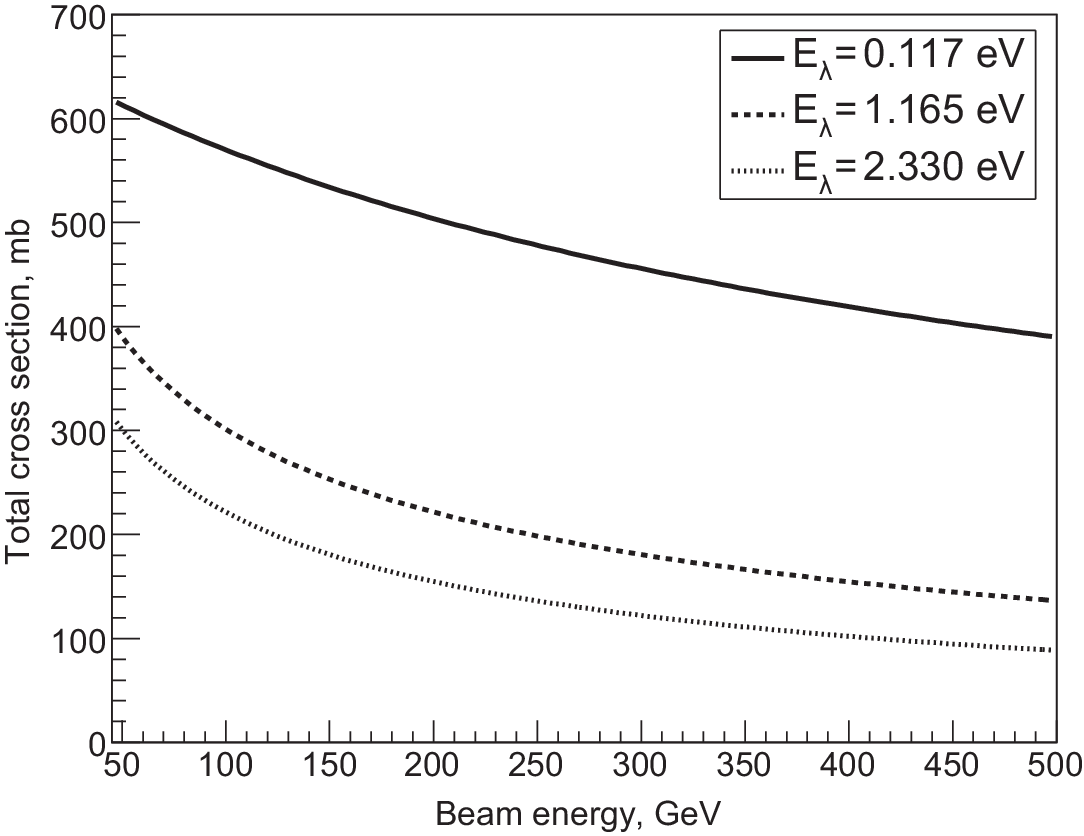}
\caption{Compton backscattering cross section versus beam energy for three
  laser energies.}
\label{fig:total_cross_section}
\end{figure}

\vspace{3mm}
\noindent
Fig.~\ref{fig:total_cross_section} shows the unpolarized Compton cross section 
as a function of the beam energy for three laser energies, $E_{\lambda}$ = 0.117,
1.165 and 2.33 eV. At all incident energies,
the $CO_2$ laser with $E_{\lambda}$ = 0.177 eV 
provides the largest cross sections,
while the $Nd:YAG$ laser (with $E_{\lambda}$ = 1.165 or 2.33 eV) cross sections
are significantly smaller. For example, at 250 GeV the $CO_2$ cross section is
more than two times larger than the $Nd:YAG$ laser values.

\begin{figure}[ht]
\center
\includegraphics[height=8cm,width=12cm]{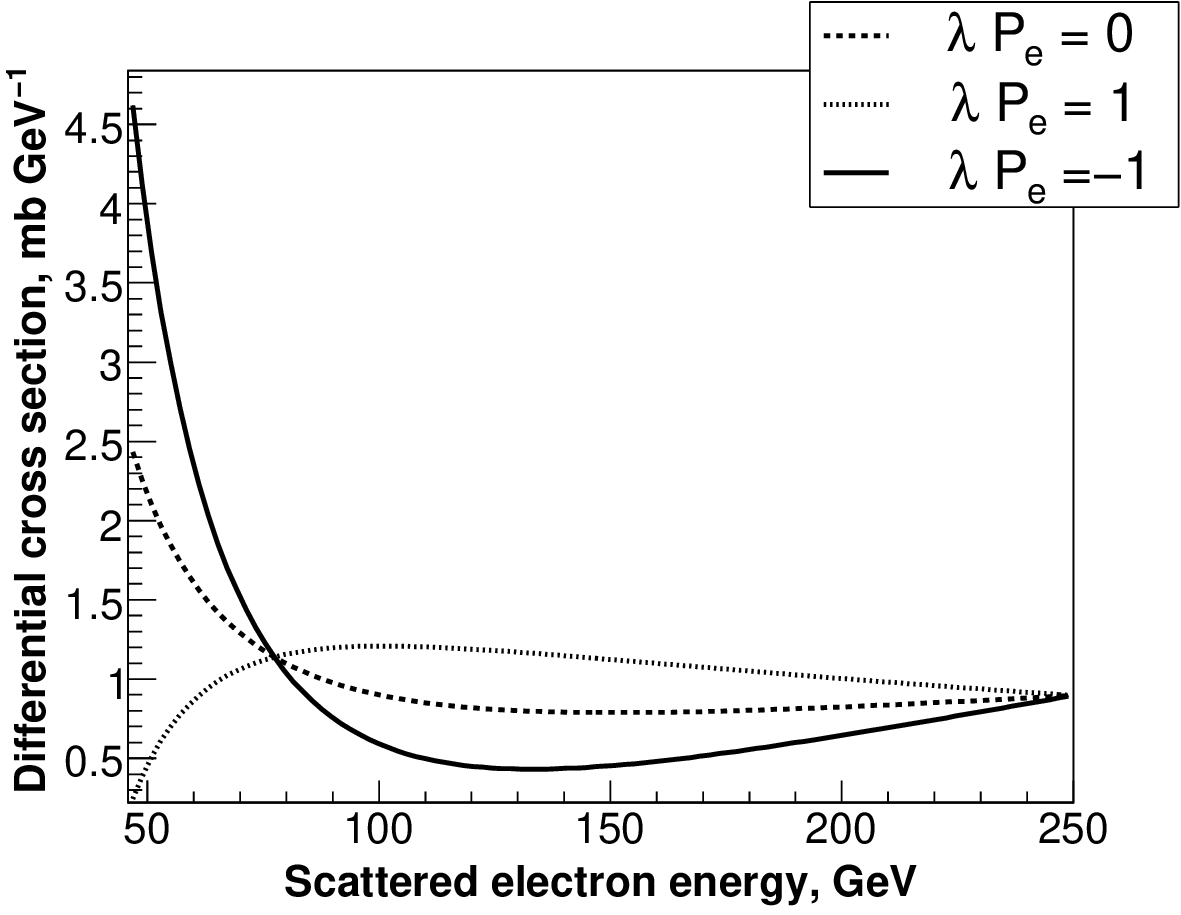}
\caption{Compton backscattering cross section for three polarization configurations
  versus scattered electron energy for an infrared $Nd:YAG$ laser at 250 GeV.}
\label{fig:polarization_cross_section}
\end{figure}

\vspace{3mm}
\noindent  
We also note that for the polarization configuration
$P_e\lambda$ = -1, the cross section close to the electron's kinematic endpoint
is enhanced by typically a factor two,
while for the configuration $P_e\lambda$ = +1 the edge Compton cross section
vanishes. This behavior is shown in Fig.~\ref{fig:polarization_cross_section},
where for the three cases, $P_e\lambda$ = -1, $P_e\lambda$ = +1 and unpolarized,
the cross section is plotted as a function of the scattered electron
energy for the infrared $Nd:YAG$ laser at 250 GeV.
For polarized electrons
the favored spin configuration $P_e\lambda$ = -1 can always be achieved
by adjusting the laser helicity $\lambda$.

\subsection {\boldmath Properties of the Final State Particles \unboldmath}

\begin{figure}[ht]
\begin{minipage}[b]{0.45\linewidth}
 \centering
  \includegraphics[height=6cm,width=8.0cm]{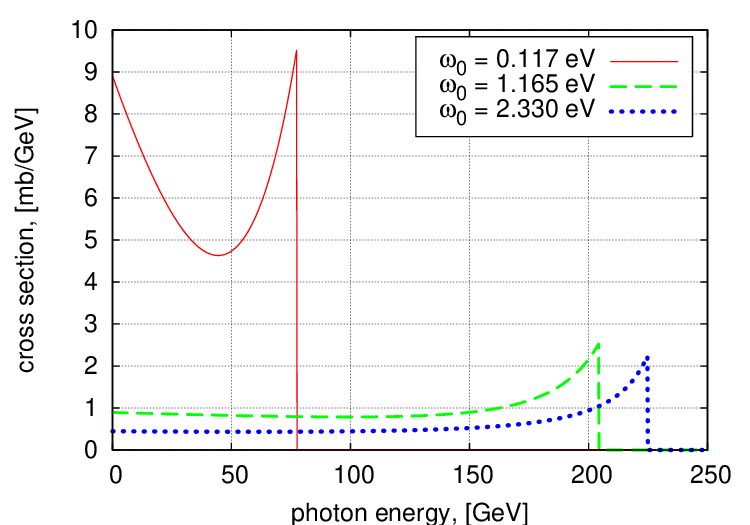}
\end{minipage}
\begin{minipage}[b]{0.5\linewidth}
 \centering
  \includegraphics[height=6cm,width=9.0cm]{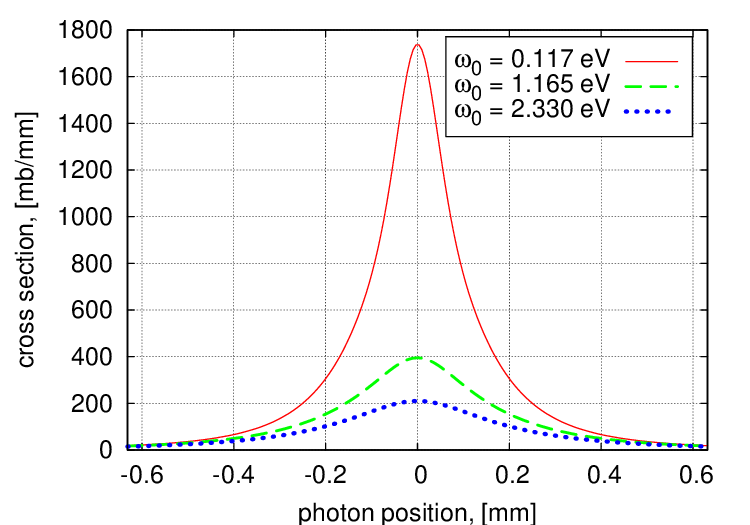}
\end{minipage}
\caption{Energy and x-position of backscattered photons
  for three laser energies, $\alpha$ = 8 mrad and $E_b$ = 250 GeV.
  The photon position is determined at a plane 50 m downstream of the Compton IP.}
\label{fig:photon_energy_position}
\end{figure}

After scattering, the angles of the Compton scattered photons and electrons relative
to the incoming beam direction are 
\begin{eqnarray}
 \theta_{\gamma} = \frac{m}{E_b} \cdot \sqrt{\frac{x}{y} - (x+1)}~, 
 ~~~~~~\theta_e = \theta_{\gamma} \cdot \frac{y}{1-y}~,  \label{equ:theta_gamma_theta_e}
\end{eqnarray}
and the $\gamma$-ray emerges with an energy of
\begin{equation}
 E_{\gamma} = E_{\lambda} \cdot \frac {1 - \beta \cos \alpha}
   {1 - \beta \cos \theta_{\gamma} + \frac {E_{\lambda}
   (1 - \cos (\theta_{\gamma}-\alpha))} {E_b}}
\end{equation}
at small angle $\theta_{\gamma}$, with $\beta$ the beam electron velocity
divided by the speed of light and
$\alpha$ the angle between the laser light and the incident beam.
$E_{\gamma}$ ranges from zero to some maximum value 
\begin{equation}
 E_{\gamma,max}=\frac{E_b^2}{E_b+\frac{m^2}{4\omega_0}}~,~~~
    ~~~~\omega_0 = E_{\lambda} \cdot \cos^2(\alpha/2)~.            \label{equ:e_gamma_max}
\end{equation}
Fig.~\ref{fig:photon_energy_position} illustrates 
the energy and x-position of the scattered photons at a plane 
located 50 m downstream of the Compton IP for three laser energies,
$\alpha$ = 8 mrad and $E_b$ = 250 GeV.
According to eq.(\ref{equ:e_gamma_max}), $\gamma$-rays
with highest energy travel exactly forward.

\vspace{3mm}
\noindent
The energy of the Compton electrons is determined
by energy conservation. The maximum energy of the Compton photon
is related to the minimum (or edge) energy of the scattered electron, $E_{edge}$, 
via
\begin{eqnarray}
 E_{edge} = E_b + E_{\lambda} - E_{\gamma,max} = 
            \frac{E_b}{1 + \frac{4 E_b \omega_0}{m^2}}~,                 \label{equ:edge}
\end{eqnarray}
if $E_{\lambda}$ is neglected.
The electron scattering angle $\theta_e$,
given in eq.(\ref{equ:theta_gamma_theta_e}), approaches zero as $\theta_{\gamma}$
becomes smaller. Thus, in the region of smallest electron energy,
the region of our interest, both the scattered
electrons and photons are generated at very small angles.

\begin{figure}[ht]
\center
\includegraphics[height=7cm,width=12cm]{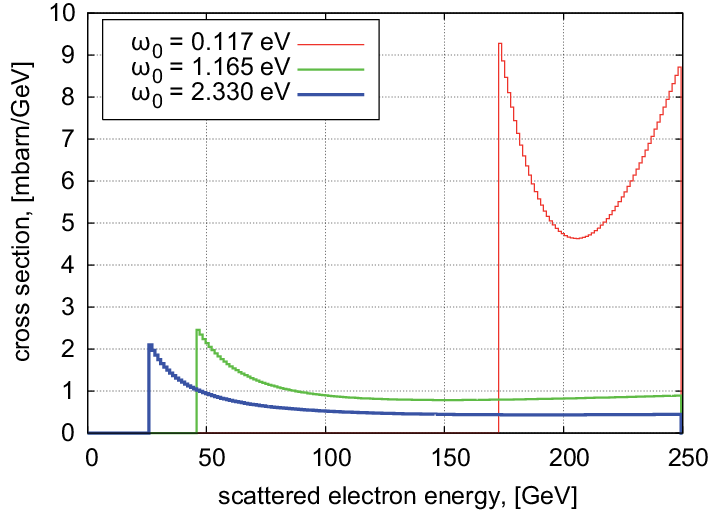}
\caption{Compton backscattering cross section versus electron energy for three
  laser wavelengths at 250 GeV.}
\label{fig:cross_section}
\end{figure}

\vspace{3mm}
\noindent
Fig.~\ref{fig:cross_section} shows the unpolarized Compton cross section as a function
of the scattered electron energy for three laser energies at 250 GeV.
The $CO_2$ laser (with an energy of 0.177 eV)
provides the most pronounced edge cross section,
while the $Nd:YAG$ laser (with $\omega_0$ = 1.165 or 2.33 eV)
cross sections are significantly smaller. 
At the electron's edge position, $E_{edge}$,
both $Nd:YAG$ lasers provide cross sections of similar size,
with edge energy values relatively close to each other.

\begin{figure}[ht]
\center
\includegraphics[height=8cm,width=12cm]{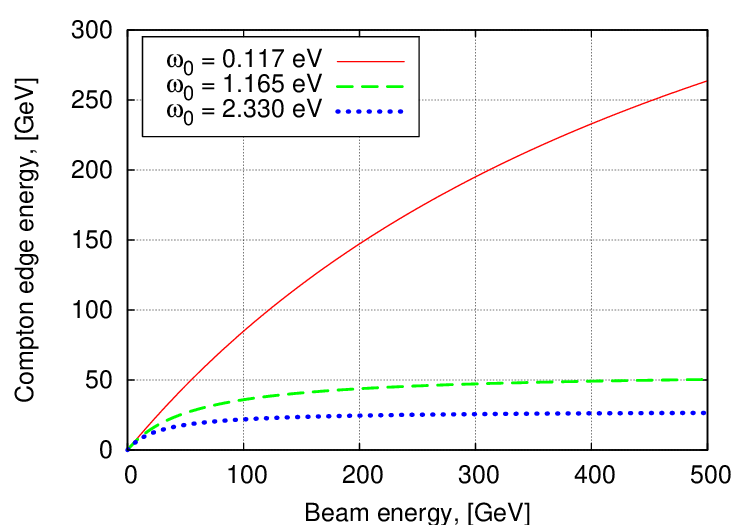}
\caption{Edge energy of Compton electrons as a function of $E_b$
  for a $CO_2$, infrared and green laser.}
\label{fig:laser_edge}
\end{figure}

\vspace{3mm}
\noindent
Since one of the proposed methods for measuring the beam energy utilizes
the variation of the edge energy on $E_b$, see eq.(\ref{equ:edge}),
we present in Fig.~\ref{fig:laser_edge} the edge energy dependence on $E_b$
for three laser wavelengths.
As can be seen, the derivative $dE_{edge}/dE_{\lambda}$
or the slope, respectively, sensitivity of the edge energy on $E_b$ decreases
with increasing laser energy. In particular, for an infrared or green laser,
the sensitivity is very small,
which suggests to employ lasers with large wavelengths, such as a $CO_2$ laser,
for this method.

\vspace{3mm}
\noindent
From these discussions we can draw first conclusions relevant
for beam energy determinations:
\begin{itemize}
\item
the electron edge energy, $E_{edge}$, depends on the beam energy (eq.(\ref{equ:edge})),
on which one of the proposals for measuring $E_b$ relies;
\item
if this method will be utilized,
low energy lasers are advantageous because of large Compton cross section
and high endpoint $E_b$ sensitivity;
\item
backscattered electrons and photons are predominantly scattered in the direction of the
incoming beam;
\item
photons associated with the edge electrons have largest energy and point
towards $\theta_{\gamma}$ = 0;
\item
the unpolarized Compton cross section peaks at $E_{edge}$ which results in beam energy
determinations with small statistical errors;
\item
for polarized electrons, choose the polarization configuration $P_e\lambda$ = -1;
the unfavored configuration $P_e \lambda$ = +1 spoils any $E_b$ determination.
\end{itemize} 
So far, the cross section formulas and backscattered particle properties
were discussed in Born approximation.
Possible modifications due to multiple scattering, \ee pair background,
higher order corrections and nonlinear effects were partially discussed in \cite{HJS}
and are further studied in Sect.3.10.

\subsection {\boldmath Luminosity of Compton Scattering \unboldmath}

To turn from cross sections to number of Compton events,
the luminosity of $e^-\gamma$ collisions has to be known.
In principle, there are two cases to consider:
collisions of beam electrons with a continuous laser or a pulsed laser
that matches the pattern of the incident electron bunches at the ILC.
In the following we assume that the particle densities in both beams are
of Gaussian-shape.
\begin{itemize}
\item Continuous laser
\end{itemize}
The luminosity of a continuous laser with a pulsed electron
of round transverse profile ($\sigma_x = \sigma_y$) can be expressed as
\cite{Suzuki:1976xe}

\begin{eqnarray}
 L_{cont} =
 \frac{1+\cos \alpha}{\sqrt{2\pi}\sin\alpha} \cdot \frac{N_eP_L}{c E_{\lambda}}
   \cdot \frac{1}{\sqrt{\sigma _{x\gamma} ^2 + \sigma _{xe} ^2}}~,
\end{eqnarray}
where $N_e$ is the number of electrons per bunch,
$P_L$ the average power of the laser with energy $E_{\lambda}$,
and $\alpha$ the crossing angle of the two beams.
The horizontal beam sizes are characterized by $\sigma _{x\gamma}$
and $\sigma _{xe}$. Although the ILC beam is not actually round as assumed,
it does not matter here, since usually $\sigma_{x\gamma} > \sigma_{xe}$.

\vspace{3mm}
\noindent
If the crossing angle $\alpha$ becomes zero, the expression for
the luminosity explodes. If, however, the electron bunch is completely
contained within the laser spot, as is  normally the case,
the luminosity is restricted by the finite laser beam emittance $\varepsilon _\gamma$
\begin{equation}
 L_{cont,max} = \frac{N_eP_L}{c E_{\lambda}}\cdot\frac{1}{\varepsilon _\gamma}~.
\end{equation}
For a perfect laser, the best possible emittance is limited by the laws of optics
and depends on the wavelength $\varepsilon _\gamma = \lambda /4\pi$. The associated
maximum possible luminosity is then determined as
\begin{equation}
 L_{cont, max} = 4\pi \cdot \frac{N_e P_L}{hc^2}~,
\end{equation}
where $h$ is the Planck constant and $c$ the speed of light.

\begin{itemize}
\item Pulsed laser
\end{itemize}
For a pulsed laser, the luminosity per bunch crossing is \cite{Suzuki:1976xe}
\begin{eqnarray}
 L_{pul} = N_{\gamma} \cdot N_e \cdot g~,        \label{equ:pulse_lumi}
\end{eqnarray}
with $N_\gamma$ the number of photons per laser pulse and $N_e$ the number
of electrons per bunch. 
With no loss of generality, the geometrical factor $g$ for vertical beam
crossing\footnote{For horizontal crossing, the roles of x and y have to be interchanged.}
is well approximated by
\begin{equation}
 g = \frac{\cos^2 \alpha/2}{2\pi} \cdot
   \frac{1} {\sqrt{\sigma^2_{xe} + \sigma^2_{x\gamma}}} \cdot
   \frac{1}{\sqrt{(\sigma^2_{ye} + \sigma^2_{y\gamma}) \cos^2 (\alpha/2) +
   (\sigma^2_{ze} + \sigma^2_{z\gamma}) \sin^2(\alpha/2)}}~,   \label{equ:g_factor}
\end{equation}
where $\alpha$ is the crossing angle and the transverse laser profile
is assumed to be constant. Note that
the vertical, respectively, longitudinal bunch sizes $\sigma_{y\gamma}$, $\sigma_{ye}$ and
$\sigma_{z\gamma}$, $\sigma_{ze}$
of the interacting beams contribute.

\vspace{3mm}
\noindent
For small $\alpha$ and transverse dimensions of the electron beam compared 
to the laser focus, i.e. $\sigma_{xe} < \sigma_{x\gamma}$ and $\sigma_{ye} < \sigma_{y\gamma}$,
which is generally valid at the crossing point,
the geometrical factor reduces to
\begin{equation}
 g = \frac{1} {2\pi \sigma_{x\gamma} \sigma_{y\gamma}~
    \sqrt{1 + (0.5\alpha \cdot \sigma_{z\gamma}/\sigma_{y\gamma})^2}}~.
\end{equation}

\vspace{3mm}
\noindent
For given $\sigma_{x\gamma}$, $\sigma_{y\gamma}$ of the laser focus,
the bunch related luminosity reaches a maximum for small crossing angles
and short laser pulses:
\begin{eqnarray}
 L_{pul,max} =\frac{N_{\gamma} \cdot N_e}{2\pi\sigma_{x\gamma} \sigma_{y\gamma}}~.
\end{eqnarray}
This formula is very similar to the expression given for
the luminosity of the colliding beams at the physics $e^+ e^-$ interaction point.


\section {\boldmath Overview of the Experiment \unboldmath}

\subsection {\boldmath Basic Experimental Conditions \unboldmath}

Within the so-called single-event regime, individual Compton events originate from
separate accelerator bunches. 
As was realized in experiments at storage rings \cite{Taiwan, BESSY, Novosibirsk},
recording the maximum energy of the scattered photons out of many events
enables to infer the beam energy.

\vspace{3mm}
\noindent
The experimental conditions at the ILC with large bunch crossing frequencies
and high particle intensity require to operate with short
and intense laser pulses so that high instantaneous event rates are achieved.
As a result, the detector signals for a particular bunch crossing
correspond to a superposition of multiple events.
In such a regime, single photon detection cannot be realized and
the signal will likely be an energy weighted integral
over the entire photon spectrum.
The number of Compton interactions should, however, be adjusted such that
neither the incident electron beam will be disrupted
nor the Compton event rate degrades the performance of the detectors.

\vspace{3mm}
\noindent
It is also worth to note that it might be useful for e.g. calibration
purposes to operate occasionally in the single-event regime, either with
reduced pulse power of the laser or even with CW lasers.

\begin{figure}[ht]
\center
\includegraphics[height=8cm,width=14.5cm]{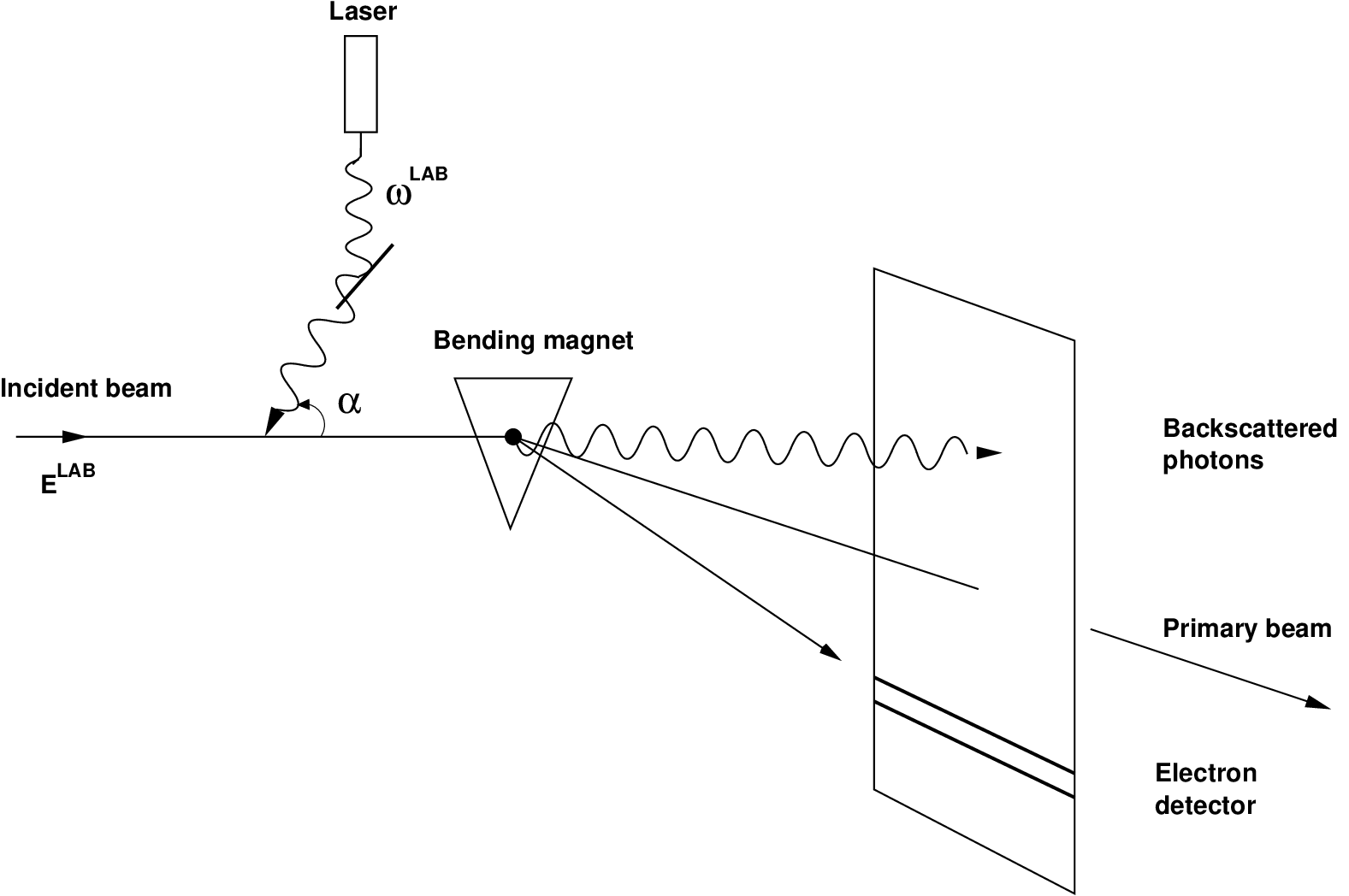}
\caption{Scheme of the proposed energy spectrometer based on Compton backscattering.}
\label{fig:basic_setup}
\end{figure}

\vspace{3mm}
\noindent
The concept of a possible Compton energy spectrometer is shown in Fig.~\ref{fig:basic_setup}.
Downstream of the laser crossing point,
a bending magnet is positioned which is followed by a dedicated particle detection system.
This system has to provide precise position information of the backscattered photons
and electrons close to the edge and, employing an alternative method,
the position of the unscattered beam.

\vspace{3mm}
\noindent
The vacuum chamber between the Compton IP and the detector plane needs some special design
to accommodate simultaneously the trajectories of the photons, the degraded
backscattered electrons and the non-interacting beam particles. 
In order to ensure large luminosity, the crossing angle should be very small
and, for reasons of reduced radiation exposure to the optical elements
above and below the electron beamline, vertical beam crossing is suggested.

\vspace{3mm}
\noindent
The dipole magnet located about 3 m downstream of the 
crossing point separates the particles
coming from the IP into the undeflected
backscattered photons, the Compton electrons and the beam particles
with smallest bending angle. The B-field integral should be scaled
to the primary beam energy, so that
beam particle deflection occurs always at 1 mrad. Thus, one BPM
with fixed position is sufficient to record the beamline position at all energies.
The photon detector is located in the direction of the original beam,
while the electron detector has to be adjusted horizontally
according to Compton scattering kinematics and the magnetic field\footnote{Whether
such a setup can be realized at highest energies needs
careful studies in order not to spoil the beam emittance too much.}.

\vspace{3mm}
\noindent
The laser system should consist on a pulsed laser, while
a continuous laser might only be occasionally used for special tasks
such as detector calibration or operation at the $Z$-pole. At ILC energies
Compton scattering with typical continuous lasers in the 1-10 Watt range
takes some fraction of an hour to collect enough statistics 
for precise $E_b$ determination. Thus, in order to perform
bunch related energy measurements the default laser system should be
a pulsed laser with a pattern that matches the peculiar
pulse and bunch structure of the ILC, i.e. at 250 GeV 
an inter-bunch spacing of $\sim$300 ns within 1 ms long pulse trains at 5 Hz.
In order to collect typically $10^6$ Compton events per
bunch crossing, the pulse power of the $CO_2$ laser should be about 5 mJ
\footnote{The laser power estimation assumes
electron and laser beam parameters as discussed in Sect.3.9.},
while for an infrared laser with $E_{\lambda}$ = 1.165 eV,
the smaller Compton cross section will be partially compensated
by a smaller spot size, a power of 30 mJ is needed.
A laser in the green wavelength range with 2.33 eV photon energy
requires a pulse power of 24 mJ for $10^6$ Compton interactions.
For $Z$-pole running, the laser power can be somewhat smaller, 
but it has to be increased for 1 TeV runs.
Since at present lasers with such exceptional properties are not commercially
available, R\&D is needed to achieve the objectives,
see e.g. \cite{omori, will, schreiber}.

\vspace{3mm}
\noindent
To maximize the $e\gamma$ luminosity,
the crossing angle $\alpha$ should be small, in our case 8-10 mrad,
and the laser spot should be larger than the horizontal electron beam size,
which is expected to be in the range of 10-50 $\mu$m
within the beam delivery system (BDS)\footnote{The vertical
beam size is much smaller and will not exceed few micrometers, resulting to
an horizontal/vertical aspect ratio of typically 10-50 within
the BDS of the ILC.}. For a well aligned laser it should be practicable
to keep possible horizontal and vertical relative displacements of the electron
and laser beams small enough, so that  permanent overlap is ensured
even in cases of beam position jitter.

\vspace{3mm}
\noindent
The choice of a suitable laser system is determined by several constraints.
Basically, lasers with large wavelengths such as a $CO_2$ laser with
$\lambda = 10.6~\mu$m provide high event rates due to large Compton
cross sections and best beam energy sensitivity of the endpoint position,
see Fig.~\ref{fig:laser_edge}. Lasers in the infrared region such as $Nd:YAG$
or $Nd:YLF$ lasers, however, provide at present a better reliability,
in particular with respect to the bunch pattern and pulse power \cite{schreiber}
and would relax geometrical constraints of the spectrometer setup
due to substantially smaller electron edge energies, see Fig.~\ref{fig:cross_section}.
Green laser R\&D is ongoing within the ILC community to develop
laser-wire diagnostics \cite{Blair} and high energy polarimeters \cite{Schueler}.

\begin{figure}[ht]
\center
\includegraphics[height=8cm,width=12.0cm]{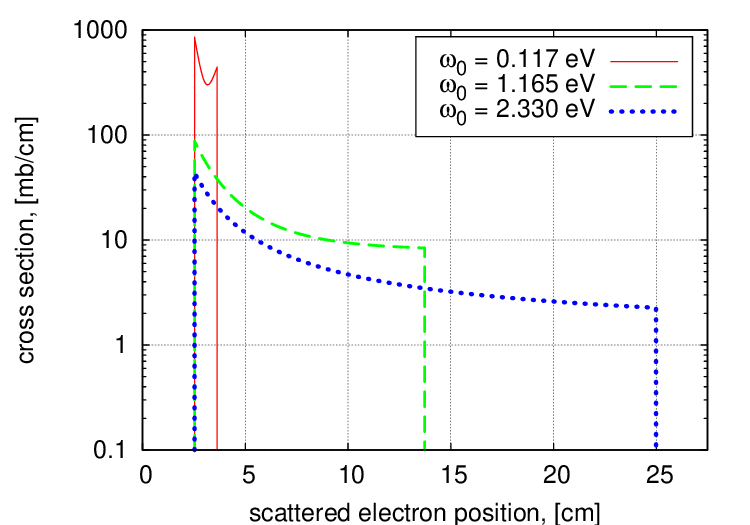}
\caption{Scattered electron positions for $E_b$ = 250 GeV, a B-field of 0.28 T 
       and three laser energies. The detector is placed 25 m downstream 
       of the spectrometer magnet.}
\label{fig:end_point}
\end{figure}

\vspace{3mm}
\noindent
Fig.~\ref{fig:end_point} shows for three wavelengths and
a particular setup (with a B-field of 0.28 T and a detector 25 m
downstream of the magnet) the horizontal or x-position of the Compton electrons.
The position of electrons with highest energy coincides
with the beamline position independent
of the laser, whereas the positions of the edge electrons with largest deflection
are very distinct. They are smaller for larger laser wavelength. 
For a $CO_2$ laser at 45.6 GeV, the edge electrons are separated
by only 2.2 mm from the beamline,
while they are displaced from the backscattered $\gamma$-rays by about 2.6 cm.
Such space conditions would prevent the use of a $CO_2$ laser for 
$Z$-pole calibration runs. An increased B-field and/or a larger drift distance
could somewhat relax the situation.

\vspace{3mm}
\noindent
Lasers in the green or infrared wavelength region have some disadvantages. They provide
smaller Compton cross sections and hence smaller event rates, 
which might only be 
compensated by higher laser power and/or smaller but limited spot sizes. 
Also, the smaller sensitivity of the edge position on  $E_b$
(Fig.~\ref{fig:laser_edge}) and the generation
of additional background at large $\sqrt{s}$ due to \ee pairs 
from Breit-Wheeler processes\footnote{These are
$\gamma-\gamma$ interactions, where one $\gamma$ stems from the 
Compton process and the other from the laser.} might disfavor their application.
As soon as the variable x of eq.(\ref{equ:x_variabel})
exceeds 4.83, which is for example the case at 250 GeV and a green laser,
\ee pair production is kinematically 
possible\footnote{The threshold of \ee pair creation is $E_m E_{\lambda} = m^2 c^4$, with
$E_m = x \cdot E_b/(x +1)$, which gives $x = 2(1+\sqrt{2}) \simeq4.83$.}.
Whether this source of background is tolerated will be studied in Sect.3.10.
Some of the disadvantages discussed are of less relevance if an alternative
method, called method B in the following, will be employed 
for beam energy determination.

\subsection {\boldmath Method A \unboldmath}

One approach to measure the ILC beam energy by Compton backscattering relies
on precise electron detection at the kinematic endpoint.
In particular, endpoint or edge energy measurements 
are performed, from which via eq.(\ref{equ:edge}),
the beam energy is accessible. In particular, the Compton edge electrons
are momentum analyzed by utilizing a dipole magnet and recording their
displacement downstream of the magnet.

\vspace{3mm}
\noindent
The conceptual detector design consists of
a component to measure the center-of-gravity
of the Compton backscattered $\gamma$-rays\footnote{The
center-of-gravity of the photons resembles precisely
the position of the original beam at the crossing point.}
and a second one to access the position of the edge electrons.
The distance $D$ of the center-of-gravity to the edge position
and the well known drift space $L$ between the dipole and the
detector determine the bending angle $\Theta$ of the edge electrons, 
which, together with the B-field integral, fixes
the energy of the edge electrons:
\begin{equation}
      E = \frac{c \cdot e}{\Theta} \int\limits_{magnet} B dl~.
\end{equation}
Here, c is the speed of light and e the charge
of the particles. Thus, 
for sufficient
large drift space the edge electrons are well separated from
the Compton scattered photons which pass the magnet undeflected.

\vspace{3mm}
\noindent
A demanding aspect of this approach is the precision
for the displacement, $\Delta D$, which is related to the beam energy
uncertainty as   
\begin{equation}
  \frac{\Delta E_b}{E_b}  =  (1+\frac{4 E_{\lambda} E_b}{m^2})
   \sqrt{ \left( \frac{\Delta B}{B} \right)^2 + \left( \frac{\Delta L}{L} \right)^2
   + \left( \frac{\Delta D}{D} \right)^2}~.       \label{equ:error_energy-1}
\end{equation}
This relation follows from eqs.(\ref{equ:e_gamma_max}), (\ref{equ:edge})
and
\begin{equation}
 \frac{\Delta E_{edge}}{E_{edge}} = \frac{E_{edge}}{E_b}
  \cdot \frac{\Delta E_b}{E_b}~
\end{equation}
as well as
\begin{equation}
  \left( \frac{\Delta E_{edge}}{E_{edge}} \right)^2 =
    \left( \frac{\Delta\Theta}{\Theta} \right)^2 +
      \left( \frac{\Delta B}{B} \right)^2~
\end{equation}
together with $D = \Theta \cdot L$ from the geometry of the setup.
Synchrotron radiation effects on $\Delta E_b/E_b$, estimated to be significantly smaller
than any term in (\ref{equ:error_energy-1}), were omitted.
\begin{figure}[ht]
\center
\includegraphics[height=6.5cm,width=14cm]{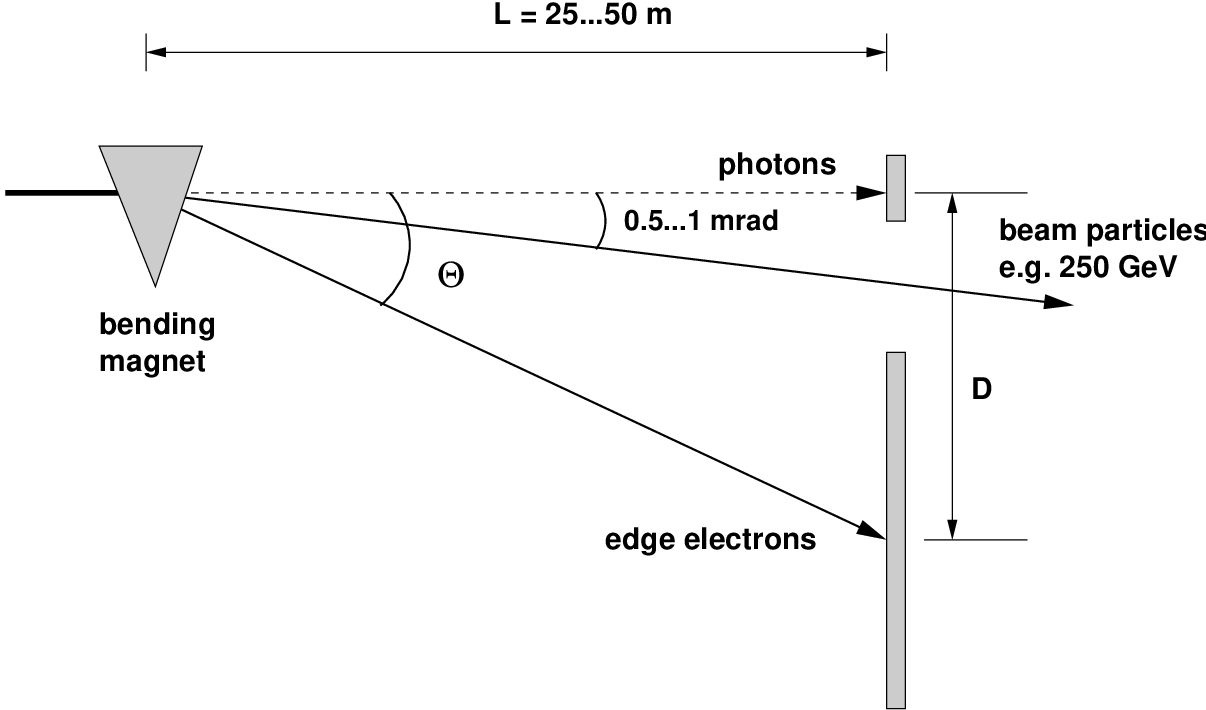}
\caption{Sketch of an experimental layout.}
\label{fig:exampel_1_setup}
\end{figure}                                        
One notices from eq.(\ref{equ:error_energy-1})
that smallest beam energy uncertainties are achievable for lasers
with large wavelengths, such as a $CO_2$ laser.

\begin{figure}[ht]
\center
\includegraphics[height=14cm,width=15cm]{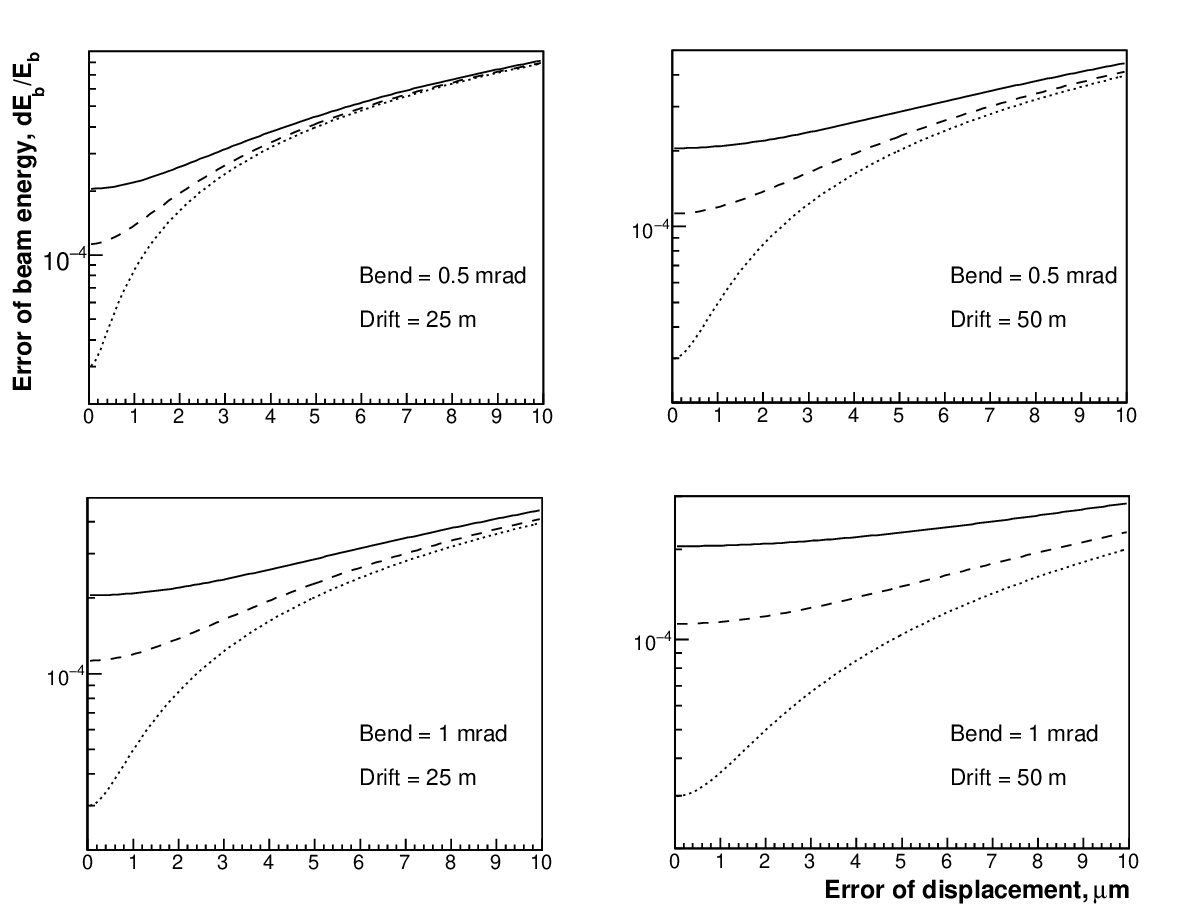}
\vspace{-3mm}
\caption{Beam energy uncertainty as a function of the edge electron displacement error
  for the green laser (full curve), the infrared laser (dashed curve) 
  and the $CO_2$ laser (dotted curve) for two values of drift space and
  bending angle. The beam energy is 250 GeV.}
\label{fig:beam_error_displ}
\end{figure}

\vspace{3mm}
\noindent
Assuming a relative error of the field integral of $2 \cdot 10^{-5}$
and for $\Delta L/L = 5 \cdot 10^{-6}$,
$\Delta E_b/E_b$ values as a function of $\Delta D$ are displayed
in Fig.~\ref{fig:beam_error_displ} for three laser options at 250 GeV.
Drift distances of either 25 or 50 m and 0.5 or 1.0 mrad for the bend angle
were assumed.
Clearly, in order to achieve a precision of $\Delta E_b/E_b = 10^{-4}$,
$\Delta D$ has to be smaller than a fraction of a micrometer for a green laser,
even for a drift distance of 50 m and 1 mrad bending power.
In contrast, a $CO_2$ laser allows for less stringent demands of the displacement
error: $\Delta D$ might be in the order of few micrometers.

\vspace{3mm}
\noindent
Since the displacement is determined by the
center-of-gravity of the recoil $\gamma$-rays
and the position of the electron edge, the displacement error
$\Delta D$ is given by the corresponding uncertainties as
$\sqrt{\Delta X_{\gamma}^2 + \Delta X_{edge}^2}$.
The edge position accuracy $\Delta X_{edge}$ can be estimated as
\begin{equation} 
   \Delta X_{edge}  =  \sqrt{ \frac{2 \cdot \sigma_{X_{edge}}}
                           {\frac{dN}{dx}(X_{edge})}}~,     \label{equ:error_x-edge}
\end{equation}
where $\frac{dN}{dx}$ is the scattered electron density at the detector plane and
$\sigma_{X_{edge}}$ the width of the edge.
After passing the spectrometer magnet the edge electrons 
are displaced from the beam electrons by an amount of $A \cdot \frac{4\omega_0}{m^2}$,
with $A \sim L \cdot \int B dl$, $\omega_0$ as given in (\ref{equ:e_gamma_max})
and m the electron mass (see also eq.(\ref{equ:edge_pos})),
with a width practically identical to that
of the beam. $\sigma_{X_{edge}}$ is uniquely determined
by linac parameters such as the beam size, energy spread, divergence, etc.
Neglecting correlations between initial state parameters
the width of the edge at the detector can be written as
\begin{equation}
  \sigma_{X_{edge}} \simeq \sqrt{\sigma_{x}^2+{(\sigma'_{x}\cdot L )}^2
     +{\left( X_{beam}\cdot\frac{\sigma_{E}}{E}\right)}^2}~,  \label{equ:sigma_edge}
\end{equation}
with $\sigma_x$ the horizontal bunch size at the electron-laser crossing point,
$\sigma_x'$ the beam divergence,
$L$ the distance to the detector and $\frac{\sigma_E}{E_b}$ the relative energy spread
of the beam. As can be realized, eq.(\ref{equ:sigma_edge}) does not involve
laser parameters because their contributions to $\sigma_{X_{edge}}$ are much smaller
or negligible. Using beam values as discussed in Sect.3.8,
$\sigma_{X_{edge}}$ is estimated to be in the range of 70-90 $\mu$m.
In our approach, see below, the edge distribution is assumed to be described
by a convolution of a Gaussian  with a step function, 
but any other ansatz may be taken into account. 

\vspace{3mm}
\noindent
For $10^6$ Compton scatters, $\Delta X_{edge}$ turns out to be in the order of 6 $\mu$m
for an infrared laser, so that together with $\Delta X_{\gamma}$ = 1 $\mu$m
(Sect.3.7.4), the displacement error
is close to 7 $\mu$m, and somewhat larger for a green laser.
Therefore, if the approach of measuring the energy of edge electrons is followed,
the use of a $CO_2$ laser is favored and excludes
(with high confidence) operation of lasers with smaller wavelengths.
A stronger B-field would noticeably improve $\Delta E_b/E_b$
only at 45.6 GeV, while better knowledge of $\Delta B/B$
of e.g. $1 \cdot 10^{-5}$ only provides minor improvements at all energies.

\begin{figure}[ht]
\center
\includegraphics[height=15.0cm,width=17.0cm]{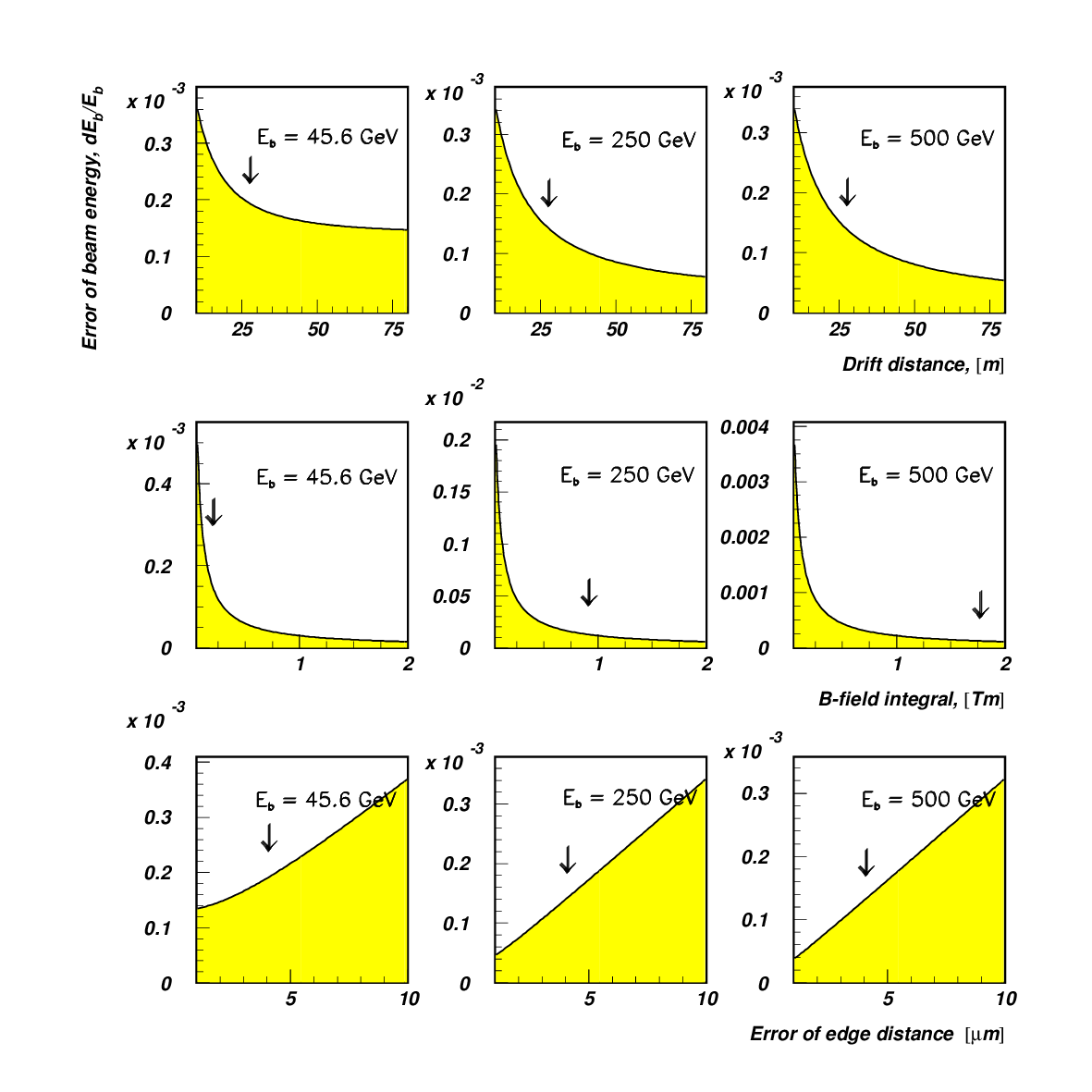}
\vspace{-5mm}
\caption{Beam energy uncertainty as a function of drift distance, 
         B-field integral and edge displacement error for the $CO_2$ laser
         and three beam energies. The arrows in the histograms show
         the default values of the corresponding variable.
         For the variables not shown, the default values are used.
         The detector is placed 25 m downstream of the magnet.}
\label{fig:energy_errors}
\end{figure}

\vspace{3mm}
\noindent
In the present BDS \cite{BDS_present}, free drift space  allows
for lever arms of about 25 m and together with $\Delta L/L = 5\cdot 10^{-6}$,
a dipole bending power of 1 mrad for beam particles,
an uncertainty of $\Delta B/B $ = 2$\cdot 10^{-5}$ and an error for
edge displacements of 4 $\mu$m as default values\footnote{These
values are considered to be feasible.},
Fig.~\ref{fig:energy_errors} shows beam energy uncertainties
as a function of the drift distance, the integrated B-field 
and the edge displacement error for the $CO_2$ laser..
The arrows indicate the default values of the corresponding variable.
As can be seen, using the default values, as an example,
the beam energy can be determined to 1.88 (1.40, 1.31) $\cdot 10^{-4}$
at 45.6 (250, 500) GeV, with room for improvements. 
In particular, at 45.6 GeV
a stronger B-field would improve ${\Delta E_b}/{E_b}$ substantially,
while at 250 and 500 GeV
an improved edge displacement measurement or a larger drift space
or some better knowledge of the B-field strength results
to less significant improvements of $\Delta E_b/E_b$.

\vspace{3mm}
\noindent
A peculiar problem which we have to account for is the amount
of synchrotron radiation generated when the beam electrons pass through 
the dipole magnet and its possible impact on precise position measurements.
This will be discussed in Sect.3.7.1.

\subsection {\boldmath Method B \unboldmath}

Beam and Compton scattered electrons with energy $E$
propagate to the detector
such that their transverse position is well approximated by
\begin{equation} 
    X(E)  =  X_0 + \frac{A}{E}~,                \label{equ:electron_pos}
\end{equation} 
where $A \sim L \cdot \int B dl$ and $X_0$
the position of the original beamline extrapolated to the detector plane,
which is given by the center-of-gravity of the
backscattered $\gamma$-rays, $X_{\gamma}$.
Note that in (\ref{equ:electron_pos}) small effects related to synchrotron radiation
are omitted.

\vspace{3mm}
\noindent
According to eqs.(\ref{equ:edge}) and (\ref{equ:electron_pos}), the positions
of the beam and edge electrons can be expressed as
\begin{eqnarray}
    X_{beam} \equiv X(E_{beam}) = X_{\gamma} + A /E_{beam}    \label{equ:beam_pos}
\end{eqnarray}
\begin{eqnarray}
    X_{edge} \equiv X(E_{edge}) = X_{beam} + A \cdot \frac{4\omega_0}{m^2}~.  \label{equ:edge_pos}
\end{eqnarray}
Hence, the beam energy can be deduced from
\begin{eqnarray}
   E_b = \frac{m^2}{4\omega_0} \cdot \frac{X_{edge}-X_{beam}}
                                {X_{beam}-X_{\gamma}}~.          \label{equ:ratio}
\end{eqnarray}
Thus, instead of recording the energy of the edge electrons, the beam energy
can be accessed from measurements of three particle positions, the position
of the forward going backscattered $\gamma$-rays, the position of the edge electrons and
the position of the beam particles. 
The position $X_{beam}$ can be measured by a beam position monitor (BPM),
while recording $X_{edge}$ and $X_{\gamma}$ needs dedicated
high spatial resolution detectors very similar to the demands of method A.

\vspace{3mm}
\noindent
Besides the limitation to a $CO_2$ laser for the concept of edge energy
measurements (method A), the demand of
$2 \cdot 10^{-5}$ for the field integral uncertainty
is rather challenging, and less stringent requirements
would be of great advantage. In method B, $E_b$ determination does not depend on
the field integral, the length of the magnet as well as the distance
to the detector plane. In particular, the independence on the integrated B-field
only requires rather coarse $\Delta B/B$ monitoring.
It is, however, necessary to ensure that both the beam and the edge electrons
have to pass through the same  B-field integral,
i.e. the magnetic field has to be uniform across the large bending range.
Also, the distance $X_{edge}-X_{beam}$
in (\ref{equ:ratio}) which involves as a product
the integrated B-field and the sum of the drift distance and the length of the magnet
\cite{Muchnoi}, does not depend on the beam energy.
Possible variations of this distance may only be caused by rather slow processes
of environmental nature. 
Thereby, by accumulation of many bunch related
$X_{edge}-X_{beam}$ measurements, high statistical precision can be achieved
for this quantity. This implies the option to operate the spectrometer with lasers
of less pulse power, which is of great advantage
since the laser pulse power is a critical issue for method A.
The novel approach of recording three particle positions (the three-point concept)
seems therefore a very promising alternative\footnote{Also,
vice versa, knowing $X_{edge}-X_{beam}$ with high precision,
the B-field integral can be deduced with similar accuracy.}.

\vspace{3mm}
\noindent
Also, eq.(\ref{equ:ratio}) reveals that due to the proportionality between the beam energy
and the distance $X_{edge}-X_{beam}$, which is larger 
as smaller the wavelength of the laser, 
best beam energy values are obtained for high energy lasers,
a situation which is opposite to that of method A.

\begin{figure}[t]
\center
\includegraphics[height=14cm,width=15.0cm]{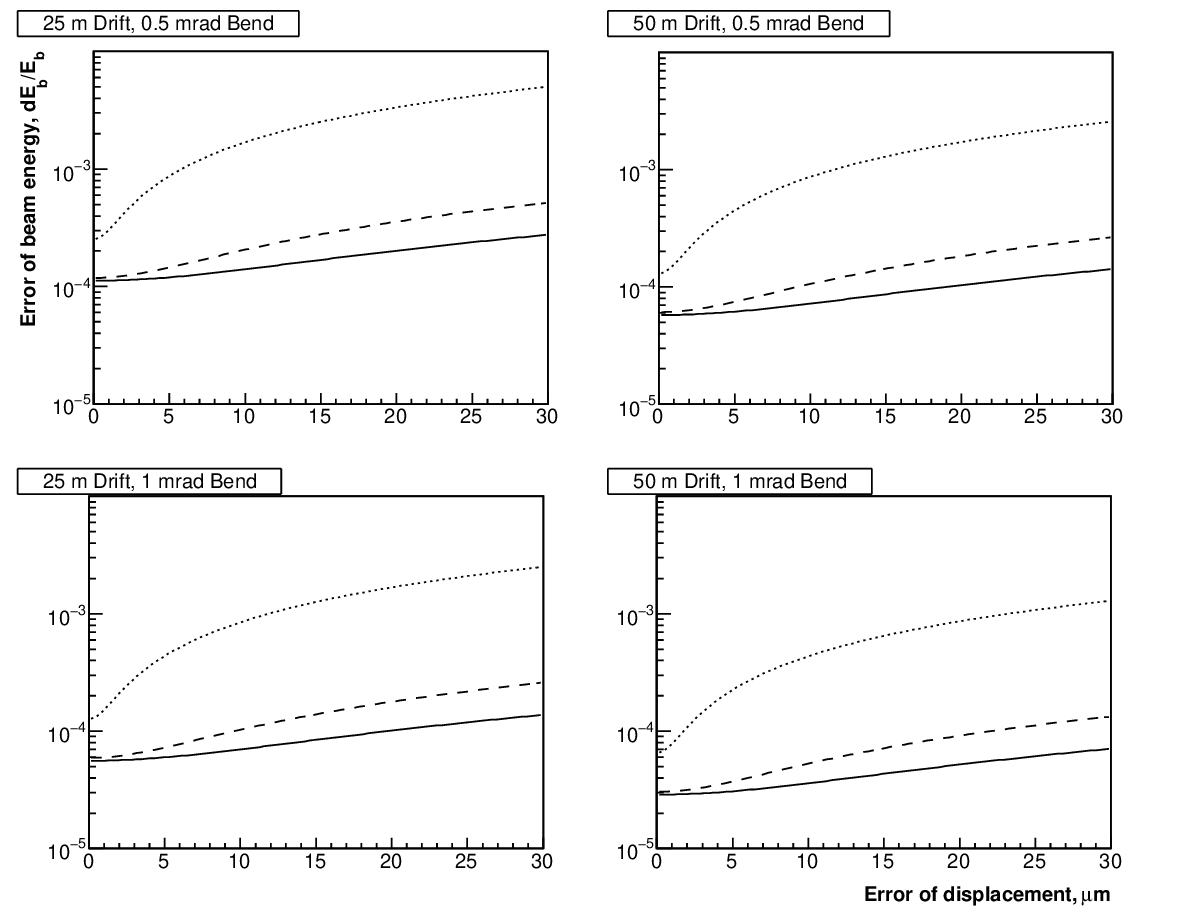}
\caption{Beam energy uncertainty as a function of the edge position error
  for the green laser (full curve), the infrared laser (dashed curve) 
  and the $CO_2$ laser (dotted curve) for $\Delta X_{beam}$ = $\Delta X_{\gamma}$ 
  = 1 $\mu$m,  two values of the drift space and
  bending angle. The beam energy is 250 GeV.}
\label{fig:beam_error_edge}
\end{figure}

\vspace{3mm}
\noindent
The precision of the beam energy can be estimated as
\begin{eqnarray}
  \frac{\Delta E_b}{E_b} = \frac{X_{edge}} {X_{edge}-X_{beam}}
   (\frac{\Delta X_{edge}} {X_{edge}}) \oplus \frac{X_{edge}} {X_{edge}-X_{beam}}
   (\frac{\Delta X_{beam}} {X_{beam}}) \oplus \frac{\Delta X_{\gamma}} {X_{beam}}~.
   \label{equ:error_ratio}
\end{eqnarray}                                                                               
Here, the three terms have to be added in quadrature.
Assuming for the crossing angle 10 mrad and (achievable) values for
$\Delta X_{beam}$ = 1 $\mu$m and $\Delta X_{\gamma}$ = 1 $\mu$m, expected
beam energy uncertainties are shown in Fig.~\ref{fig:beam_error_edge} against
the edge position error, $\Delta X_{edge}$, for the $CO_2$, infrared and green lasers
at 250 GeV, in analogy to Fig.~\ref{fig:beam_error_displ}.
Drift distances of 25 or 50 m and beam bend angles of 0.5 or 1 mrad
are supposed. Clearly, for edge position errors of 10 $\mu$m 
and a limited drift range of 25 m,
$\Delta E_b/E_b$ = 10$^{-4}$ can only be achieved by employing 
an infrared or a green laser. A $CO_2$ laser should
not be considered as an option for this approach since
$\Delta E_b/E_b$ exceeds very quickly the anticipated limit of $10^{-4}$
if $\Delta X_{edge}$ becomes few micrometers.
Even for a perfect edge position measurement,
i.e. for $\Delta X_{edge}$ = 0, the precision of the beam energy
is often larger than $10^{-4}$.

\begin{figure}[t]
\center
\includegraphics[height=15.0cm,width=17.0cm]{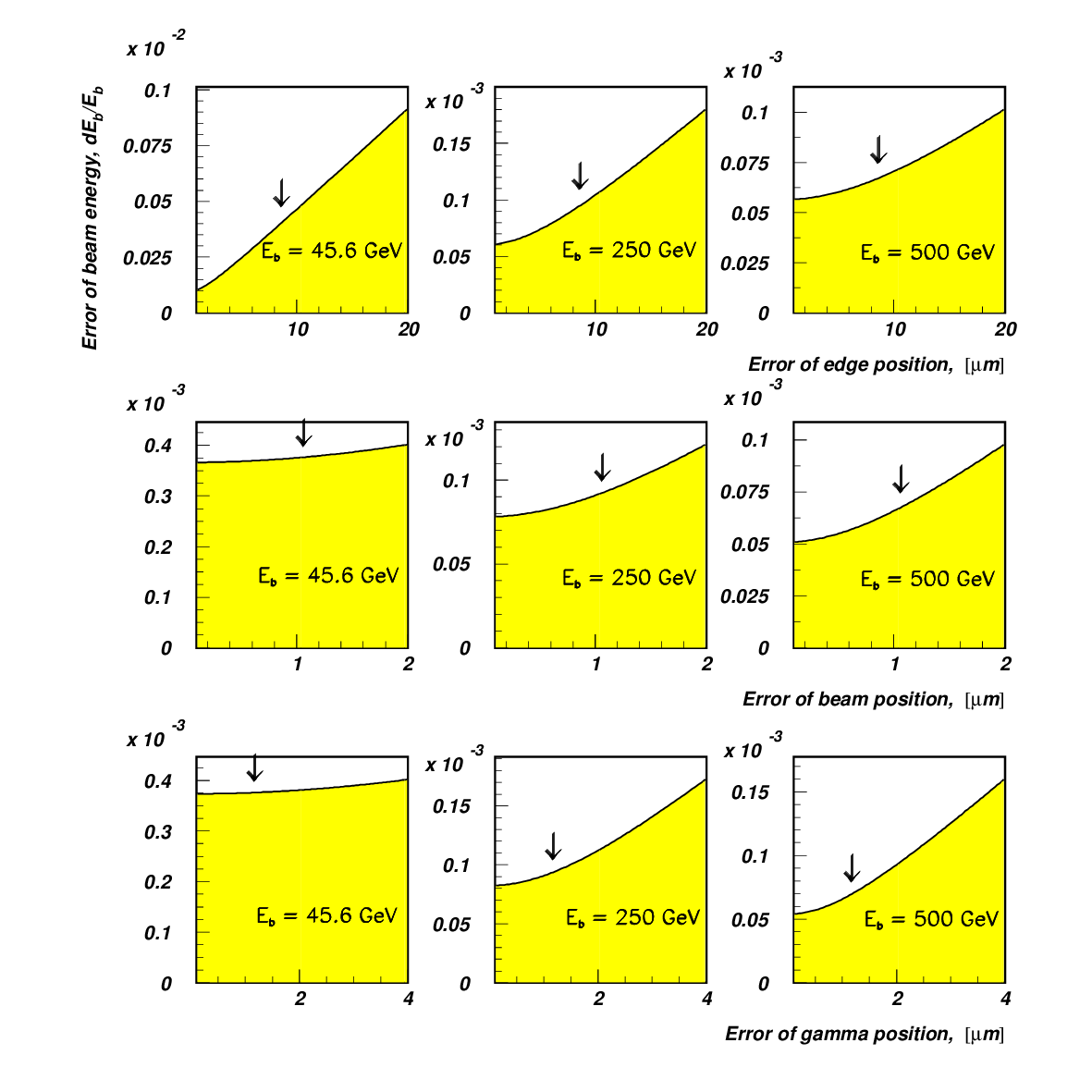}
\vspace{-5mm}
\caption{Beam energy uncertainty as a function of the errors of the edge,
         beam and Compton photon positions for the infrared (1.165 eV) laser
         and three beam energies. The arrows in the histograms show
         the default values of the corresponding variable.
         For the variables not shown, the default values are used.
         The detector is placed 25 m downstream of the magnet.}
\label{fig:error_bpm_all}
\end{figure}

\vspace{3mm}
\noindent
In Fig.~\ref{fig:error_bpm_all}, $\Delta E_b/E_b$ values are plotted against
the accuracies of the edge, beam and $\gamma$-ray positions for the infrared laser,
a 25 m drift distance and a bend angle of 1 mrad for three beam energies.
We also assume $\Delta X_{edge}$ = 8 $\mu$m, $\Delta X_{beam}$ = 1 $\mu$m
and $\Delta X_{\gamma}$ = 1 $\mu$m as default values\footnote{The position
of the beam can be well measured with few hundred nanometer accuracies
using modern cavity
beam position monitors, see e.g. \cite{KEK_BPM, SLAC_BPM, ESA_BPM}.}.
Utilizing these values, $\Delta E_b/E_b$ results to
3.74 (0.91, 0.66) $\cdot 10^{-4}$ at 45.6 (250, 500) GeV in good agreement
with the demands.
Improvements for the $Z$-pole value are possible by employing e.g. a green laser and/or better
$X_{beam}$ and $X_{\gamma}$ position measurements.
%
\vspace{3mm}
\noindent


\subsection {\boldmath The Vacuum Chamber \unboldmath}

In order to maximize the Compton signal, the location of the laser crossing point
should be close to a waist of the electron beam.
Having such a position found, the usual round electron beam pipe with typically
20 mm diameter will be
replaced by a rectangular vacuum chamber with entrance and exit windows
for the laser beam. Crossing of the two beams is assumed
to occur at the center of the chamber.
All particles generated at the IP should be conveniently accommodated by the chamber
without wall interactions.

\vspace{3mm}
\noindent
We plan vertical crossing of the laser light, utilizing
a non-zero but small crossing angle of 8-10 mrad. Small crossing
angles avoid $e \gamma$ luminosity loss. For lasers with
short ($\simeq$10 ps) pulses, the degree of sensitivity of the luminosity
to the relative timing of the two interacting beams and the laser pulse length itself
is less critical. However, the benefits of a small crossing angle must be balanced
against possible luminosity loss associated with an enlarged laser focus.
A quantitative analysis must consider the wavelength dependent emittance
of the laser, the pulse length and time jitter together with the geometry
of the vacuum chamber and the laser beam optics ( see Sect.3.6 for some details).

\begin{figure}[ht]
\center
\includegraphics[height=11cm,width=15cm]{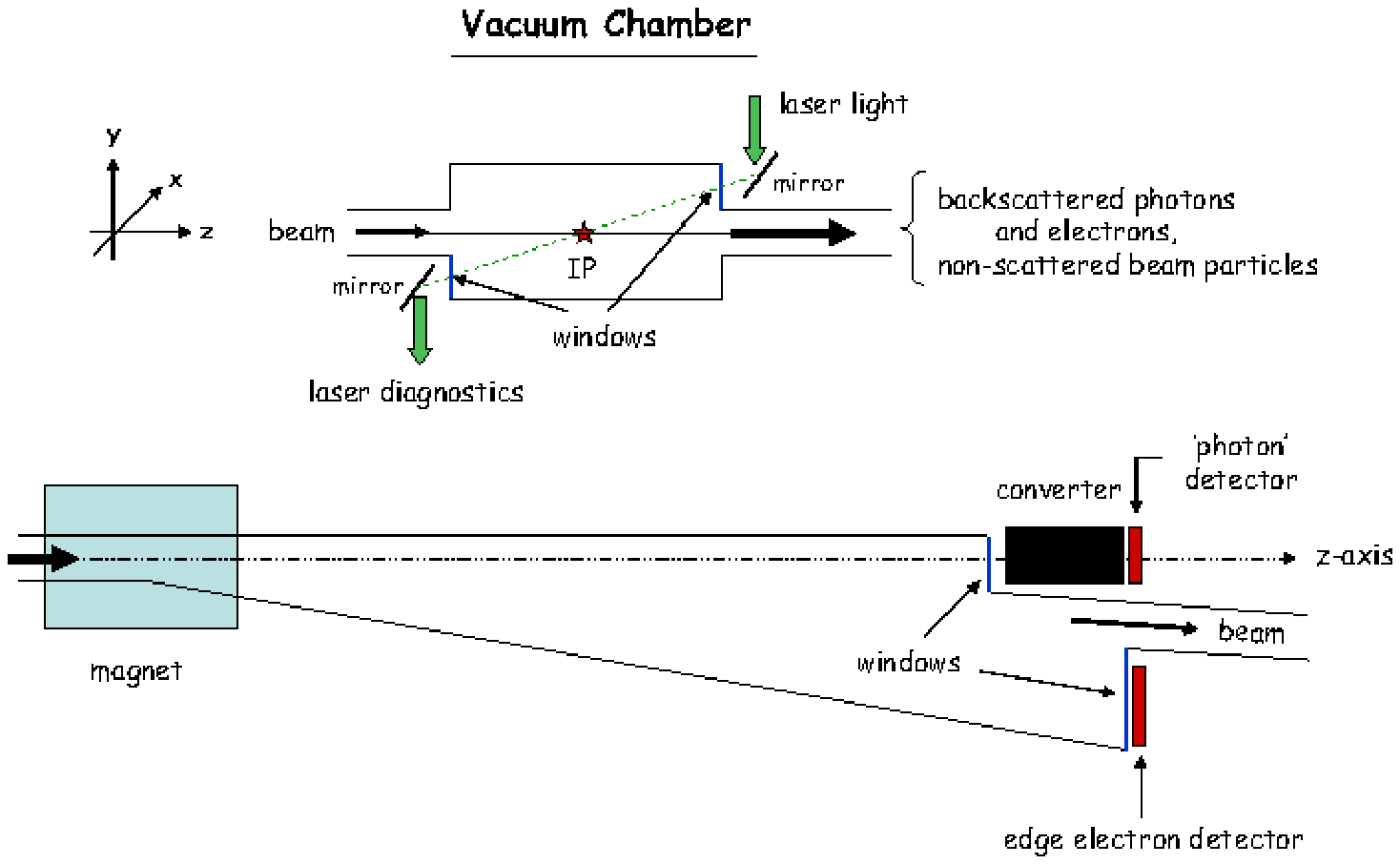}
\vspace{-15mm}
\caption{Top: side view of the laser beam crossing vacuum chamber.
  Bottom: top view of the vacuum chamber downstream of the magnet.}
\label{fig:vacuum_chamber}
\end{figure}

\vspace{3mm}
\noindent
The form and size of the vacuum chamber are mainly dictated by the trajectories
of the unscattered beam, the Compton scattered particles and the laser properties.
We propose to replace the original round beam pipe near the IP by a
6 m long vacuum chamber with rectangular cross section of $x \times y = 60 \times 60$ mm$^2$
in order to accommodate both beams conveniently\footnote{Whether such a vacuum chamber
causes non-acceptable beam emittance dilution needs further studies.}.
The laser beam will be, after passing through the entrance window,
focused by a parabolic mirror with high reflectivity
to the interaction region, as sketched in the top part of
Fig.~\ref{fig:vacuum_chamber}. The window might be a vacuum-sealed $ZnSe$ coated window
that introduces the laser light into the vacuum. It is
mounted about 3 m off the IP nearly perpendicular to the beam direction with
a vertical offset of 25 mm from the beamline\footnote{It might be worthwhile
to mount two windows for redundant vacuum isolation.}. This geometry ensures
almost head-on collision of the laser light with the incident electrons.

\vspace{3mm}
\noindent
After passing through the IP, the laser beam leaves the chamber
through the exit window. After some redirection by a second mirror,
the laser light enters a powermeter for monitoring the power or a wavemeter
to control the spectral position of the laser line. The chamber
does not require internal vacuum mirrors
since optical components installed in
the vacuum are susceptible to be damaged by the beam or synchrotron
radiation. For this reason it is proposed to mount the mirrors
outside the vacuum at positions as indicated in Fig.~\ref{fig:vacuum_chamber}.

\vspace{3mm}
\noindent
Near the position of the entrance window
the vertical dimension of the vacuum chamber is reduced to 20 mm, so that
the cross section becomes 60$\times$20 mm$^2$. In this way,
the entrance (and exit) mirror together with small mounts and adjustment
devices can be placed close to the beamline. The rectangular shape of the chamber
is continued up to the center of the magnet and increases from here
continuously towards the deflection direction, 
as indicated in the bottom part of Fig.~\ref{fig:vacuum_chamber}.
The vertical chamber size of 20 mm will be kept up to the detector plane.
Thus, particles with different deflection angles
are well accommodated and tracked in ultra-high vacuum up to their recording
by the detectors.
Also, in order to minimize wake field effects,
variations of transverse dimensions of the chamber should be smooth.
For a fixed bending power of 1 mrad, the actual horizontal size of the chamber
varies strongly with the laser wavelength.
Tab.~\ref{tab:chamber_sizes} collects the horizontal extensions of the chamber
with respect to the incident beam direction, $x_{right}$ and
$x_{left}$, for three laser and beam energies
at the exit of the magnet and the detector plane located 50 m further downstream.
A safety margin of 5 mm toward negative x-values has always been added.
Note, a $CO_2$ laser needs smallest chamber sizes due to
largest edge electron energies.
Near the detector position, the vacuum chamber is
largely modified and reduced to the usual round beam pipe with 20 mm diameter.
Here, the BPM for beamline position measurements has to be incorporated.
Large exit windows (of e.g. 0.5 mm Al)
in front of the photon converter and edge detector
allow the Compton scattered particles to leave the vacuum.
\begin{table}[htb]
\begin{center}
\begin{tabular}[l]{|c|c|c|c|c|}
\hline
Beam energy, & Laser energy, & Edge energy, &   x-values  &  x-values  \\
    GeV      &    eV         &   GeV        & at magnet exit, mm  &  at detector plane, mm \\
\hline \hline
45.6   &  0.117  &  42.15  &  10 / -7  &  10 / -32   \\
       &  1.165  &  25.14  &  10 / -8  &  10 / -53   \\
       &  2.330  &  17.35  &  10 / -9  &  10 / -75   \\
\hline
250.0  &  0.117  &  172.6  &  10 / -7  &  10 / -43   \\
       &  1.165  &  45.77  &  10 / -13 &  10 / -150  \\
       &  2.330  &  25.19  &  10 / -20 &  10 / -268  \\
\hline
500.0  &  0.117  &  263.70 &  10 / -8  &  10 / -55   \\
       &  1.165  &  50.39  &  10 / -20 &  10 / -268  \\
       &  2.330  &  26.53  &  10 / -33 &  10 / -505  \\
\hline 
\end{tabular}
\end{center}
\caption{Extensions of the vacuum chamber in x with respect to the incident beam direction,
  $x_{right}$ and $x_{left}$, for three laser
  and beam energies at the exit of the magnet and the detector plane.
  A safety margin of 5 mm towards the bending direction has been added.
  The detector is assumed to be located 50 m downstream of the magnet.}
\label{tab:chamber_sizes}
\end{table}


\subsection {\boldmath The Magnet \unboldmath}

In this note we propose, as a first step, to employ the magnet as discussed
in Ref.\cite{ILC_spec}. The magnet has a wide gap
of $170 \times 35$ mm$^2$ to simultaneously accommodate all particle trajectories
over a wide range in energy and magnetic field monitoring devices.
The bend angle for beam electrons between 45 and 500 GeV,
specified to be 1 mrad, results in a field integral of 0.84 T$\cdot$m at 250 GeV.

\begin{figure}[ht]
 \begin{center}
 \epsfig{file=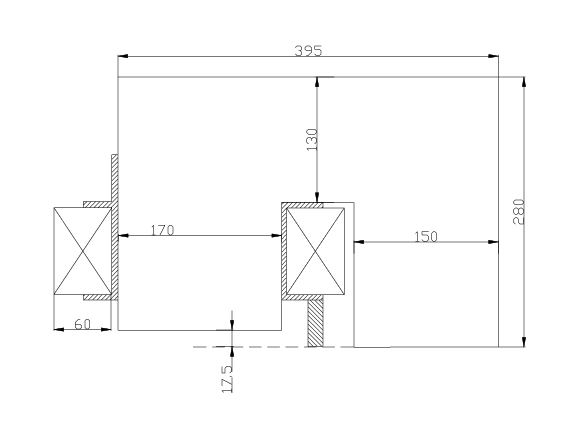,width=0.7\textwidth,height=7cm}
 \vspace{-5mm}
 \caption{ \label{fig:magnet} Cross section of the spectrometer magnet (1/2 part).}
 \end{center}
\end{figure}

\vspace{3mm}
\noindent
Estimation and optimization of the parameters for the magnet were 
performed by a series of 2D and 3D computer model calculations \cite{morozov_1, 
morozov_2, morozov_3, morozov_10, morozov_13}.
The proposed C-type solid iron core magnet has a length of 3 m.
Mirror end plates are installed to contain the fringe fields. 
The magnet proposed 
facilitates vacuum chamber installation and maintenance as well as simplifies
magnetic field measurements. The transverse cross section of the magnet is
shown in Fig.~\ref{fig:magnet} and
its main characteristics are listed in Tab.~\ref{tab:magnet}.
\begin{table}[htb]
 \begin{center}
 \begin{tabular}[l]{|l|c|}
\hline
& spectrometer magnet \\
\hline
Magnetic field (min/max)(T) & 0.05/0.55 \\
\hline
Pole gap (mm) & 35 \\
\hline
Yoke type & C \\
\hline
Yoke dimensions (mm) & 395x560x3000 \\
\hline
Yoke weight (t) & 4.51 \\
\hline
A$^*$turns (1 coil)(max) & 6335 \\
\hline
Number of turns (1 coil) & 6$^*$4=24 \\
\hline
Conductor type, sizes (mm) & Cu, 12.5x12.5, ${\oslash}7.5$ \\
\hline
Conductor weight (t) & 0.36 \\
\hline
Coil current (max)(A) & 264 \\
\hline
Current density (max)(A/mm$^2$) & 2.4 \\
\hline
Coil voltage (max)(V) & 13.3 \\
\hline
Coils power dissipation (max)(kW) & 3.5 \\
\hline
Number of water cooling loops & 6 \\
\hline
Length of cooling loop (m) & 56 \\
\hline
Water input pressure (Bar) & 6 \\
\hline
Water input temperature (deg C) & 30 \\
\hline
Maximal temperature rise of the & 1.4 \\
cooling water (deg C) & \\
\hline
 \end{tabular}
 \end{center}
\caption{ \label{tab:magnet} Basic technical parameters of the spectrometer magnet.}
\end{table}
The magnet iron core is divided into only two parts  by a horizontal symmetry
 plane. This decision gives confidence for tight tolerances of the
 parallelism of the magnet poles and would decrease substantially field 
distortions from the joining elements. The coils of the magnet are proposed to be
 made from 12.5 mm$^2$ copper conductors with water cooled channels of 7.5 mm
diameter. Each pole coil consists of  three double pancake coils (4 turns in 
two layers). According to 3D field simulations,
a field integral uniformity of 20 ppm was found over almost 20 mm
for the anticipated beam energies.

\vspace{3mm}
\noindent
More details of the magnet are discussed in \cite{ILC_spec},
which includes production tolerances, demands for the materials, fringe field
limitations, temperature stabilization and cooling system, zero-field adjustment,
power supplies and the control system. The overall objective of the field integral
uncertainty of $2 \cdot 10^{-5}$ might be achievable by accounting for all
these aspects. If the field integral uniformity region is,
due to manufacturing errors, somewhat reduced,
a fraction of beam energy-laser energy combinations
in Tab.~\ref{tab:chamber_sizes} has to be reconsidered.
Whether a redesign of the magnet
is necessary depends on its final properties and
the choice of the laser. If beam energy determinations will be performed
by means of precise edge electron measurements (method A),
the uniformity region with 20 ppm uncertainty has to be adjusted
such that the path of the edge electrons is properly covered by the B-field.

\vspace{3mm}
\noindent
The uncertainty of the field integral  
$\Delta B/B = 2 \cdot 10^{-5}$, a demanding request, needs
careful design and production of the magnet, accurate field calibration and monitoring.
Thorough mapping of the field in the laboratory under a variety of conditions
that are expected during operation is essential
and monitoring standards should be calibrated with sufficient accuracy.
We propose two independent, high precision methods to measure the field integral
as well as the field shape of the magnet: (i) the moving wire technique as e.g. described
in \cite{levi} and 
(ii) the moving probe technique, where the field integral is obtained by driving
NMR and Hall probes along the length of the magnet
in small steps.

\vspace{3mm}
\noindent
When the magnet is installed in the beamline, 
absolute laboratory measurements should be used
to simultaneously calibrate three independent, transferable standards for
monitoring the field strength: (i) a rotating flip coil, (ii) stationary NMR probes
and (iii) a current transductor \cite{levi}. Since a $2 \cdot 10^{-5}$ 
field integral precision is envisaged, performance
of the magnet and the monitors, in particular
the stability of the power supply current and the magnet temperature,
have to be investigated.

\vspace{3mm}
\noindent
In addition to the field of the spectrometer dipole itself, other sources
of fields are expected in the ILC tunnel which might affect
the path of the Compton electrons.
The earth's magnetic field, for example,
should be measured and corrected for. Also fields produced due to currents
to drive magnets in the beamline might be non-negligible and time-dependent.
Therefore, the ambient field strength in the tunnel has to be explicitly monitored
and corrections applied to avoid spurious bends on the Compton electrons
while they travel to the detector.

\vspace{3mm}
\noindent
The requests for the magnet are less demanding for the alternative method B 
where the positions of the Compton edge electrons and photons as well as of the beam
particles are recorded.


\subsection {\boldmath The Optical Laser System \unboldmath}

\subsubsection{\boldmath General Aspects \unboldmath}

In order to achieve the necessary $e \gamma$ luminosity
and rate of Compton events the laser system should provide
pulse energies, duration and repetition rates as required.
The initial parameters of the beam and its optical quality
should drive the design of an adequate laser transport system.
The basic scheme of the laser source contains a master oscillator
which provides the initial laser pulse  pattern that matches that
of the incident electron bunches.   
Additional amplification might be needed to achieve the necessary pulse energy.

\vspace{3mm}
\noindent
Propagation of laser light is usually considered in the framework
of the Gaussian beam optics, and by definition, the transverse intensity profile
of a Gaussian beam with power $P$ can be described as \cite{rp-photonics}
\begin{equation}
I(r,s)=\displaystyle\frac{P}{\pi w(s)^2 / 2}
\exp\Biggl\{-2\displaystyle\frac{r^2}{w(s)^2}\Biggr\}\;,
\label{gbp}
\end{equation}
where the beam radius $w(s)$ is the distance from the beam axis 
to the $1/e^2$-intensity drop, and $s$ denotes
the coordinate along beam propagation.
It is important to note that this definition 
of the beam radius is twice as large as the usual Gaussian 'sigma', $w(s)=2\cdot\sigma(s)$.
In practice, the transverse intensity profile of lasers, operating in the TEM$_{00}$ mode,
is only close to but not exactly a Gaussian.
A pure Gaussian beam has as lowest possible beam parameter
product the quantity $\lambda/\pi$ (with $\lambda$ the laser wavelength),
whereas for real beams the beam parameter product
is defined as the product of the beam radius (measured at the beam waist)
and the beam divergence half-angle (measured in the far field). 
The ratio of the real beam parameter product to the ideal one is called $M^2$,
the beam quality factor.

\vspace{3mm}
\noindent
In free space, the beam radius varies along 
the traveling direction according to
\begin{equation}
w(s)=w_0 \cdot \sqrt{1+
\Biggl(\displaystyle\frac{M^2 \lambda s}{\pi w_0^2}\Biggr)^2}\;,
\label{equ:gbs}
\end{equation}
with $w_0=w(s=0)$ as the beam radius at the waist. The radius of curvature $R$
of the wavefronts evolves as
\begin{equation}
R(s)=s\cdot \Biggl[ 1+\Biggl(\displaystyle\frac{\pi w_0^2}{M^2 \lambda s}\Biggr)^2 
\Biggr]\;
\label{equ:gbr}
\end{equation}
and the beam status at a certain position $s$ can be specified by a complex parameter $q$:
\begin{equation}
\displaystyle\frac{1}{q(s)} = \displaystyle\frac{1}{R(s)} + 
\displaystyle\frac{i M^2 \lambda}{\pi w(s)^2}\;.
\label{equ:gbq}
\end{equation}
The passage of the beam  through optical elements may be characterized by transforming
$q$ utilizing an $ABCD$ matrix for each element \cite{rp-photonics, matrop}:
\begin{equation}
q' = \displaystyle\frac{Aq+B}{Cq+D}\;,
\label{equ:gbabcd}
\end{equation}              
and by multiplying all matrices the whole system is described.

\subsubsection{\boldmath Final Focus Scheme \unboldmath}

\begin{figure}[h]
\centering
\includegraphics[height=7cm,width=15cm]{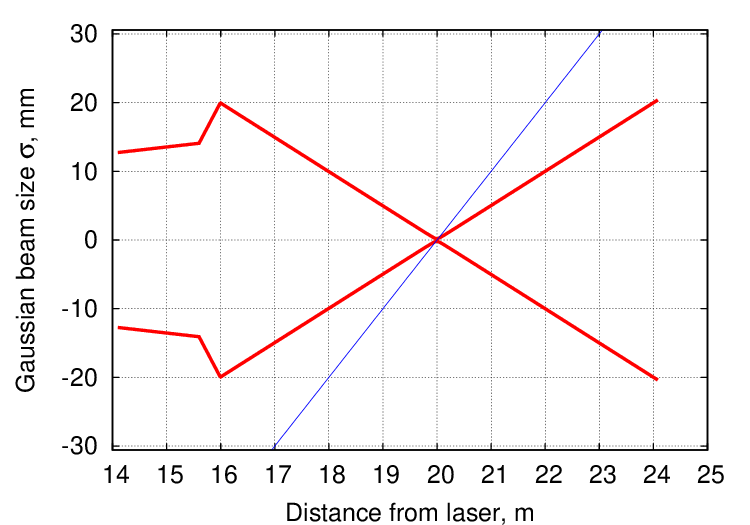}
\caption{ 1$\sigma$ laser beam size near the crossing region with $\alpha$ = 10 mrad.
  The thick line represents the laser beam, 
  while the thin line shows the incident electron trajectory.}
\label{fig:laser_geom}
\end{figure}

For largest $e \gamma$ luminosity, the laser beam delivery system
should provide the lowest possible waist size at the crossing point.
But due to alignment uncertainties and possible relative laser 
and electron beam position jitters, options to adopt best waist sizes 
have to be foreseen. This requirement can easily be achieved
when a short-focus lens doublet is used for the final focusing system 
close to the interaction area. Fig.~\ref{fig:laser_geom} shows for a particular laser optics
and a crossing angle of 10 mrad, irrespective of the laser wavelength,
the 1$\sigma$ beam size of the laser near the crossing region.
The waist is positioned 20~m away from the
laser exit aperture\footnote{In reality, this optics could be changed
to an appropriate configuration by adding more lenses to the laser beam delivery system.}.
The beam is focused by two lenses, $L_1$ and $L_2$, with focal length $f_1$ =-1.0~m,
respectively, $f_2$ = 1.0~m. In order to avoid an additional waist
between the two lenses, the first focal length has to be negative.
The lenses are positioned at 15.600~m and 15.992~m,
see Fig.~\ref{fig:laser_geom}, where the laser beam enters the vacuum chamber 17~m
downstream from the laser exit aperture, and 3~m prior to the interaction point.
The position of one of the lenses is supposed to be accurately adjustable 
by precise mechanics. The thin line in Fig.~\ref{fig:laser_geom}
indicates the corresponding electron beam line. Since the optical system 
was designed for a crossing angle of 10~mrad, limits are imposed
on the laser beam divergence, $LD$, after the final focus system.
The divergence (at 1$\sigma$ transverse laser beam size) has to be at least two times 
smaller than $\alpha$. Thus, for strict Gaussian beams, the acceptance 
of the beam delivery system should be larger than $2\sigma$,
which is the reason for the assumed laser angular divergence 
of 5~mrad in Fig.~\ref{fig:laser_geom}.

\vspace{3mm}
\noindent
The laser waist size is coupled to the laser beam angular divergence via 
\begin{equation}
\sigma_{waist}=\frac{M^2\lambda}{4 \pi LD}~,
\label{equ:wmin}
\end{equation}
which is derived from (\ref{equ:gbs}).
Minimal possible waist sizes and $M^2$ values so obtained are summarized 
in Tab.~\ref{tab:waist_sizes}.
$M^2$ varies with $E_{\lambda}$ according to typical parameters
of the laser sources. The assumed laser spot sizes
at the crossing point of 200, 100 and 50 $\mu$m for the $CO_2$, respectively,
infrared and green laser are in accord with the numbers given in Tab.~\ref{tab:waist_sizes}.
\vspace{4mm}
\begin{table}[h]
\centering
\begin{tabular}{c|c|c|c|c}
Laser & $E_{\lambda}$ & $M^2$ & $\sigma_{waist}$ ($LD=\alpha/2$) & $\sigma_{waist}$ ($LD=\alpha/3$)\\
\hline
$ CO_2 $    & 0.117~eV &  1.1  & 186 $\mu$m & 280 $\mu$m\\
$ Nd:YAG $  & 1.165~eV &  1.2  & 20  $\mu$m & 30 $\mu$m\\
$ Nd:YAG $  & 2.330~eV &  1.3  & 11  $\mu$m & 17 $\mu$m\\
\end{tabular}
\caption{Possible minimal laser waist sizes and $M^2$ values for different laser energies
 and a crossing angle of 10~mrad.}
\label{tab:waist_sizes}
\end{table}

\vspace{3mm}
\noindent
Since the beam and the laser widths are of similar size, central collisions
of both beams is essential in order to avoid systematic shifts 
of the center-of-gravity of the Compton photons.  
To ensure such collisions we propose to install a partially transmitting mirror
close to the vacuum entrance window so that most of the laser light is employed
for Compton collisions and only a small fraction
hits a CCD camera or an avalanche photodiode (APD). The camera,
respectively, the APD is used in the alignment procedure to permanently steer the laser
onto the electron beam. Spot size and position of the laser can so be monitored.
A feedback system allows to adjust
the focus by the last mirror in the laser beamline, which
might be a deformable or segmentable one. Smallest displacement
of both beams from one another can be maintained by performing a scan
that samples across successive electron bunches for highest Compton event rate.
The required laser pointing stability should be $\sim$10 $\mu$m
which seems to be achievable \cite{PITZ}.
Furthermore, upstream and downstream of the collision chamber
beam position monitors may be needed to monitor 
the position of the electron beam.
More discussions about requirements for central collisions can be found in Sect.3.11.


\subsection {\boldmath Electron and Photon Detection \unboldmath}
                                                           
The detector assembly is supposed to be located at least 25 m downstream of the magnet.
Since we plan to operate the spectrometer with  
an energy independent fixed bending angle of 1 mrad,
the distance of the backscattered $\gamma$-ray
centroid to the beamline 25 m downstream of the dipole is 26 mm for all $E_b$ values,
while the displacement
of the edge electrons depends on $E_b$ and $E_{\lambda}$.
This displacement in the range of a few centimeters to about a quarter of a meter
requires high stability of the detector assembly and
its adjustment to micrometer accuracy.
Therefore, the individual detector components should be connected rigidly
and installed on a vibration damped table
that can be moved horizontally (and vertically) and controlled with high precision.

\vspace{3mm}
\noindent
After leaving the vacuum chamber,
the Compton scattered electrons near the edge traverse 
a position sensitive detector with high spatial resolution.
We propose to employ either a diamond micro-strip
or an optical quartz fiber detector.
Such detectors, frequently applied in particle physics experiments,
have demonstrated their ability to achieve micrometer
spatial resolution within an intense radiation field,
see e.g. \cite{Diamond, Gorodetzki}.

\vspace{3mm}
\noindent
The center-of-gravity of the Compton $\gamma$-rays might be recorded
by employing one of the two following concepts. One concept consists in measuring
high energy electrons and positrons from photon interactions
in a converter placed closely in front of the tracking device.
According to simulations, a tungsten converter\footnote{Tungsten
with its large atomic number of 74 and high density of over 19 g/cm$^3$ is an attractive
material for small converters. However, pure material is difficult
to cast or machine, but powder metallurgy processes can produce a sintered form
of tungsten, with a density only slightly below that of the pure metal.}
of sufficient radiation lengths seems to be suitable. Such a scheme, however,
constitutes some trade-off between large conversion rates and
accurate photon position determinations, which might be altered
by multiple scattering of the forward collimated $e^{\pm}$ particles within the converter.
As a position sensitive detector a  quartz fiber detector similar to that
for edge position measurements is proposed and, as simulation studies revealed,
submicrometer precisions of the original photon position are achievable.

\vspace{3mm}
\noindent
An alternative for measuring $X_{\gamma}$, respectively, the undeflected beam position
consists in monitoring one of
the edges of the synchrotron radiation (SR) generated in the dipole magnet of the spectrometer.
A detector sensitive to SR and 'blind' with respect to high energy Compton photons
would be appropriate for this task. For this option, a converter
is not needed and all $\gamma$-rays are incident on the edge position detector.

\subsubsection {\boldmath Synchrotron Radiation \unboldmath}
                                                           
Synchrotron radiation will be generated by electrons passing through
the magnet. For the magnet as described in Sect.3.5,
about five photons per beam particle with an average energy of 3.8 MeV 
are generated, resulting to a total number of $10^{11}$ $\gamma$'s per bunch.
They are concentrated within the cone of
the forward produced Compton scattered photons
and the bent beam. If a tungsten converter of e.g. 16 radiation lengths ($X_0$)
in front of the $X_\gamma$ detector is inserted,
it also serves as an effective shield against SR.
However, the huge amount of such photons 
(plus a minor fraction from Compton scattered electrons) may preclude perfect SR protection.
Possible low energy electrons and positrons from SR showers
are expected to enter the detector
and could modify the response and eventually the center-of-gravity
of the primary Compton photons. The impact of this background 
(together with machine related background)
has to be taken into account in procedures of precise $X_{\gamma}$ determinations.
Properties of particles leaving the converter and prescriptions
addressed to eliminate center-of-gravity distortions are discussed in Sect.3.8.
Discussions on whether recording the incident beam position
by means of SR is superior to the conventional converter approach
are also included in this section.

\subsubsection {\boldmath Diamond Strip Detectors \unboldmath}

A potential candidate for the high spatial resolution tracking device is
the diamond strip detector (DSD). Chemical vapor deposition strip detectors
indicate, due to their inherent properties, that they are very radiation resistant.
They are a promising, radiation hard alternative to silicon
detectors. In addition, diamond is favored over silicon due to its smaller 
dielectric constant, which yields a smaller detector capacity and, thereby
a better noise performance. It is also an excellent thermal conductor
with thermal conductivity exceeding e.g. that of copper by a factor of five.

\vspace{3mm}
\noindent                                                           
When a minimum ionizing particle traverses the diamond,
36 electron-hole pairs are created per micrometer due to Coulomb interaction, 
Bremsstrahlung and scattering with electrons along its path.
Per electron-hole pair, a mean energy deposit of 13 eV is needed.
The electric field in the volume causes a drift of the electrons and holes 
across the diamond to the positive, respectively, negative electrode.
The induced current produces a signal, which can be amplified and integrated resulting in a
voltage signal proportional to the total charge.

\vspace{3mm}
\noindent                                                          
The spatial resolution of DSD's
is obtained by segmentation of the anode (p$^+$) into
so-called micro-stripes. The micro-stripes might only be ten micrometer
apart and this pitch determines the detector resolution.
Employing the charge division method, the spatial resolution for single-particle
passage can be further improved compared to the binary resolution
of $pitch/\sqrt{12}$.
In this way, the resolution of large scale diamond strip detectors
with a pitch of e.g. 50 $\mu$m was found to be in the range of 7-15 $\mu$m
\cite{Barbero, Adam, Krammer, Zoeller} which is better or close
to the binary resolution of 14.4 $\mu$m. Also, excellent linearity
of the detector system over four decades of incident particles was observed \cite{Barbero}.

\vspace{3mm}
\noindent
Diamond strip detectors were also used as beam monitors \cite{Bol, Barbero, Krammer}
to access the cross-sectional beam profile online for single bunches.
In particular, Ref.\cite{Bol} proposed to perform such measurements for the
TESLA linear collider with $2\cdot10^{10}$ electrons per bunch. Tests in 
heavy ion and electron beams with up to $3\cdot10^{10}$ particles/bunch
were successfully performed although the precision of the measurement
was difficult to estimate.
In our approach, the number of instantaneous particles incident per readout pitch
is at most few hundred for endpoint position 
(or thousands for $X_{\gamma}$) measurements and hence
orders of magnitude smaller than for bunch profile measurements.
The spatial resolution in cases of high occupancy is, however, expected 
to be slightly worse than for single-particle crossing mainly due to $\delta$-electrons
and spreading of charge carriers inside the active volume of the detector,
especially if the electric field inside the sensor breaks down.
For example, a resolution of 23 $\mu$m was measured for a 
50 $\mu$m pitch detector \cite{Adam}.
Reduction of the thickness of the sensor to e.g. 80 $\mu$m
and shorter strips should improve the resolution.

\vspace{3mm}                                      
\noindent
The main parameters of a DSD are the thickness which the ionizing particles cross,
the strip pitch and its width.
The typical bias depletion voltage is 1 V/$\mu$m.
More details of such a device will be discussed in Sect.3.8.

\subsubsection {\boldmath Quartz Fiber Detectors \unboldmath}

In view of the properties of a detector for precise edge position
and $\gamma$-ray centroid measurements a suitable option consists
in a detector of quartz fibers. This option is driven by several aspects
such as high spatial resolution, fast signal collection such that all charges
associated with one bunch crossing are collected before the next bunch crossing,
very high radiation hardness and the insensitivity to induced activation
and possible consequences on measurements. In addition,
tracking detectors based on quartz fibers (QFD's)
are simple in construction and operation. They do not need 
any internal calibration and can work at very high flux. The availability
of square fibers today allows to construct a detector 
of e.g. 100 or even 50 $\mu$m fibers with excellent spatial resolution.

\vspace{3mm}
\noindent                 
In quartz, the signals are caused by Cerenkov light production 
for which quartz is transparent, predominantly for ultraviolet light 
within the 300 to 400 nm wavelength region.
Cerenkov radiation is intrinsically a very fast process with a typical
time constant of less than 1 ns. Instrumental effects
(e.g. those caused by light detection devices) may broaden the signal,
but still the overall charge collection time is less than 10-20 ns.
The fibers are readout by photodetectors which are usually placed
as close as possible to the sensitive layer. 

\vspace{3mm}
\noindent
The so-called lightguide condition in optical fibers together
with the fact that Cerenkov light emitted inside the fiber has a specific angle
with respect to the particle direction leads to an angle dependent
light output at which the particles traverse
the fiber. The production of Cerenkov light is maximum for particles
passing the quartz fiber axis at angles of incidence of 40$^0$-50$^0$ .

\vspace{3mm}
\noindent
A potential drawback of a quartz fiber detector constitutes to the low light yield
for single-particles. One expects typically 1-3 photoelectrons/GeV
incident energy [p.e./GeV],
but yields of 10 p.e./GeV were reported \cite{Gorodetzki}. We expect,
however, due to the large number of Compton scattered particles per fiber
no limitations of photoelectron statistics compared
to other sources of fluctuations.

\vspace{3mm}
\noindent
Many of quartz fiber detectors are calorimeters, see e.g. \cite{Akchurin}.
Quartz fibers were chosen as active material, with diameters ranging
from $\sim$800 to 270 $\mu$m, and often both the energy and impact position
of particles are measured. Spatial resolutions of typically a fraction
of a millimeter were achieved. Other applications consist in beam diagnostics systems
in harsh radiation environments \cite{Goettmann}
and in tracking and vertexing in HEP experiments \cite{H1-Coll}.
Recently, the ATLAS collaboration \cite{ATLAS} proposed a fiber tracker for luminosity
measurements with a spatial resolution of approximately 15 $\mu$m.
However, fiber trackers for precise particle profile measurements
as anticipated in this study were, to our knowledge, not employed.

\vspace{3mm}
\noindent
Our baseline configuration of a quartz fiber detector utilizes square fibers
with a size of 50 $\mu$m having the advantage that their effective thickness
is roughly the same for all traversing particles. 
Due to the small fiber length of few centimeters
geometrical constraints for precise micrometer measurements are 
of no concern. A cladding thickness of 5 $\mu$m
results in an active fiber core of 40 $\mu$m. Despite of
the high occupancy sufficient position resolution is expected, 
in particular for a staggered layer arrangement.
Since in our case practically
all electrons pass the detector with $90^0$ angle of incidence,
little light emission is expected.
Therefore, we propose to incline the detector by $45^0$ with 
respect to the vertical direction so that large signals are obtained
which can be conveniently extracted and transported to the shielded location
for the readout electronics. Fiber ends 
are coupled through an air lightguide to a photomultiplier tube (PMT).
Whether it is worthwhile to polish the opposite end of the fibers 
to enhance the light reflection needs further studies.
Typical solutions for QFD readout use PMT's
with multi-anode structure. Such PMT's are well established and robust,
and crosstalk between channels is at the level of only 2-3\%.

\vspace{3mm}
\noindent
For both the DSD and QFD detector schemes
the sensitive region of the device can be small, in the order
of $1 \times 1$ cm$^2$, since only the position of electrons at or
close to the edge, respectively, the center-of-gravity 
of the forward produced Compton photons is of interest.
Thereby, a relative small number of readout channels is needed,
and, together with some fast and robust data processing, the system should provide
position information of micrometer resolution. It is advantageous to house
the detector assembly inside a Roman Pot.
In the case of a quartz fiber detector,
$\mu$-metal shielding for PMT's is required in the presence of stray magnetic fields
in excess of 10 Gauss in order to maintain the gain and hence the detection efficiency.
The output signal can be readout by a relatively simple binary electronics chain,
for which an example is given in \cite{ATLAS}. Even for a relative small single fiber
detection efficiency of 70 to 80\%, excellent overall performance
of the detector is expected.

\subsubsection {\boldmath Photon Detector Options \unboldmath}

One possibility to perform $X_{\gamma}$ measurements
consists in using a quartz fiber detector in conjunction with a closely  
placed converter of adequately chosen radiation length. Compton backscattered
photons will be affected during their propagation through the converter
by several processes such as ($e^+, e^-$)-pair creation and Compton collisions.
Once $e^{\pm}$ particles are created, they are subject to multiple scattering,
ionization, and $\delta$-ray production, bremsstrahlung
and annihilation of positrons. After some tracking, the particles
either stop, interact or escape the converter. The converter, e.g. tungsten
of 16 $X_0$, primarily aims to convert the high energy Compton $\gamma$-rays
to $e^{\pm}$ particles, since only charged particles generate Cerenkov
light within quartz fibers. The position of the strongly forward collimated
photons is maintained by the $e^{\pm}$ shower profile
when escaping the converter, as demonstrated by simulation 
in the next section. SR photons
constitute some background and, due to their asymmetry with respect to x = 0,
they can disturb the original position of the Compton photons after pair creation.
Therefore, the converter should absorb most of these photons and $X_{\gamma}$
position measurements have to account for some possible residual asymmetric
detector response. The converter is supposed to have a cross section of $2 \times 2$ cm$^2$
and a length of 16 $X_0$. The transverse dimension of the converter is mainly dictated
by the small displacement of the beam particles 25 m downstream
of the spectrometer magnet. A converter of e.g. 26 $X_0$ with more efficient
SR removal results to less precise $\gamma$-centroid measurements and
is considered to be less favored.

\vspace{3mm}
\noindent
A completely different way to record 
the undeflected beam position relies on monitoring the edge of
SR light at x = 0, without a converter in front of the position
device. Dedicated and novel SR devices were suggested in \cite{SR_paper}.
In this paper, we propose to employ the plane-parallel avalanche detector with
gas amplification. SR light which passes a $10 \times 10$ mm$^2$
entrance window of 1 mm beryllium\footnote{ The beryllium foil
also acts as the high-voltage cathode plane.}
generates an avalanche in xenon gas at 60 atm over a range of 1.5 mm,
the gap between the anode and cathode. The transverse size
of the avalanche is expected to be close or below 1 $\mu$m, and due to the
amplification process, a large number of electrons is produced
and generates a sufficiently strong output signal \cite{SR_paper}.
The anode plane of the detector consists of 1 $\mu$m nickel layers with 2 $\mu$m
NiO dielectric separation in between. Such a geometry matches very well
the transverse size of the avalanche and permits 
submicrometer access of the position of the SR edge.
Since no converter is planed in this scheme, the $10^6$ high energy
Compton photons are now background.
Their impact on the accuracy of the SR edge
is negligible as will be shown below.

\subsection {\boldmath Simulation Studies \unboldmath}

A full Monte Carlo simulation based on the GEANT toolkit\cite{GEANT} 
\footnote{At the beginning of the study GEANT3 (version 3.21/14) has been used,
while later on GEANT4 (version 4.8.2) was applied.} has been developed
to analyze the basic properties of the Compton spectrometer
and to evaluate design parameters for the detectors. Bunches of $2 \cdot 10^{10}$ electrons
are colliding with unpolarized or
circular polarized infrared or green laser pulses of 10 ps duration 
by a Compton generator\footnote{Operating with a $CO_2$ laser requires larger drift space
than available in the present BDS. Therefore, no simulation results are presented
for such a laser.}.
The generator accounts for 
an internal electron bunch energy spread of 0.15\% which is slightly larger
than the values given in \cite{RDR}  \footnote{ The ILC Reference Design
Report lists for the relative energy
spread 0.14 and 0.10\% for the electrons, respectively, positrons.
The larger value for the electrons is due to their passage
through a long undulator.},
a transverse bunch profile of 20 $\mu$m and 2 $\mu$m in 
horizontal, respectively, vertical direction
and a 300 $\mu$m extension along the beam direction, all of Gaussian shape.
An angular spread of 1 and 0.5 $\mu$rad in x-,
respectively, y-direction has been assumed. Such input parameters are in accord with
ILC beam properties within the BDS. A high-power pulsed
laser with either $E_{\lambda}$ = 1.165 eV or 2.33 eV is focused onto 
the incident beam with a crossing angle of 8 mrad.
The transverse spot size of the laser at the Compton IP
is set to 100 (50) $\mu$m for the infrared (green) laser,
and the laser angular spread was assigned to 2.50 (1.25) mrad.
Also, perfect laser pointing stability and instantaneous laser power are assumed.
As default event rate, $10^6$ Compton scatters are generated for single bunch crossing.

\vspace{3mm}
\noindent
Compton recoil electrons and photons as well as non-interacting beam particles
are tracked through the spectrometer and recorded by the detectors.
A special vacuum chamber as sketched in Fig.~\ref{fig:vacuum_chamber} ensures
negligible Coulomb scattering. The magnet
provides a fixed bend of 1 mrad for all beam energies anticipated.
At the nominal energy of 250 GeV, the magnet rigidity
corresponds to 0.84 Tm for a magnet length of 3 m.
The simulation also includes a 1\% integrated B-field fraction 
for the fringe field.
Synchrotron radiation with properties as discussed in \cite{SR_paper}
is enabled when electrons pass through the magnet. On average, a beam particle radiates
about 5 photons with an average energy of 3.8 MeV and an energy spectrum 
that peaks below 1 MeV. 

\vspace{3mm}
\noindent
The position sensitive detectors which perform $X_{\gamma}, X_{beam}$ and $X_{edge}$
measurements are located 25 m downstream of the spectrometer magnet.
For the edge electrons, we assume either a 
diamond strip or a quartz fiber detector\footnote{Due to 
the large radiation dose expected, a silicon strip detector will not
be considered here unless very radiation hard Si detectors
become available.}. Both detector options have a
transverse size of $1 \times 1$ cm$^2$. For the 100 $\mu$m thick diamond detector
a pitch of 50 $\mu$m and a strip width of 15 $\mu$m were chosen.
A crosstalk of 2\% and a 99\% detection efficiency were assumed.
When passing through a thin layer of matter, charged particles
lose energy which follows in good approximation a Landau distribution.
Thereby, in rare cases the electron transfers a large amount of energy
within the sensor which implies a large charge signal.
A code based on GEANT has been written that simulates
all physical processes taking place in the DSD and calculates
the energy deposited along the particle track in the detector\footnote{ In general 
the charge signal depends on the energy deposited along the track 
rather than the energy loss. Some of the energy lost by the particle is carried away
by secondary electrons or by Cerenkov radiation.}. The resulting deposited energy is used
to weight each electron and, after summation over all entries in a given channel,
the total signal is shown in the corresponding figures.

\vspace{3mm}
\noindent
For the quartz fiber detector,
Compton electrons are measured by a single layer
of 50 $\mu$m square fibers. A cladding thickness of 5 $\mu$m
on each side results in an active fiber core of 40 $\mu$m.
Crosstalk between fibers was set to 3\%.
Largest response of the detector is obtained
when the angle of particle incidence corresponds to the Cerenkov angle of 46$^o$.
Therefore, the quartz fiber detector was inclined by 45$^o$ with respect to the
vertical direction. Since only a fraction of typically a few percent
of the light produced in the fibers is trapped and transported to the light detector,
the small probability to detect a minimum ionizing particle
is to great extent compensated by the large number of electrons traversing a single fiber.
Therefore, despite a small
single-particle light yield, a detection efficiency for individual fibers
of 95\% was assumed. The quartz fiber response was simulated by counting the number
of light photons generated by each electron along its path through the detector.
The sum over all such photons within a fiber is proportional to the output
signal and is plotted in the figures.

\begin{figure}[ht]
\begin{center}
\hspace{-3mm}
\includegraphics[height=13.0cm,width=17.0cm]{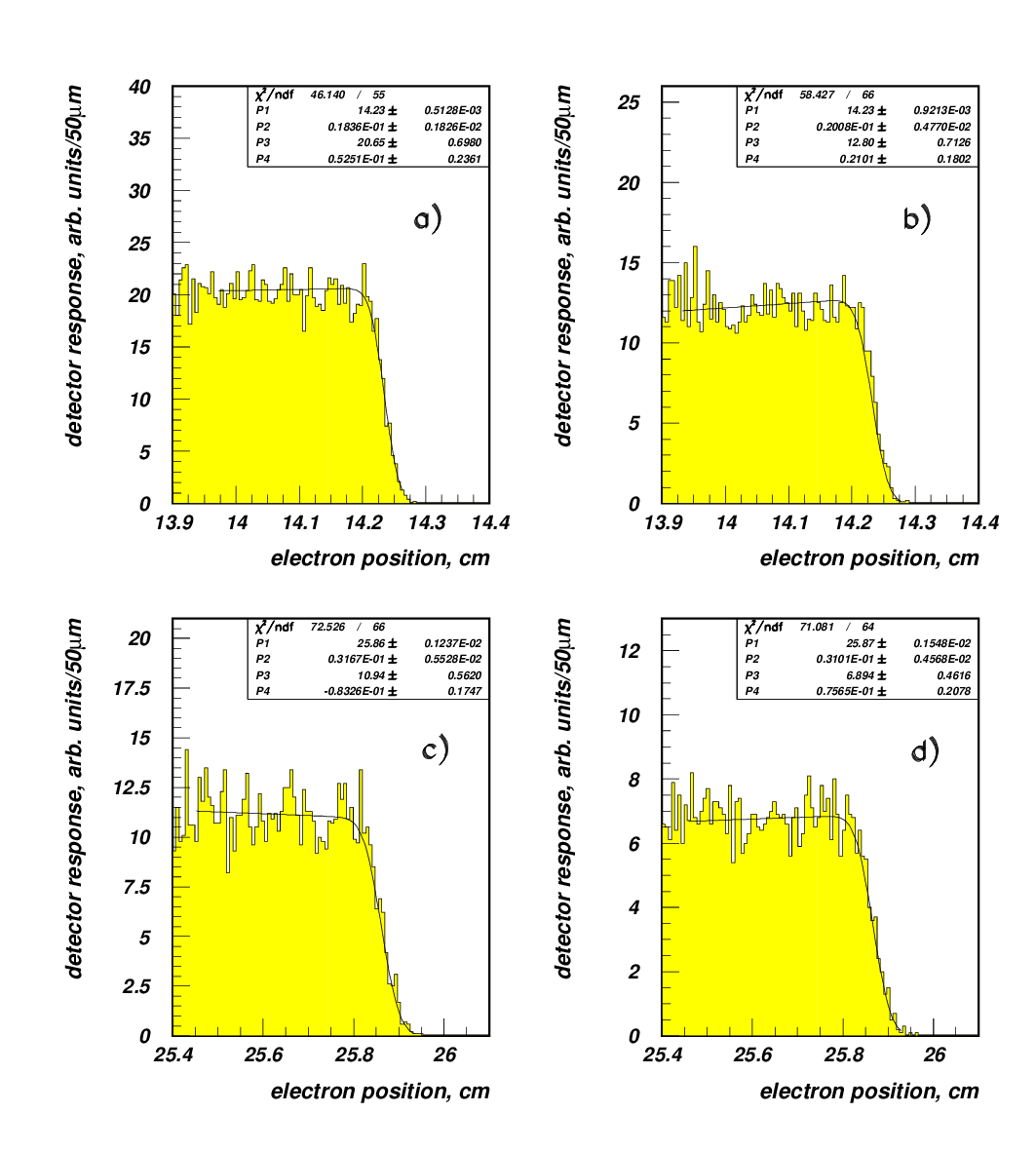}
\caption{ \label{fig:edge_positions}Infrared laser electron edge position simulations
   for a) a diamond strip detector and b) a quartz fiber detector. For the green laser
   the corresponding spectra are shown in c) and d). The incident beam energy is 250 GeV
   and the number of events were normalized to $10^6$ primary Compton scatters.
   The lines represent the result of the fits.}
\end{center}
\end{figure}

\vspace{3mm}
\noindent
The profile of scattered electrons measured 
by both detectors considered is shown in Fig.~\ref{fig:edge_positions}.
For an incident beam energy of 250 GeV, Figs.~\ref{fig:edge_positions} (a) and (b) plot
examples of simulated edge spectra for the diamond strip and quartz fiber detectors
utilizing the 1.165 eV infrared laser.
For the green laser with $E_{\lambda}$ = 2.33 eV, analogous spectra
are displayed in parts (c) and (d) of the figure.
All spectra are normalized to $10^6$ primary Compton events assumed for single bunch crossing.
As can be seen, the expected sharp edges of the spectra are somewhat diluted,
mainly due to the energy spread of the beam particles, angular dispersions,
beam spot size, detector position resolution
and crosstalk.
The edge positions of the spectra were obtained by a fit of a function
which results from a step-function plus a (uniform) background
folded by a Gaussian as proposed in e.g.
\cite{BESSY, Novosibirsk}:
\begin{eqnarray}
  G(x,p_{1...6}) &=& \frac{1}{2}(p_{3}+p_{4}(x-p_{1}))
      \cdot \mbox{erfc} \left[ \frac{x-p_{1}}{\sqrt{2} p_{2}} \right]
\nonumber \\
      & &-\frac{p_2p_4}{\sqrt{2\pi}}\cdot \mbox{exp}\left[
      \frac{(x-p_{1})^2}{2 p_{2}^2} \right] + p_5 + p_6(x-p_1)~.  \label{equ:fit_procedure}
\end{eqnarray}
The edge position $p_1$, the edge width $p_2$,
the amplitude of the edge $p_3$, the slope $p_4$,
the background level $p_5$ and its slope $p_6$ were treated
as free parameters.
Assuming $p_5$ = $p_6$ = 0 in our particular case, 
the errors of the edge positions were found in the range of 5 to 15 $\mu$m,
with values of 5 and 9 (12 and 15) $\mu$m for the infrared (green) laser.
These numbers are in 
accord with the endpoint position demands shown in e.g. Figs.~\ref{fig:beam_error_edge}
and \ref{fig:error_bpm_all} for the approach of recording three particle positions,
$X_{\gamma}$, $X_{beam}$ and $X_{edge}$. Similar uncertainties
were obtained if 100 $\mu$m fibers 
with 10 $\mu$m cladding were utilized. It is also evident
that method A based on direct edge energy measurements
(by means of precise B-field integral and edge displacement
information) seems to be nonfavored: precisions of
edge electron displacements of a fraction of a micrometer up to only few micrometers
(see Fig.~\ref{fig:beam_error_displ}) are difficult to achieve without additional effort.

\vspace{3mm}
\noindent
In principle, the beam polarization may affect the endpoint $p_1$ which might be coupled
with the slope of the energy spectrum $p_4$ in the vicinity of the edge position
as indicated in Fig.~\ref{fig:polarization_cross_section}.
By Compton simulation of 80\% polarized electrons of 250 GeV with circular polarized
infrared laser light we found that the edge position differs by less than 1 $\mu$m
with respect to the case of unpolarized electrons.
Thereby, Compton scattering of polarized beams will not noticeably
modify $p_1$ and hence the beam energy measurement.

\vspace{3mm}
\noindent
The assumption of a Gaussian internal energy spread 
relies on ongoing machine design studies. As long as collective effects 
as intra beam scattering (IBS) or interactions with the vacuum chamber impedance
are negligible the energy spread is expected to be of Gaussian shape.
Since at present a final design of the vacuum chamber to minimize 
the beam impedance and IBS effects is not completed a realistic shape 
of the energy distribution is missing.
Deviations from a Gaussian, if any, are however expected to be small \cite{Susanna}.
Preliminary accelerator simulations reveal
that the energy spread is close to a Gaussian distribution \cite{Latina}
and support our assumption.
This holds for the electrons as well as the positrons despite  
different sizes of the relative energy spread.
If it will be demonstrated by measurements that the energy spread is not Gaussian
distributed, the fitting function (\ref{equ:fit_procedure}) has to be modified
according to the findings.

\vspace{3mm}
\noindent
For the diamond detector, the number of electrons per 50 $\mu$m detector pitch
is about 200 (110) for the infrared (green) laser. The deposited energy
amounts to $4.0 (1.6) \cdot 10^{-5}$ W, of which 90\% is due to the current
induced in the diamond and 10\% due to ionization. 
The associated heat load is expected
to be of no concern since the thermal conductivity of diamond is very high.
The heat, locally induced, can propagate very quickly away before the next bunch arrives.

\vspace{3mm}
\noindent
Using the density of diamond (3.5 g/cm$^3$), the deposited energy as given above,
a bunch crossing rate of $15 \cdot 10^3$ Hz and $10^7$ seconds for a year of
data taking, a radiation dose of 1.3 (0.5) MGy (with an uncertainty of about 30\%)
is expected. This level is considerably
below irradiation level investigations by the RD42 collaboration \cite{Diamond}
ensuring survivability of the detector.

\vspace{3mm}
\noindent
For the quartz fiber detector, about 120 (70) scattered electrons\footnote{ These numbers
are corrected for 20\% detector inefficiency.}
cross a single fiber. Most of the energy loss of the electrons is caused by
ionization, while emission of Cerenkov light constitutes only a minor contribution.
The released energy within 70 $\mu$m fiber
pathlength is approximately $2.6 (0.8) \cdot 10^{-6}$ W, which together with
the density of quartz ($SiO_2$) of 2.2 g/cm$^3$ yields a radiation dose of 0.054 (0.021) MGy
per year. Again, these levels are associated with an uncertainty of 30\%.
Since absorbed doses up to few hundred MGy were measured in quartz fibers
without serious degradation\footnote{For ultra-pure quartz, a limit
has not yet been seen.} \cite{Gorodetzki_2}, radiation damage
of a quartz fiber detector for edge electron measurements will not matter at all.

\begin{figure}[ht]
\begin{minipage}[b]{0.55\linewidth}
 \centering
  \includegraphics[height=7cm,width=8.50cm]{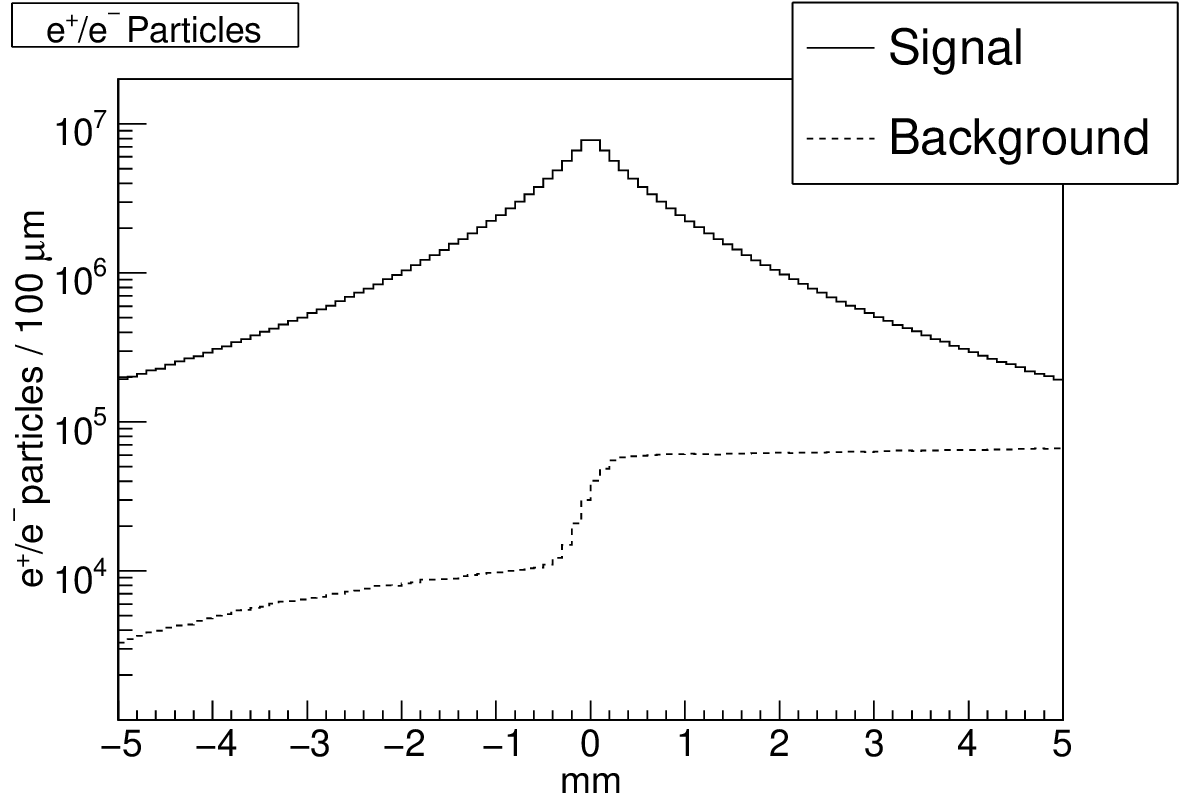}
\end{minipage}
\begin{minipage}[b]{0.45\linewidth}
 \centering
  \includegraphics[height=7cm,width=8.50cm]{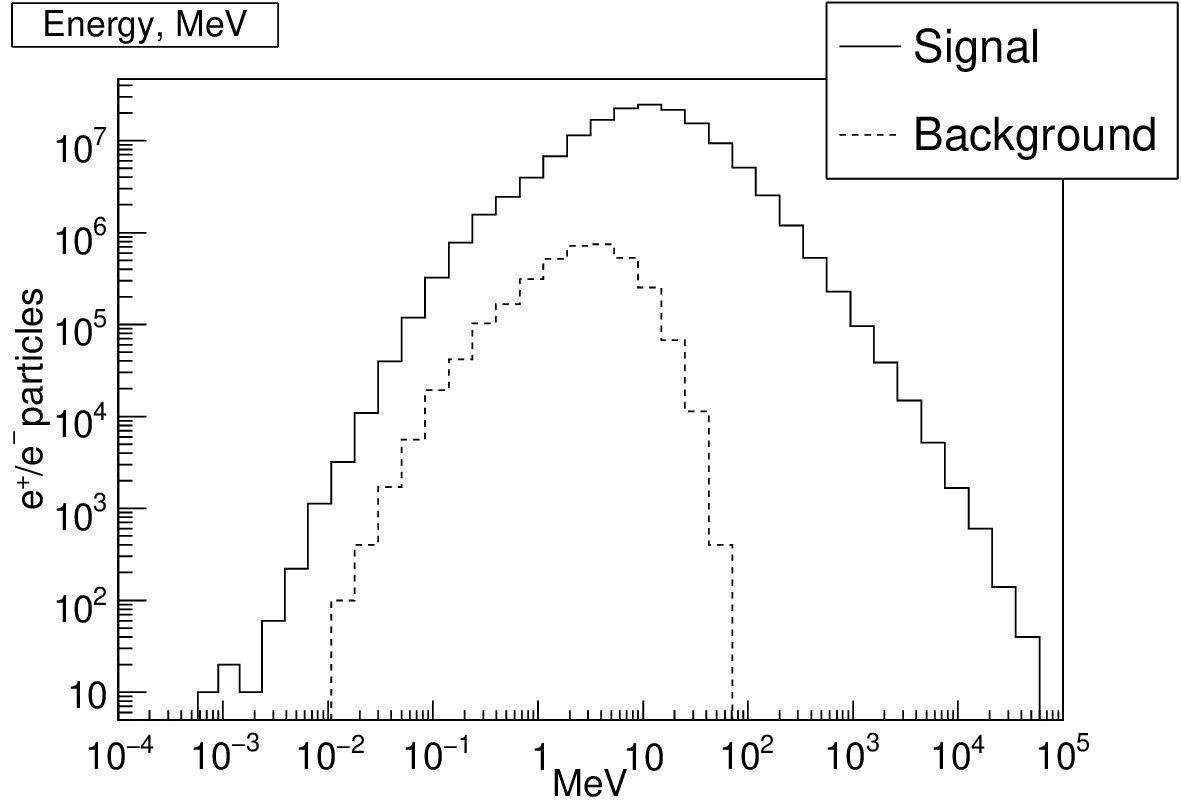}
\end{minipage}
 \caption{Left: Number of charged particles escaping the 16 radiation lengths tungsten
 converter as a function of x.
 Right: Energy distribution of charged particles escaping the converter.
 The 'signal' spectra are normalized to $10^6$ Compton scatters, while the
 'background' spectra are normalized to $2 \cdot 10^{10}$ beam particles
 within a bunch. }
 \label{fig:charged_particle_x_energy}
\end{figure}

\vspace{3mm}
\noindent
Within the approach of measuring $X_{\gamma}$,
the center-of-gravity of the Compton scattered $\gamma$-rays is measured
indirectly via conversion to electrons and positrons within
a 16 radiation lengths tungsten converter. When entering the converter,
the photons are concentrated within a spot of approximately 250 $\mu$m r.m.s.,
a size which is dominated by the $\sim$1/$\gamma$ angular distribution
of the Compton process.
After a first estimate of the thickness of the converter, a full simulation
of the 56 mm long conversion material has been performed. In particular,
the process of converting the $10^6$ Compton photons together with the 
SR photons along with the trajectories of the resulting electrons and positrons
through the converter and into the fiber detector was simulated.
Despite the small transverse extension of the converter,
the core of the shower particles caused by
Compton photons is assumed to maintain the initial $\gamma$-centroid position
(being at x = 0.0 in the simulation). Directly after the converter
the quartz fiber detector array of 50 $\mu$m fibers has been placed 
in order to measure the $e^{\pm}$ shower particles
from which the $\gamma$-centroid position has to be deduced.
Fig.~\ref{fig:charged_particle_x_energy}(left)
shows the number of charged particles escaping the converter as a function
of x, while their energy behavior is shown on the
right-hand side\footnote{ Analogous spectra are
obtained for the  vertical direction as well as if the infrared laser 
is replaced by the green laser.}.
The spectra indicated as 'Signal'
are $e^{\pm}$ particles from Compton photons, whereas those
marked as 'Background' are from synchrotron radiation.
We expect $1.5 \cdot 10^8$ charged particles from $10^6$ Compton events,
with an average energy of 25.8 MeV. Their density distribution, $dN/dx$, 
clearly peaks at x = 0.

\begin{figure}[ht]
\begin{minipage}[b]{0.55\linewidth}
 \centering
  \includegraphics[height=7cm,width=8.50cm]{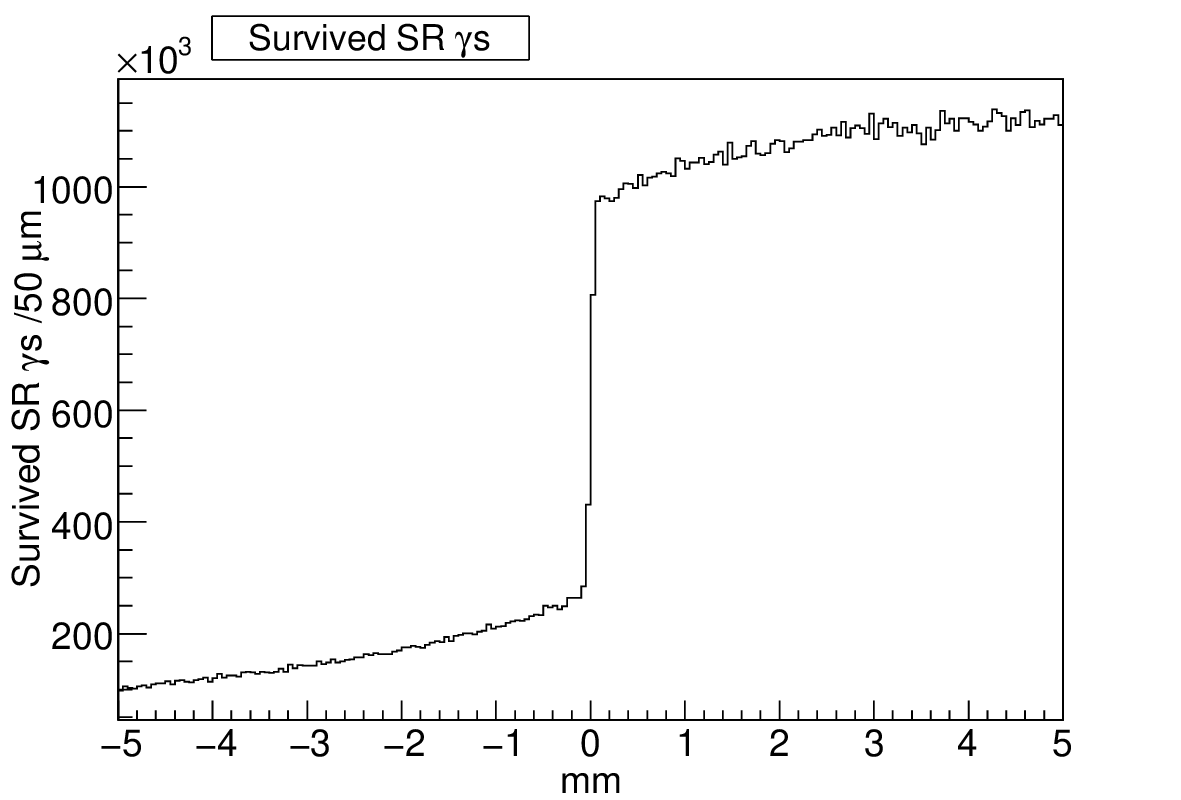}
\end{minipage}
\begin{minipage}[b]{0.45\linewidth}
 \centering
  \includegraphics[height=7cm,width=8.50cm]{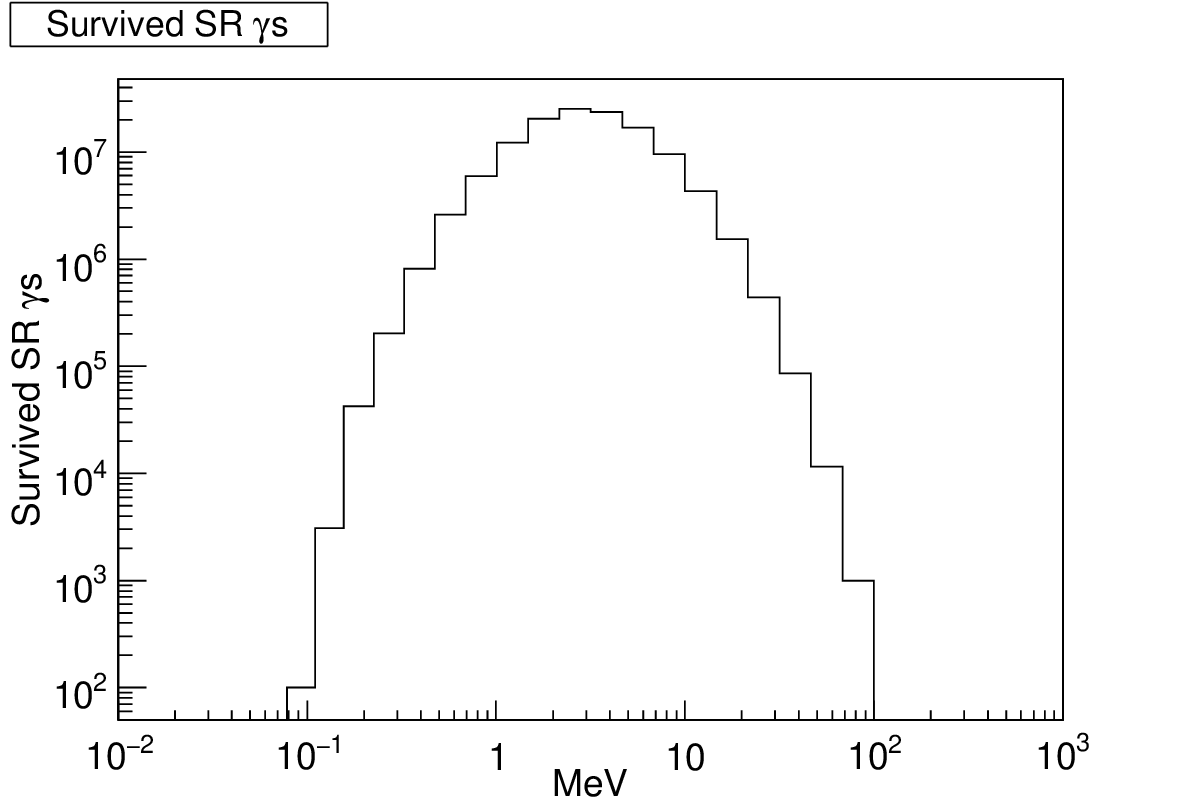}
\end{minipage}
\caption{Left: Number of SR $\gamma$-rays which escape the
  16 radiation lengths tungsten converter as a function of x.
   Right: Energy distribution of SR $\gamma$-rays escaping the converter.
   Both spectra are normalized to $2 \cdot 10^{10}$ beam particles within a bunch. }
 \label{fig:gamma_x_energy}
\end{figure}

\vspace{3mm}
\noindent
Besides of charged particles, photons also escape the converter.
They are either generated within electromagnetic showers
from Compton scattered and SR $\gamma$-rays or are SR photons
which pass the converter without interaction. A fraction of less than 2\%
of the original SR yield with an average energy of 3.9 MeV survives.
Their $dN/dx$ and energy spectra are shown in Fig.~\ref{fig:gamma_x_energy}.
The overwhelming fraction of the SR photons is converted to $e^{\pm}$ pairs and some of them
($3.5 \cdot 10^6$) escapes the converter, see Fig.~\ref{fig:charged_particle_x_energy}.
They are expected to affect the $\gamma$-centroid position
and have to be accounted for in any $X_{\gamma}$ determinations.

\begin{figure}[ht]
 \centering
  \includegraphics[height=7.0cm,width=10.0cm]{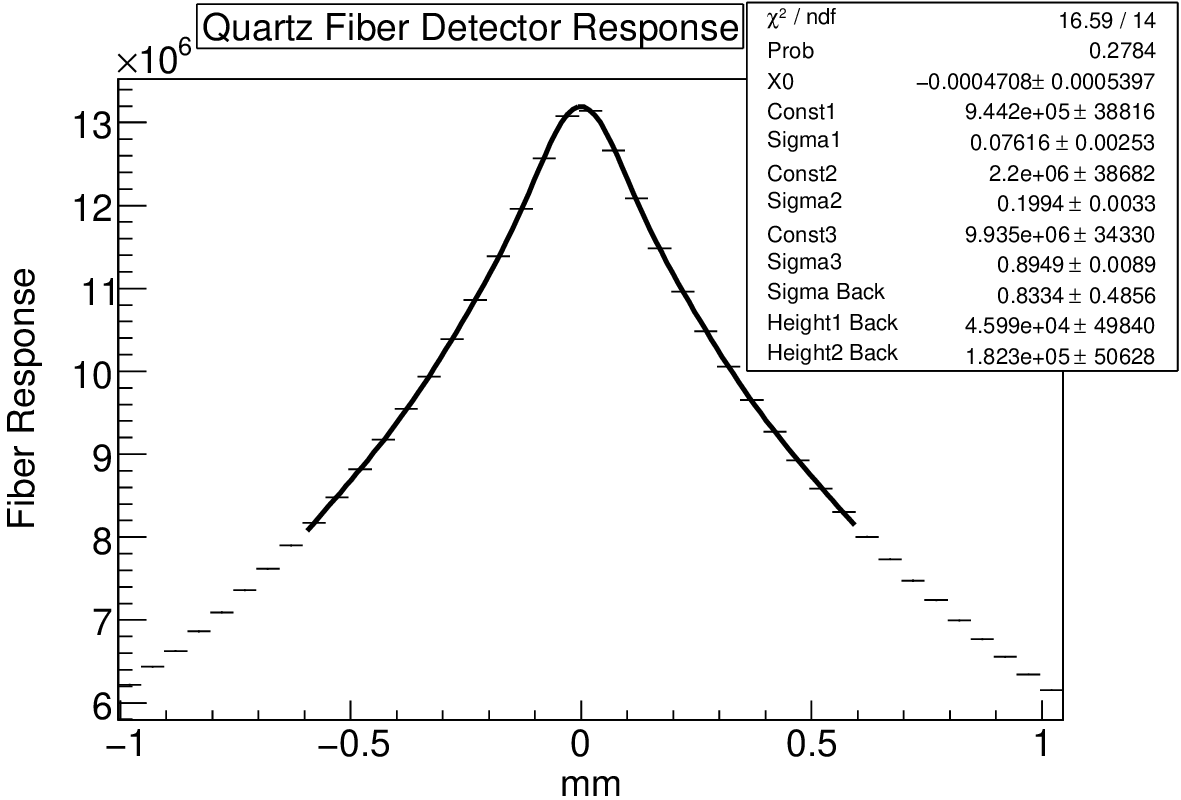}
  \caption{Cerenkov light response of all charged particles passing 
   the quartz fiber detector.
   The curve is the result of a fit of the sum of three Gaussian distributions
   and the step function in eq.(\ref{equ:fit_procedure}) with $p_4$ = $p_6$ = 0.}
 \label{fig:gamma_det_response}
\end{figure}

\vspace{3mm}
\noindent
For the position sensitive $X_{\gamma}$ device, 
a single layer of quartz fibers is supposed
with properties identical to those for the edge electron detector.
Basically, this detector should have
a large sensitivity to charged particles from pair production of Compton photons
within the converter and 'blind' with respect to background (SR) $\gamma$-rays.
In Fig.~\ref{fig:gamma_det_response}, the response function of the detector in terms
of the amount of Cerenkov light generated from all $e^{\pm}$ particles within a fiber
is shown together with the result of a fit.
An electron energy detection threshold of 0.6 MeV for Cerenkov light production is
included. The fit result is based on a two-step procedure. First,
due to an a priori unknown precise $\gamma$-centroid position,
$X_{\gamma}$ is approximately determined by a simple algorithm \cite{Morhac},
which fixes the peak position within about $\pm$25 $\mu$m. Then,
selecting a fitting range of some $\pm$600 $\mu$m around this preliminary
centroid, an empirical fit of the sum of three Gaussians and the step function  
in eq.(\ref{equ:fit_procedure}), with $p_4$ = $p_6$ = 0, provides the ultimate
peak position of $X_{\gamma}$ = -0.47 $\pm$0.54 $\mu$m
with a $\chi^2/NDF$ = 16.59/14,
corresponding to 27.8\% probability\footnote{ If the fit is performed
with the sum of only two Gaussians and the step function, the $\chi^2/NDF$
is significantly worse.}.
The fit range chosen excludes particles which are less sensitive 
to the peak position but sensitive to the background.
The peak value found is in good agreement with the expectation
of zero and its error is less than the anticipated limit of $\sim$1 $\mu$m.
The spectrum in  Fig.~\ref{fig:gamma_det_response} resembles the response
of all escaping $e^{\pm}$ particles generated
from $10^6$ Compton photons and the appropriate fraction of SR,
after normalization to $2 \cdot 10^{10}$ beam electrons.
The latter causes a slight asymmetry with respect to x = 0 and is the reason
to include the step function within the fit.
As a consequence, a rather complicated response behavior is obtained and
after some trials the spectrum was reasonably described by the selected ansatz.
If instead of a 50 $\mu$m fine segmented detector an array
of 100 $\mu$m quartz fibers is utilized the centroid position
and its error are found to be in agreement with the values quoted above.
Irrespectively of the details for the final design of the converter-fiber detector system,
this option seems to be capable to meet the requirements,
in particular if instead of only one
fiber layer several layers with some staggering are employed.
%

\begin{figure}[ht]
 \centering
  \includegraphics[height=7.0cm,width=15.0cm]{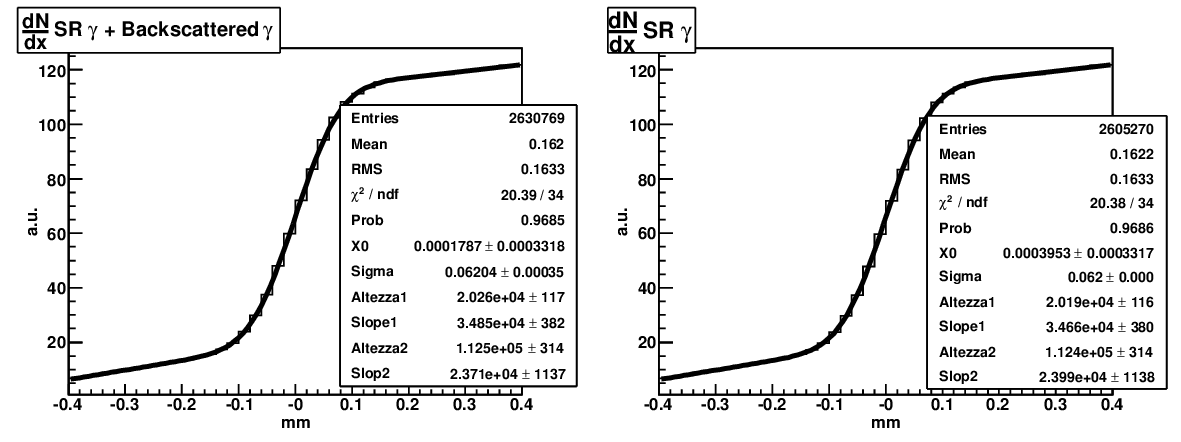}
  \caption{Left: Response function of the avalanche SR edge detector for all
   signal and background photons. Right: Response function of the same detector
   for only SR signal photons.
   The curves are the results of a fit 
   of eq.(\ref{equ:fit_procedure}) supplemented by an additional background tail.}
 \label{fig:avalanche_det_response}
\end{figure}

\vspace{3mm}
\noindent
Basically, a different approach to record the incident beam direction
consists of using a SR edge detector.
The avalanche detector of Ref.\cite{SR_paper} with xenon
being in a superfluid state with a density of 3.05 g/cm$^3$ is proposed
to perform SR edge position measurements around x = 0. A detector acceptance
of $\pm$5 mm will be exposed by some 20\% of the $10^{11}$ SR photons
and all $10^6$ Compton recoil $\gamma$-rays, which are considered now as background.
Photons traversing the detector interact with the xenon
so that electrons are created via e.g. the photoelectric effect
or pair production.
These electrons drift towards the anode and in collisions with xenon atoms
they liberate further electrons. This process is accompanied by loss of energy
of the electrons and deflection from their incident direction.
The response of such a detector was simulated
and the x-position of each electron-atom collision weighted 
by the corresponding released energy is plotted
in Fig.~\ref{fig:avalanche_det_response} for all photons (left) and only 
the SR $\gamma$-ray (right). Clearly, the SR edge at x = 0 is
well recognized and a fit using eq.(\ref{equ:fit_procedure})
provides $X_\gamma = 0.18 \pm$ 0.33 $\mu$m. This number is, despite of the crudeness
of the simulation, in perfect agreement with the demands and indicate that
the response of Compton photon as background is not important.
Hence, $X_\gamma$ position measurements
can be performed with an avalanche SR detector as proposed in \cite{SR_paper}.
Presently, R\&D for such a detector is ongoing and first results are expected
in 2008/09 \cite{avalanche_det}.

\subsection {\boldmath Laser Power \unboldmath}
                               
So far we assumed $10^6$ Compton interactions per crossing
regardless of the laser type used. To achieve such an event rate,
the required laser power is estimated as follows. 
Utilizing for the incident electron beam
transverse bunch sizes of $\sigma_x$ = 20 $\mu$m,
$\sigma_y$ = 2 $\mu$m and 300 $\mu$m in longitudinal direction at the Compton IP,
for the transverse laser spot size 100 (50) $\mu$m in the case of an infrared (green) laser,
a pulse duration of 10 ps, a crossing angle of 8 mrad
and $2 \cdot 10^{10}$ electrons per bunch, the infrared laser $e\gamma$ luminosity per crossing
is according to eq.({\ref{equ:pulse_lumi}) evaluated to 0.166 per millibarn
and $\mu$J,
while a green laser provides 0.307 mb$^{-1}\mu$J$^{-1}$ \footnote{Shortening
the pulse duration to 5 ps increases
the luminosity by only 0.6\% (2.3\%) for infrared (green) laser operation.}.
If these luminosities are combined with
the corresponding Compton cross section of $\sigma$ = 197.9 mb, respectively,
137.7 mb, a bunch related laser power of 30 or 24 mJ is obtained.
At present, such lasers that match the pattern of the incident electron bunches
are not commercially available. But the FLASH collaboration
\cite{will, schreiber}, employing a laser in the infrared region with good
reliability, and ongoing R\&D for green lasers within the ILC community \cite{Blair}
will set milestones in the future, from which this proposal could greatly benefit.

\subsection {\boldmath Potential Background Processes of Electron to Photon Conversion \unboldmath}

Usually, the characteristics of Compton scattering are calculated
within the Born approximation, see Sect.2 as an example.
Compton scattering processes
at the ILC with large bunch densities, large laser flash energies
and small pulse lengths ensure sufficient $e\gamma$ luminosity, which is important
for precise $E_b$ determination. 
When the thickness of the laser target is about one collision length as at the ILC,
each electron may undergo multiple Compton scattering within the crossing region \cite{Telnov}.
The probability might not be small because, after a large energy loss
in a first collision, the Compton cross section increases       
and together with the high particle densities of the colliding bunches
further collisions can be caused. Such multiple scattering leads also to a low energy tail
in the energy spectrum of the scattered electrons and could modify
the sharp edge behavior. Using the program package CAIN \cite{CAIN}
the rate of electrons which scatter more than once compared to single scatters
has been conservatively evaluated to $\sim$0.7$\cdot 10^{-4}$,
utilizing default beam parameters and a $CO_2$ laser with
a pulse power of 1 mJ. Thus, out of $10^6$ Compton scatters
only a small fraction undergoes multiple scattering.
The disturbed energy spectrum is displayed in Fig.~\ref{fig:E_multiple_scattering},
while the position distribution of the electrons 50 m downstream of the magnet
in Fig.~\ref{fig:x_multiple_scattering}. No significant distortion of the
recoil electron spectra is expected.
\begin{figure}[ht]
\begin{minipage}[b]{0.45\linewidth}
 \centering
  \includegraphics[height=5.0cm,width=7.0cm]{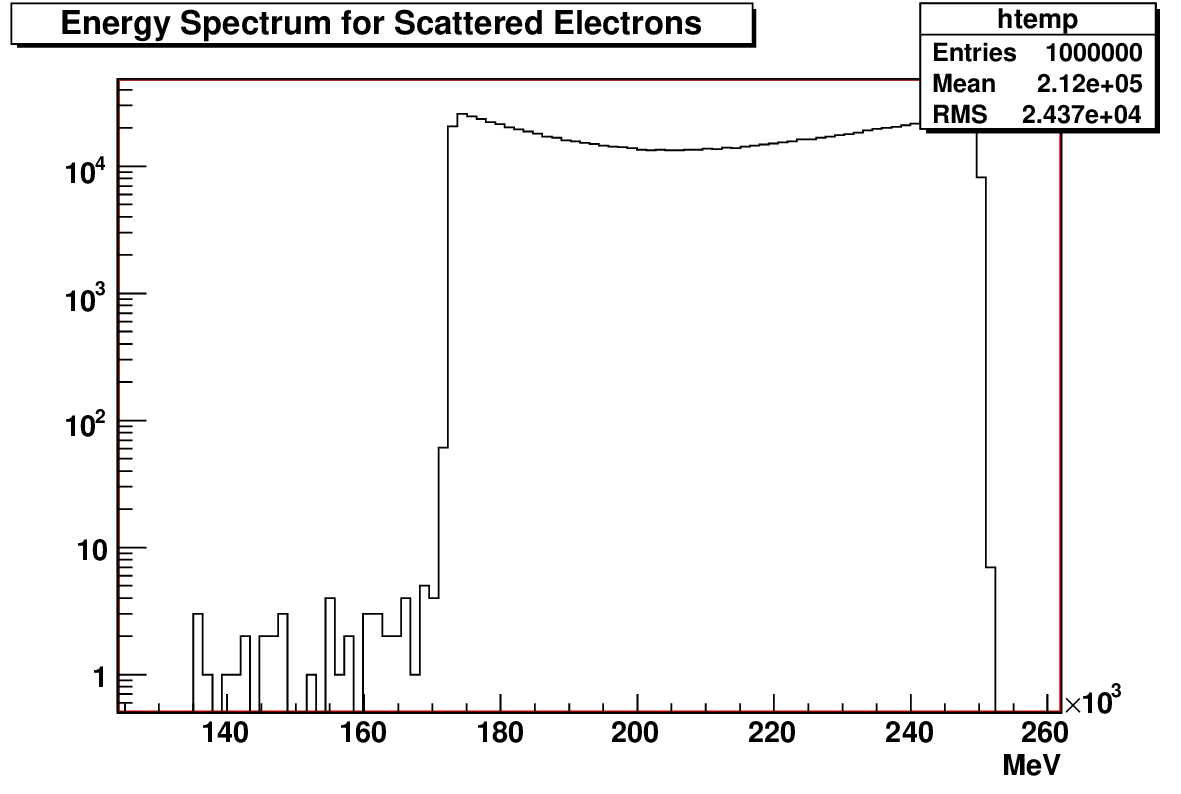}
 \end{minipage}
\hspace{1cm}
\begin{minipage}[b]{0.5\linewidth}
 \centering
  \includegraphics[height=5.0cm,width=7.0cm]{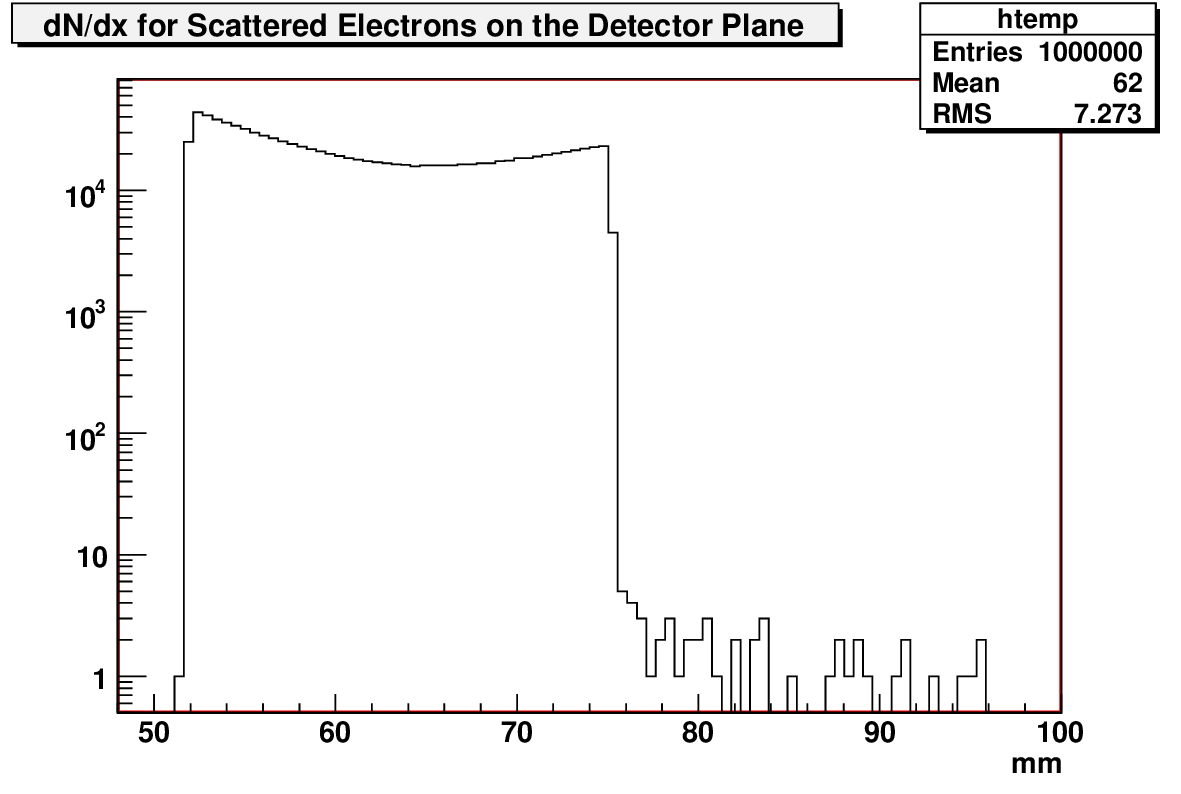}
\end{minipage}
\begin{minipage}[t]{0.45\linewidth}
 \caption{Energy spectrum of Compton electrons including multiple scattering
   for a $CO_2$ laser at 250 GeV beam energy.}
 \label{fig:E_multiple_scattering}
\end{minipage}
\hspace{1.5cm}
\begin{minipage}[t]{0.45\linewidth}
 \caption{Position spectrum of Compton electrons 50 m downstream of the default magnet
  for a $CO_2$ laser at 250 GeV beam energy.}
\label{fig:x_multiple_scattering}
\end{minipage}
\end{figure}

\vspace{3mm}
\noindent
For the calculation of the $e \rightarrow \gamma$ conversion efficiency, 
one has besides geometrical
properties of the laser and the Compton effect also to consider so-called
nonlinear effects in the scattering process. Since the field in the laser wave
at the crossing region
can be very strong, electrons have a chance to interact simultaneously 
with several laser photons
(called nonlinear QED effects). These nonlinear effects
are characterized by the parameter \cite{Berestetskii}
\begin{equation}
   \xi^2 = \frac{2n_{\gamma} \cdot r_{e}^2 \cdot \lambda} {\alpha_{fsc}}~,
\end{equation}
where $n_{\gamma}$ is the density of the laser photons, $r_e$ the classical
electron radius, $\lambda$ the laser wavelength and $\alpha_{fsc}$ the fine structure constant.
At $\xi^2 \ll 1$, the electron scatters on only one laser photon,
while at $\xi^2 \gg 1$ on several.

\vspace{3mm}
\noindent
The transverse motion of an electron in the electromagnetic wave leads to
an effective increase of the electron mass and the maximum energy 
of the scattered photon decreases as $E_b \cdot x / (1+x+\xi^2)$,
with x given by eq.(\ref{equ:x_variabel}).
Thus, with growing $\xi^2$ the energy spectrum of the Compton electrons will be
modified in two respects, (i) the spectrum is shifted to higher energies
and (ii) higher harmonics appear.
Simulations with CAIN showed that for a $CO_2$ laser
with 1 mJ pulse power, $\xi^2 = 1.04 \cdot 10^{-5}$, so that within $10^6$
Compton events about 10 electrons absorb two photons at the same time.
The relative shift of the edge energy is estimated to $3.2 \cdot 10^{-6}$, a value
practically not accessible by any of the detection systems proposed.

\vspace{3mm}
\noindent
Besides nonlinear QED effects  higher order QED corrections
may also affect the electron endpoint behavior. In order to study such corrections
the Compton electron energy cross section has been calculated
for the complete order-$\alpha_{fsc}^3$ approximation and compared
with the Born cross section in Fig.~\ref{fig:QED_corrections}. The computer code
used relies on Ref.\cite{Swartz}. The spectra shown assume
Compton scattering of 500 GeV polarized electrons with green laser pulses
of $P_e\lambda$ = -1. Such conditions
allow for largest higher order contributions. As can be seen,
the $e \gamma \rightarrow e \gamma$ Born approximation (black histogram)
and the Born plus order-$\alpha_{fsc}^3$ correction cross section
(open histogram) are very close to each other. The inclusion
of the process $e \gamma \rightarrow e \gamma \gamma$ enhances the spectrum 
by about 5\% (shaded histogram) without, however, a measurable shift
of the endpoint value. The application of the code for the $e^- e^+ e^-$
final state indicates that the minimum electron energy is about
34.4 GeV, i.e. $e \gamma \rightarrow e e^+ e^-$ contributions
are expected far outside of the region of interest.

\vspace{3mm}
\noindent
In addition, with an increase of the variable x
(eq.(\ref{equ:x_variabel})), $e^+ e^-$ pair creation
by high energy Compton photon collisions with laser photons leads to further background,
which has the potential to disturb edge electron characteristics. 
If x is larger than 4.83 which happens when e.g.
250 GeV electrons collide with green laser light, associated $e^\pm$ pair background
is generated. For beam parameters and laser pulse power as mentioned above,
CAIN provides about 18 $e^+ e^-$ background pairs for $10^6$ Compton events.
Besides this negligible event rate, the energy of such $e^\pm$ background particles
is far away from the energy of the edge electrons.

\vspace{3mm}
\noindent
These preliminary results indicate that
effects as discussed will not affect the properties
of the edge electrons and the backscattered $\gamma$-rays
in a measurable manner. 
\begin{figure}[ht]
\begin{center}
\vspace{3mm}
\includegraphics[height=8.0cm,width=16.0cm]{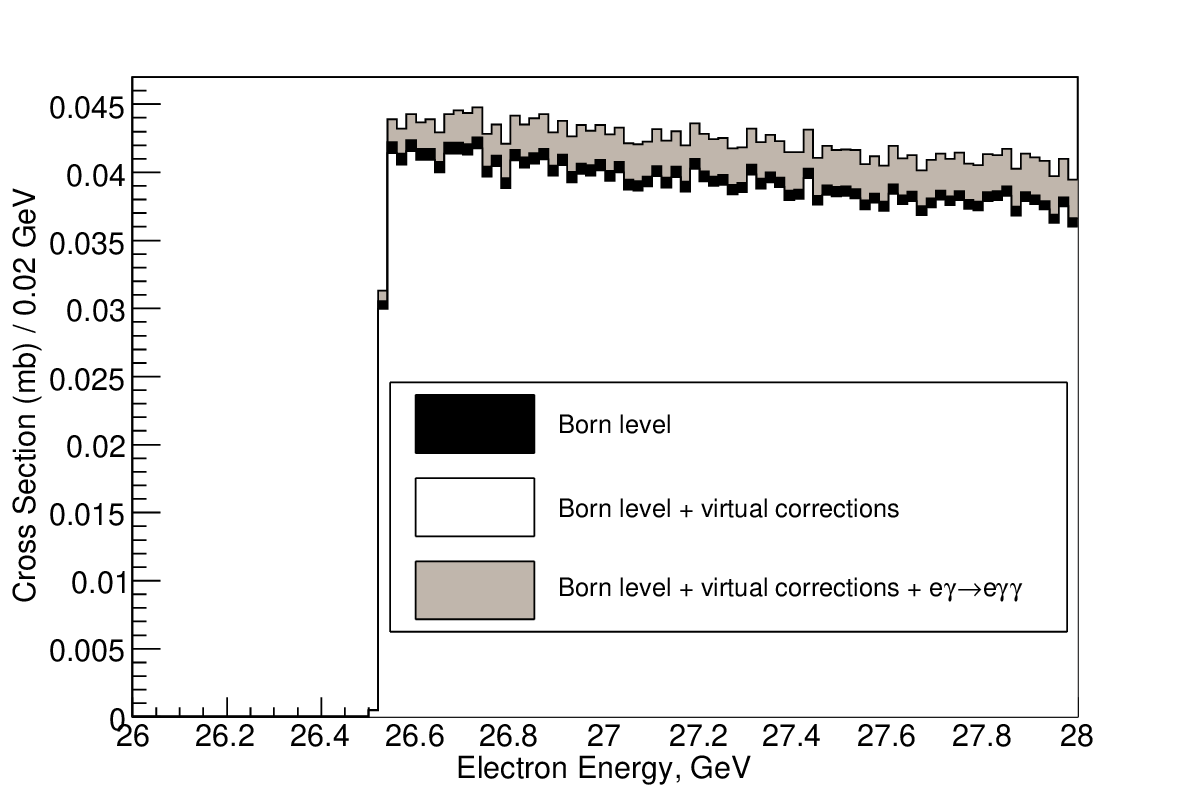}
\caption{ \label{fig:QED_corrections}Born cross section of the Compton process 
  (black histogram),
  Born plus order-$\alpha_{fsc}^3$ correction cross section (open histogram)
  and the Born plus order-$\alpha_{fsc}^3$ correction cross section
  including the reaction $e \gamma \rightarrow e \gamma \gamma$
  (shaded histogram) with $P_e\lambda$ = -1 at 500 GeV.}
\end{center}
\end{figure}

\subsection {\boldmath Error Discussion and Calibration Issues \unboldmath}

Concerning the statistical error, it has been shown that with $10^6$ Compton
events per bunch crossing the accuracy on the beam energy $\Delta E_b/E_b$
can be brought well below the requirement of $10^{-4}$.

\vspace{3mm}
\noindent
Different sources of potential systematic errors may affect
the measurement of $E_b$ and are discussed below. As outlined in Sect.3.10,
Compton processes beyond Born approximation such as multiple scattering,
nonlinear effects in the
$e \rightarrow \gamma$ conversion process, higher order QED corrections
or $e^+ e^-$ pair creation will not significantly modify the scattered photon
and edge electron behavior. As shown by simulation, their effect on
$\Delta E_b/E_b$ is negligible even after summation over all non-Born
approximations discussed.
                                                
\vspace{3mm}
\noindent
Variations of the transverse profile of the incident electron or laser beam 
and the angular spread
as given in Sect.3.8 by a factor 2 or assuming a rectangular beam shape instead
of a Gaussian do not modify the positions of the Compton
photons and edge electrons in a measurable manner as long as 
both beams collide centrally. This means,
$E_b$ determination by Compton scattering is rather robust
against initial state beam variations.

\vspace{3mm}
\noindent                  
Relying on method A, $\Delta E_b/E_b$ is controlled
by the accuracy of the integrated B-field of the spectrometer magnet,
$\Delta B/B$, the drift distance to the detector plane, $\Delta L/L$,
and the offset of the edge position with respect to the primary beamline, $\Delta D/D$,
see eq.(\ref{equ:error_energy-1}) and Fig.~\ref{fig:exampel_1_setup}.
The drift distance can be precisely monitored using an interferometer
\cite{interferometer}. For an accuracy of $\Delta L$ $\simeq$100 $\mu$m, which is feasible,
the relative error of L becomes few times $10^{-6}$ and hence negligible.
The required accuracy for the distance between the undeflected beam and the
endpoint in the order of few micrometers is only possible when
a $CO_2$ laser is used.
To achieve such a precision the $X_\gamma$ and $X_{edge}$
detectors should be installed on a common frame and rigidly connected
in order to avoid relative position movements. In this way and with a frame
made out of a material with a small expansion coefficient like carbon
the relative distance error between both devices can be kept below 1 $\mu$m
\cite{Barber}, even with a $\pm5^o$ change in tunnel temperature.
More important contributions to $\Delta D/D$ constitute
the uncertainties of the $X_{\gamma}$ and $X_{edge}$ position measurements themself.
As discussed in Sect.3.8, the option to perform $X_{\gamma}$ measurements
by means of a quartz fiber detector in conjunction with an adequate converter
provides a statistical precision of 0.54 $\mu$m
and the SR edge method 0.33 $\mu$m, while the total error should not exceed
$\sim$1 $\mu$m, a challenging task.

\vspace{3mm}
\noindent                  
The geometrical precision of quartz fibers is crucial for their precise
assembly into plans and stacks. Arrangements of sufficient precisions including
intrinsic fiber uncertainties are based on experience
\cite{Akchurin, H1-Coll, ATLAS} and can be kept to 2 $\mu$m
or better for small devices. The position of the fibers
can be accurately measured after the assembly and recorded in a database for use
during analysis so that final arrangement errors may be less than
few tenth of a micrometer. Also,
any tilt of the detector with respect to the vertical
direction is important. For example, a misalignment error of 1 mrad
could result to an $X_{\gamma}$ position shift up to 5 $\mu$m. Hence, the detector
has to be aligned to better than 0.1 mrad in order to keep this bias contribution
$\lsim$0.2 $\mu$m.

\vspace{3mm}
\noindent                  
Possible errors caused by the fit procedure of the escaping $e^{\pm}$ position
distribution have been checked by varying the fit region 
within reasonable values or by rebinning the spectrum or by omitting
the first step of the two-step fit procedure.
There is no evidence found for a bias of the nominal fitted values,
taking their statistical precision into account. Conservatively, we assign
an error of 0.1 $\mu$m due to residual uncertainties from the fit.

\vspace{3mm}
\noindent                  
The signal uniformity of the fiber sensor is also an important issue.
Variations from fiber-to-fiber (or strip-to-strip) may have variuos reasons.
There are statistical variations due to noise and fluctuations
of the number of photons emitted, but also variations of the signal
response laterally across the detector which are caused by fluctuations
in the local properties of the sensor. Cerenkov light variations were already
addressed in the GEANT simulations. The remaining fluctuations were        
studied by some additional Gaussian channel-to-channel signal variation
of 0.5\%, 1\% and 2\% of the total signal per fiber. It was found
that the original position of interest is shifted by less than 1 $\mu$m
for a response fluctuation not exceeding 1\%. In practice, the level of uniformity
across the sensor should be measured by an appropriate uniform illumination
with particle beams and the individual channel response
accounted for in the data analysis.

\vspace{3mm}
\noindent                  
Imperfections within the fiber readout chain are difficult to estimate
at the present stage of the project. However, in order to fulfill the request
we set a limit of 0.2 $\mu$m for 
fiber-to-fiber signal variation respectively instability.

\vspace{3mm}
\noindent                  
Alternatively, recording the SR edge by means of the avalanche detector \cite{SR_paper},
the primary beamline position depends on the amount and shape of the fringe field
of the magnet. If the 1\% integrated B-field fraction for the fringe field
as used in the simulation was varied between 0.5 and 2\% and
three field shapes are considered (a simple step function,
a straight line between zero and the B-field strength and a Gaussian distribution)
it was found that the edge positions were distributed over a range
of 1.6 $\mu$m. Thereby, precise measurement of the fringe field is mandatory,
in particular upstream of the magnet.
We estimate a residual error of $\sim$0.2 $\mu$m for $X_{\gamma}$
due to surviving uncertainties of the integrated B-field which includes
imperfections of the fringe field. Also, additional errors of 0.1 and 0.2 $\mu$m
due to imperfections in the fit procedure, respectively, electronics were assigned.

\vspace{3mm}                                                     
\noindent                  
Since the beam and the laser widths are comparable in size, the laser has to be
steered onto the electrons such that both beams collide centrally.
Otherwise, a shift of the center-of-gravity of the scattered photons 
is generated.
Options for laser spot size monitoring and its stabilization are discussed
in Sect.3.6.2. Electron position and emittance 
are supplied by BPMs, respectively, wire-scanner systems distributed within the BDS.
Beam jitter studies suggest for $\sigma_{jitter}$ = (0.1-0.5)$\cdot\sigma_{x(y)}$
\cite{RDR, Seryi},
where $\sigma_{x(y)}$ is the bunch size in x(y)-direction\footnote{ Some machine
experts prefer to use the smaller number. The size of the jitter
will depend on the stability of the ILC beamline components, on energy and kicker jitter
and on the performance of train-to-train and intratrain feedback.}.
For a beam extension $\sigma_{x}$ = 20 $\mu$m for example,
the horizontal jitter is small, in the order of few micrometer,
and negligible in the vertical direction. If one restricts any shift
of $X_{\gamma}$ to be less
than 0.3 $\mu$m, constraints for the beam displacement from one another as a function
of the laser spot size can be derived. With $\sigma_{x}$ = 20 $\mu$m
and a displacement of 15 $\mu$m, 
the laser spot size has to be $\sim$150 $\mu$m, 
or the position of the laser has to be stable within 12 $\mu$m.
Larger spot sizes relax this condition, 
whereas a bigger electron bunch size aggravates the condition
considerably. For example, a 50 $\mu$m bunch requires a laser spot
of the order of 300 $\mu$m and a laser jitter of less than 10 $\mu$m in order
to maintain the photon centroid shift below 0.3 $\mu$m.
Luminosity loss due to larger laser spot sizes can be compensated by either
an increase of the laser power or an increased pulse length, or a combination
of both. If the pulse duration is substantially increased it seems of advantage
to consider horizontal instead of vertical beam crossing. In conclusion,
in order to design a laser system which restricts the shift
of $X_{\gamma}$ due to non-central collisions of electron and laser pulses
to less than $\sim$0.3 $\mu$m some R\&D effort is needed.

\vspace{3mm}
\noindent                  
Summing all contributions quadratically, the total error associated
to the $\gamma$-ray centroid position is $\Delta X_{\gamma}$ $\simeq$ 0.8 $\mu$m
for the quartz fiber detector-converter system,
while the SR edge approach provides $\sim$0.6 $\mu$m.
Both uncertainties are smaller than the required figure of $\sim$1 $\mu$m.

\vspace{3mm}
\noindent                  
Concerning the measurement of the electron endpoint, we found for $X_{edge}$
an uncertainty of about 4 $\mu$m for the $CO_2$ laser (not discussed),
while the infrared (green) laser provides values of 5 and 9 (12 and 15) $\mu$m
for the DSD, respectively, QFD detector. The differences in precision
are mainly due to different event rates per detector strip or fiber.

\vspace{3mm}
\noindent                  
For the diamond strip detector
we assume a similar alignment precision as for the $X_{\gamma}$ fiber detector
discussed above
and a bias estimate of 1.5 $\mu$m due to imperfections of the detector response.

\vspace{3mm}
\noindent                  
The yield of the Cerenkov light in quartz fibers varies
considerably in the vicinity of the Compton edge. Here, the number of
incident electrons per fiber ranges from $\sim$150 to only a few or zero.
Correspondingly, the number of photoelectrons
in the light signal detector also varies considerably, which in turn requires
high quantum efficiency at the wavelength of maximum scintillation and
excellent single-photon detection capability.
If we e.g. assume a zero-signal for fibers
with less than 10 incident electrons,
the refitted endpoint positions were found to be within $\pm$0.6 $\mu$m compared to
the original values. This suggests to assign a total uncertainty associated 
to detector effects of 2 $\mu$m.

\vspace{3mm}
\noindent                  
As emphasized in Sect.3.2, the relative error of the B-field integral should be close
to $2 \cdot 10^{-5}$ in order to reach the required beam energy precision.
Sect.3.5 summarizes aspects necessary to fulfill this challenging request,
and more details can be found in Ref.\cite{ILC_spec}. Here we point out that,
independent of the endpoint detector utilized, in addition
to the bending field provided by the spectrometer dipole itself,
several other sources of magnetic fields may be present in the ILC tunnel 
which might influence the path of the electrons.
A large effect can come from the earth's field and
other contributions might arise from cables which provide current for 
magnets. Such fields within the space between the Compton IP and
the detector plane could spoil the endpoint position
measurement, even if this space is
free of any magnetic element. The effect of e.g. the earth's field
if normal to the full edge electron trajectory will shift the impact point
25 m downstream of the magnet by approximately 12 $\mu$m.
Hence, the ambient field can be critical and
should be either shielded or measured by e.g. a fluxgate magnetometer.
Such an instrument allows to monitor any variation of the ambient magnetic field
with time and endpoint corrections should be applied. It is estimated that such a field
has to be known with a relative accuracy of (better than) 10\% 
to ensure a tolerable contribution of $\lsim$1.5 $\mu$m
to the overall $X_{edge}$ uncertainty\footnote{ For an infrared (green)
laser this error has to be correspondingly smaller since the edge electron energy
is reduced from 172.6 to 45.8 (25.6) GeV.}.

\vspace{3mm}
\noindent                  
Adding all uncertainties together, the total error of $X_{edge}$ can be expected to be
close to 5.6 or 9.3 $\mu$m (12.2 or 15.2 $\mu$m) if an infrared (green) laser is used
in the spectrometer. The dominating fraction of the error comes from
statistics so that larger data samples would decrease these uncertainties.
In general, all estimated uncertainties are very close to or less than the errors
anticipated in Sect's.3.2 and 3.3.

\vspace{3mm}                                                     
\noindent                  
If method B will be realized, precise position of the
unscattered beam, $X_{beam}$, at the detector plane is also required.
Cavity beam position monitors with single-bunch resolution
of few hundred (or less) nanometers are best suited.
To be conservative, we assume an error for  $X_{beam}$ of 1 $\mu$m 
which has to be added in quadrature with the uncertainty from possible charged particle
background expected for one of the proposed spectrometer locations (Sect.3.12).
In the worst case, the total uncertainty of $X_{beam}$ 
results to $\sim$1.2 $\mu$m which is well within the requirement.

\vspace{3mm}                                                     
\noindent                  
In addition,
if the B-field integrals for the endpoint and beam electrons are different,
${(\int B dl)}_{edge} \neq {(\int B dl)}_{beam}$, 
the expression for the beam energy (\ref{equ:ratio}) must be rewritten as
$$
E_{b} \propto
\frac{R(X_{edge}-X_{\gamma})-(X_{beam}-X_{\gamma})}{(X_{beam}-X_{\gamma})}~,
$$
with
$$
R = \frac{{(\int B dl)}_{beam}}{{(\int B dl)}_{edge}}~.
$$
Hence, the error for the beam energy as a function 
of the relative uncertainty of R is
$$
\frac{\Delta
E_{b}}{E_{b}}=\frac{(X_{edge}-X_{\gamma})}{(X_{edge}-X_{beam})}\frac{\Delta R}{R}~,
$$
where the approximation $R \approx 1$ has been implied. If the corresponding particle
positions 25 m downstream of the spectrometer magnet are taken into account,
the ratio $\frac{(X_{edge}-X_{\gamma})}{(X_{edge}-X_{beam})}$ = 1.2 (1.1)
for the infrared (green) laser. This means 
that for $\frac{\Delta R}{R} \simeq 5\cdot10^{-5}$ or better
and any value of $R$ different from 1, eq.(\ref{equ:ratio}) is needed to be modified
as indicated above. If $R$ equals 1 (within few times $10^{-5}$)
no  correction has to be applied.
With today's common B-field and $\int Bdl$ measurement techniques such
precision for $R$ can be achieved without too much efforts.

\vspace{3mm}                                                     
\noindent                  
Basically, whatever will be the final choice for the detector
more elaborated simulation studies are mandatory.
In particular, physics processes in the sensor material, basic
parameters of the associated electronics and backgrounds need to be taken
into account. Such details may affect the edge position and its shape
and could limit the performance of the spectrometer. Studies of this kind
are, however, beyond the scope of this paper.

\vspace{3mm}
\noindent                  
The idea to pulse the laser on every ILC bunch may be diluted for background studies.
If e.g. the laser is pulsed on nine out of ten bunches, every 10th pulse can be used
for background studies. Also, when the laser
fires without beams, checks of, for example, possible  pick-ups 
in the electronics can be performed.

\vspace{3mm}                                                     
\noindent                  
The experience at SLC and LEP proved that independent measurements of the
beam energy are important. The canonical method to measure $E_b$ upstream of the
$e^+ e^-$-IP consists of the BPM-based spectrometer \cite{ILC_spec}.
Both the Compton and the BPM-based spectrometers
are designed to provide an absolute measurement of the beam energy with
a relative accuracy of $10^{-4}$. Cross-calibration of the spectrometers
would provide an important and valuable control of their systematic errors.
Also, energy measurements at the $Z$-pole would provide a unique possibility
for an early calibration in a well understood physics regime.
Although $Z$-pole calibration measurements are not part of the current
ILC baseline design \cite{RDR} it is argued \cite{Exe_summary} 
that the baseline should be modified to include such reference.
In addition, physics reference channels, such as 
$e^+ e^- \rightarrow \mu^+ \mu^- \gamma$
where the muons are resonant with the known $Z$-mass, are foreseen to provide
valuable checks of the collision energy scale, but only long after the data
were recorded.

\subsection {\boldmath Suitable Spectrometer Locations \unboldmath}

Although the today's beam delivery system \cite{BDS_present} will be further developed
within the next years, basic properties are not expected to be modified.
We propose three alternatives for possible locations of the Compton spectrometer
within the BDS, while keeping major design parameters of the spectrometer unaltered.
Each of the proposals has pros and cons and the spectrometer viability
requires sometimes, depending on the location,
slight modifications of the present BDS.
An overall view of the BDS is shown in Fig.~\ref{fig:BDS_overall_view},
where also potential locations for the Compton spectrometer are indicated.
\begin{figure}[ht]
\begin{center}
\vspace{3mm}
\includegraphics[height=9.0cm,width=18.0cm]{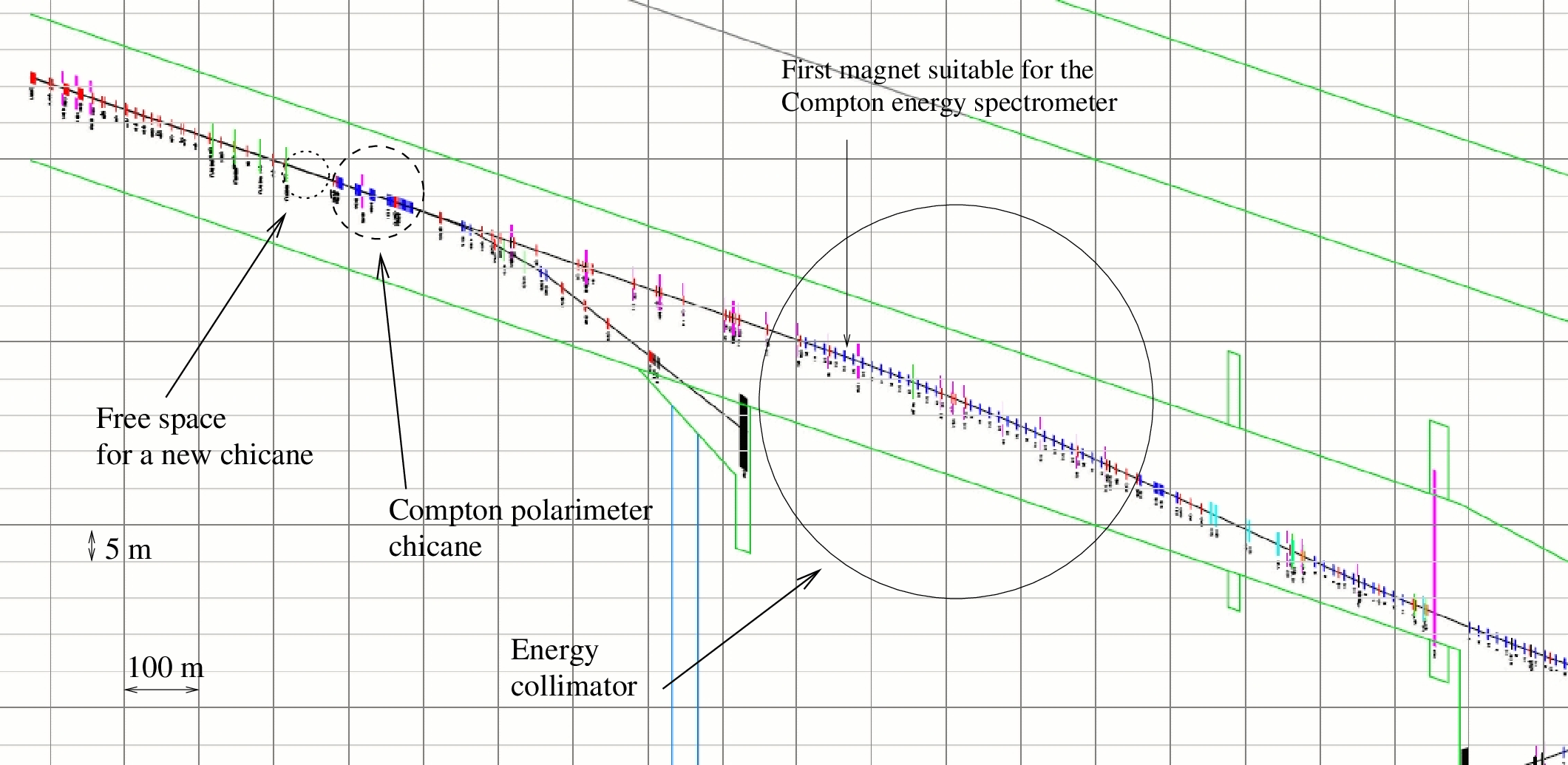}
\caption{ \label{fig:BDS_overall_view}General view of the beam delivery system with
  possible locations of the Compton spectrometer.}
\end{center}
\end{figure}

\vspace{3mm}
\noindent
Common to all alternatives is the demand to locate the spectrometer upstream
of the energy collimation system\footnote{The energy collimation system 
performs efficient removal
of halo particles which lie outside the acceptable range of energy spread.}
to avoid significant muon background excess relative to the rate
from normal collimation losses.

\vspace{3mm}
\noindent
The straight-forward approach suggests to locate the spectrometer in
an existing free-space region of the BDS. The amount of space needed
is determined by the drift distance of at least 25 m to the detector system,
the length of the magnet of 3 m and the 6 m long vacuum chamber 
upstream of the dipole in which the Compton IP is contained.
The sum of these components of 35 m has to be enlarged by additional space
to accommodate two ancillary magnets
with corresponding drift regions to compensate the bend of the spectrometer magnet.
Hence, in total 60-70 m free space is needed\footnote{It would be 
very helpful if in any new BDS design a suitable spectrometer dipole is
a priori foreseen as a standard BDS magnet. This would substantially relax
space (and other) requirements for the Compton spectrometer.}.
Far upstream of the physics $e^+ e^-$-IP such free space of some 65 m
exists, see Fig.~\ref{fig:BDS_overall_view}.
The transverse dimensions of the beam at the Compton IP
of about 20 $\mu$m versus 2 $\mu$m
perfectly match the expected spot size of the laser.
Additional muon background generated by backscattered electron interactions further downstream
was estimated and would only increase the muon rate by a small amount \cite{Keller},
independent of the laser wavelength.
This suggestion locates the spectrometer on a direct line of sight to the main linac,
which means that backgrounds in this region are likely to be significant.
In particular, charged particles off in energy may affect the position of the beamline
in the cavity BPM. Cavity beam position monitors measure the centroid
of the particle's charge distribution and, hence, particles with less energy than $E_b$
are stronger deflected by the spectrometer magnet and could shift the measured beam position.
Halo and tail generation estimates based on simulation \cite{Burkhardt}
reveal that at the exit of the linac the beam profile is superimposed
by a symmetric halo extending to about $\pm$300 (50) $\mu$m in x(y)-direction\footnote{ The
electron bunches at the end of the linac were found to be very well described 
by pure Gaussian distributions with horizontal and vertical dimensions 
of $\sigma_x$ = 39.0 $\mu$m and $\sigma_y$ = 1.80 $\mu$m, respectively.}. The fraction of
particles off in energy was estimated to be few times $10^{-5}$ with a broad energy spectrum
that sharply peaks very close to the nominal beam energy. 
A simple tracking procedure up to the BPM installed 25 m from the spectrometer magnet
indicates a shift of $X_{beam}$ of 0.65 $\mu$m which we consider
of not being catastrophic. It is also estimated that this background does not affect
the Compton endpoint position in a significant manner.
The synchrotron photons from the quadrupole fields 
within the linac and the beginning of the BDS have a $\sim10^2$ times lower critical energy
\cite{Burkhardt, Deacon} than those from the spectrometer magnet
and are considered to be of no serious issue.
However, due to the uncertainties in the charged particle background simulation 
it is favorable to locate the spectrometer after a protective bend, 
so that the beam position will be much less impacted.

\vspace{3mm}
\noindent
A major constraint for the design of the Compton spectrometer is the
synchrotron radiation emittance dilution from the additional spectrometer magnets.
Employing the magnet as discussed in Sect.3.3
and similar ancillary magnets, an emittance growth of about 0.5\%
at 250 GeV is expected, which might be considered as acceptable.
Since the emittance scales with the sixth power of the beam energy,
further studies have to reveal whether emittance dilution
at 500 GeV beam energy can be tolerated.

\vspace{3mm}
\noindent
A second option for the spectrometer location consists
in employing one or more magnets
of the present BDS as the Compton spectrometer dipole. 
Since, however, an individual magnet
with desirable properties does not exist, we suggest
to combine several consecutive bending magnets. 
At the beginning of the energy collimation section directly after the first
magnet, see Fig.~\ref{fig:BDS_overall_view},
such magnets\footnote{Each magnet has
a B-field of 291.68 Gauss, a length of 2.4 m and space in between of 12.3 m.}
might be combined to provide
the desired bending power. In particular, 
if the laser IP is located about 3 m upstream of magnet 1,
a combination of the following six magnets (magnet 1, ..., magnet 6)
provides sufficient particle separation.
For example, separation between the backscattered $\gamma$-rays 
and the beamline results to 18 mm after passing magnet 6,
while the distance of the beam to the edge electrons
is 26 mm for a $CO_2$ and 98 mm for an 1.165 eV infrared laser.
Thus, by locating the detector system close to magnet 7
convenient measurements of the positions of the Compton recoil particles
and the beamline can be performed. The transverse beam profile at the laser IP
is sufficiently small so that the beam is completely covered by the laser spot.
Additional muon background from Compton electrons
is tolerable since many of these electrons will hit either closely located magnets
or spoilers of the energy collimation system \cite{Keller}.
This option also allows to insert the laser light into the vacuum pipe downstream
of magnet 1 which makes strict head-on collisions with the beam
possible.

\vspace{3mm}
\noindent
However, the horizontal aperture of the magnets has to be continuously increased
towards the bending direction so that the edge electrons pass
in B-fields with properties as demanded. In particular,
at the exit of magnet 6 the vacuum chamber has to have a horizontal aperture
of 115 mm for infrared laser light scattering.
Furthermore, if method A is employed for beam energy measurements,
the B-field integral over all six magnets has to be known
with 20 ppm precision. Or, for method B, the three-point measurement approach,
sufficient field uniformity uncertainty
within the bending plane up to x = 115 mm
has to be ensured. Whatever method for $E_b$ determination will be realized,
the demands for the magnet system are challenging.
This alternative for the spectrometer location
is advantageous since no additional magnets are needed and, thereby,
further growth of the  beam emittance is a priori avoided.

\vspace{3mm}
\noindent
The third alternative for a location of the Compton spectrometer consists in employing
the magnetic chicane proposed for high energy polarization measurements
\cite{Schueler}. In particular, the four-magnet
polarimeter chicane with the laser IP in the mid-point 
is supposed to be supplemented by the position sensitive detector system,
which should be located upstream but close to the fourth magnet.
Also, some dedicated adjustments of space, laser and detector conditions
are needed to ensure polarization and beam energy
measurements simultaneously with precisions as anticipated.
However, the present baseline polarimeter design aims
to operate the chicane with constant field settings over a large range
of beam energies, while the Compton based beam energy spectrometer intends
to adjust the B-field to a constant bending power of e.g. 1 mrad.
Whether both approaches can be merged to a common proposal requires detailed studies.
Possible muon background increase from Compton electron interactions
was estimated to be tolerable \cite{Keller}. It is also obvious that
additional dilution of the beam emittance
caused by such  $E_b$ measurements is ruled out.

\section {\boldmath Summary and Conclusion \unboldmath}

A novel, non-invasive method of measuring the incident beam energy,
$E_b$, at the International $e^+ e^-$ Linear Collider is proposed.
Laser light scatters head-on off ILC bunches and generates
Compton electrons and photons. After the Compton IP,
the scattered particles as well as the non-interacting beam electrons
(99.9995\% of them) pass through a dipole magnet
so that further downstream access to each particle type is possible.
$E_b$ measurements can be performed continuously on a bunch-by-bunch basis
while the electron and positron beams are in collision.

\vspace{3mm}
\noindent
One approach to infer $E_b$, method A,  relies on the beam energy dependence
of the momentum of the scattered electrons at the kinematic endpoint,
the edge energy. Combining the B-field integral of the dipole
with the position of the edge electrons relative to the incident beam 
provides the energy of the edge electrons and, thereby, $E_b$. However,
integrated field uncertainties close to $2 \cdot 10^{-5}$ and
position measurements with an accuracy of at least few micrometers 
are required to achieve the anticipated value of $10^{-4}$.
The last demand is very challenging and is mainly the reason to follow a different
approach, called method B. By measuring three particle positions, 
the position of the Compton
scattered $\gamma$-rays, $X_{\gamma}$, the position of the edge electrons, $X_{edge}$,
and that of the beam, $X_{beam}$, downstream of the spectrometer magnet
allows to deduce $E_b$ with precisions of $10^{-4}$ or better.
Such precisions, however, require to measure
the distance $X_{edge}-X_{beam}$ with an accuracy of about
ten micrometer and $X_{\gamma}$ with 1-2 $\mu$m uncertainty.
Both requirements seem to be achievable,
because the distance $X_{edge}-X_{beam}$ is, in particular, beam energy independent
and accumulation over many bunches decreases its statistical error
substantially.

\vspace{3mm}
\noindent
It has been shown that effects beyond the Born approximation in the laser crossing region
are very small. They only lead to a negligible shift
of the edge electron position, $X_{edge}$.

\vspace{3mm}
\noindent
Geometrical constraints and acceptable emittance dilution
of beam particles when passing the dipole magnet
require a spectrometer length of at least 30 m.
The geometrical constraints in conjunction with free space options
within the beam delivery system preclude the usage of a $CO_2$ laser,
while an infrared (with $E_\lambda$ = 1.165 eV) or a green laser
(with $E_\lambda$ = 2.33 eV) are both suitable.
To achieve e.g. $10^6$ Compton events per bunch crossing,
a pulse power of 30 mJ, respectively, 24 mJ with a pattern that matches
the pulse and bunch structure at the ILC is needed. 
Such lasers are presently commercially not available,
but R\&D is ongoing within the ILC and other communities.

\vspace{3mm}
\noindent
For particle position measurements, detectors with high spatial resolution
have to be pursued. As a promising option for edge electron and $\gamma$-ray
center-of-gravity measurements quartz fiber detectors are suggested because they
are very radiation hard and ultrafast. An alternative to the $X_{\gamma}$
quartz fiber detector (in conjunction with e.g. a
16 radiation length tungsten converter) consists in measuring the edge position
of synchrotron radiation light generated by beam particles when passing
the spectrometer magnet, as discussed in \cite{SR_paper}.
A device based on gas amplification was considered in more details and simulations
demonstrated its reliability for our purpose. The position of the
non-interacting beam particles needs to be known with micrometer accuracy
which can be relative easily achieved by modern cavity beam position monitors.

\vspace{3mm}
\noindent
The method proposed to perform energy measurements of the incident beam
at the ILC is thought to be a complementary and cross-check approach
to the canonical concept of a BPM based energy spectrometer. Both methods
intend to achieve a precision of $10^{-4}$ on a bunch-to-bunch basis.
The method studied in this paper seems to accomplish the objective,
but more detailed studies are mandatory 
and a prove-of-principle experiment \cite{Exp_prop}
should to be performed to test the three-position measurement approach.

\subsection*{Acknowledgment}
 
We would like to thank A. Magaryan for discussions, where the idea 
of the present study originated. We would also like to thank
K. Hiller, R. Makarov, N. Morozov and E. Syresin
for helpful discussions and the sommerstudents H. Paukkunen, J. Lange
and D.A. St\"uken for various contributions. We are grateful to L. Keller 
for performing muon background calculations, A. Latina for bunch simulations
within the ILC, H. Burkhard for halo and tail information, G. Klemz for
discussions on laser system aspects and G.A. Blair for reading the manuscript.


\end {document}